\documentclass[%
 reprint,
 amsmath,amssymb,
 aps,
]{revtex4-2}

\usepackage{graphicx}
\usepackage{dcolumn}
\usepackage{bm}

\usepackage{hyperref}
\usepackage{algorithm}
\usepackage{multirow}
\usepackage{algorithmic}
\usepackage[caption=false,font=footnotesize]{subfig}
\usepackage{makecell}
\usepackage[dvipsnames]{xcolor}
\usepackage{float}

\begin{document}

\preprint{APS/123-QED}

\title{Accelerating Particle-in-Cell Kinetic Plasma Simulations via Reduced-Order Modeling of Space-Charge Dynamics using Dynamic Mode Decomposition}

\author{Indranil Nayak}
 \email{nayak.77@osu.edu}
\author{Fernando L. Teixeira}
\affiliation{%
ElectroScience Laboratory and Department of Electrical and Computer Engineering, The Ohio State University, Columbus, Ohio, USA
}%

\author{Dong-Yeop Na}
\affiliation{Department of Electrical Engineering, Pohang University of Science and Technology (POSTECH), Pohang, South Korea.
}%

\author{Mrinal Kumar}
\affiliation{LADDCS, Department of Mechanical and Aerospace Engineering, Columbus, Ohio, USA.
}%

\author{Yuri A. Omelchenko}
\affiliation{Trinum Research Inc., San Diego, California, USA.
}

\begin{abstract}
We present a data-driven reduced-order modeling of the space-charge dynamics for electromagnetic particle-in-cell (EMPIC) plasma simulations based on dynamic mode decomposition (DMD). The dynamics of the charged particles in kinetic plasma simulations such as EMPIC is manifested through the plasma current density defined {along the edges} of the spatial mesh. We showcase the efficacy of DMD in modeling the time evolution of current density through a low-dimensional feature space. Not only do such DMD-based predictive reduced-order models help accelerate EMPIC simulations, they also have the potential to facilitate investigative analysis and control applications. We demonstrate the proposed DMD-EMPIC scheme for reduced-order modeling of current density, and speed-up in EMPIC simulations involving electron beam under the influence of magnetic field, virtual cathode oscillations, and backward wave oscillator.

\end{abstract}

\maketitle



\section{Introduction}

Kinetic plasma simulations enjoy a broad range of applications ranging from the modeling of high-power microwave sources, directed energy devices, particle accelerators, terahertz devices, etc. They are also used to improve our understanding of ionospheric phenomena, magnetosphere regions, and astrophysical events~\cite{gold1997review,booske2008plasma,benford2015high,PhysRevLett.114.175002}. Historically, electromagnetic particle-in-cell (EMPIC) algorithms have been a popular choice for simulating collisionless kinetic plasmas. {Compared to magnetohydrodynamics (MHD) simulations, Particle-in-cell (PIC) algorithms can better capture intricate wave-particle interactions including electron bunching, kinetic instabilities, Landau damping, microscopic turbulence, space-charge effects, etc., ~\cite{birdsall2004plasma,eppley1988use,PhysRevE.94.053305,PhysRevE.103.L051201,DAVIDSON20151063,PhysRevE.104.055311}.} However, higher physical accuracy of the PIC simulations comes at the cost of large computational resources. A very large number of computational (super)particles are needed for an adequate sampling of the phase-space of the electrons/ions. This is one of the primary challenges preventing speeding-up of EMPIC simulations~\cite{birdsall2004plasma}. In an EMPIC simulation, the position and velocity of each superparticle must be updated individually at each time-step, covering the entire time-window of interest. 

The high-dimensionality and large computational cost of EMPIC simulations serve as motivation for developing reduced-order models capable of capturing the underlying plasma dynamics through a low-dimensional feature space. Such small sets of coherent spatio-temporal features have been shown to be effective in emulating the underlying physics in several plasma simulations \cite{pandya2016low, van2014use, byrne2017study, kaptanoglu2021physics,sasaki2019using,NAYAK2021110671}. The projection-based reduced-order methodologies such as proper orthogonal decomposition (POD) and, dynamic mode decomposition (DMD) in particular have been recently employed for modeling the field evolution in EMPIC simulations with success~\cite{julio2019,NAYAK2021110671,nayak2021dynamic,nayak2020dynamic}. However, these methods do not address the primary computational bottleneck due to the large number of particles. Authors in \cite{hesthaven2023adaptive} presented a discrete empirical interpolation (DEIM) method for reducing the computational burden of particle to mesh projections in geometric particle-in-cell (GEMPIC)~\cite{kraus2017gempic} simulations. In \cite{alves2022data}, the authors took a different approach based on sparse regression (SR) to discover the underlying partial differential equations (PDEs) from PIC simulation data. While \cite{hesthaven2023adaptive} shows the effectiveness of DMD-DEIM method for $1D-1V$ Vlasov-Poisson problems in parametric setting, it has not addressed how to provide an explicit analytical time-update scheme for particle dependent quantities such as current density which can be crucial for prediction and diagnostic purposes. The SR approach in \cite{alves2022data} strives to infer the corresponding MHD equations in integral form from the PIC time-series data, but is limited by the choice of PDE candidate terms and the cost of integration over a large volume. In this work, the computational bottleneck of a large number of particles is addressed by time-domain reduced-order modeling of the current density using dynamic mode decomposition (DMD). In a typical EMPIC setting, the charged particle dynamics manifests through temporal variation of charge and current densities defined on the spatial mesh. In the EMPIC algorithm, current density is employed for the time update of electromagnetic fields according to the Maxwell's equations. Updated fields in turn influence the motion of the charged particles, and dictate the time variation of current density. These steps are executed in a cyclic fashion (bottom left of Fig. \ref{fig:DMD-EMPIC}) for each time-step of EMPIC. In essence, if the time evolution of current density can be modeled independently of the particles, the EMPIC steps involving the particles could be completely avoided which would result in a drastic reduction in computation time.\par

Initially proposed in~\cite{schmid2010dynamic}, DMD is a data-driven method for low-dimensional surrogate modeling of high-dimensional complex dynamical systems.  Since its inception, DMD has been applied successfully to a vast array of problems including fluid-based non-linear plasma models and magnetized plasma experiments~\cite{taylor2018dynamic,kaptanoglu2020characterizing,sasaki2019using}. The ability of DMD to extract underlying dominant spatiotemporal features from self-field data of EMPIC simulation, and the effect of DMD extrapolated fields on the particle dynamics were recently investigated in~\cite{NAYAK2021110671}. In \cite{hesthaven2023adaptive}, DMD was used for approximating the electric field potential. However, DMD modeling of space-charge dynamics, especially the time evolution of {\it plasma current density} in EMPIC simulations, have not yet been explored. Since there is no analytical model available for the time evolution of current density in EMPIC simulations, a data-driven approach becomes crucial. Data-driven modeling of current density is particularly challenging due to the associated nonlinearity, non-smooth time variation pertaining to particle noise, and high-dimensionality (large number of mesh elements). The effectiveness of DMD in modeling nonlinear dynamics can be attributed to its close relation to Koopman operator theory~\cite{rowley2009spectral}. In particular, initial findings have suggested that Koopman autoencoders have good potential for reduced order modeling of currents as well~\cite{nayak2021koopman,nayak2023koopman}. However, Koopman autoencoders display the usual limitations of neural network models, namely lack of interpretability (black-box nature) and intricate training process. The use of interpretable Koopman-based data-driven reduced-order models such as DMD helps to overcome these challenges. 

{In this work, we demonstrate for the first time the application of time-series prediction in current density forecasting to accelerate EMPIC simulations.} The main contributions of the present work can be summarized as follows:
\begin{enumerate}
    \item To the best of the authors' knowledge, this work is the first instance of constructing an interpretable, reduced-order model for the space-charge dynamics in EMPIC simulations. This is achieved by DMD modeling of the current density which is essentially the manifestation of charged particle dynamics in the EMPIC simulations. While our main goal is to accelerate the EMPIC simulations, such reduced-order modeling of space-charge dynamics also helps analyze and diagnose the problem at hand. Note that using DMD-based linear reduced-order models (ROMs) of the inherently nonlinear plasma dynamics also opens the doors for leveraging control theoretic tools that already exist for linear systems. {The DMD modes and frequencies can help in analyzing the efficiency loss due to harmonic generation in high-power microwave devices. By analyzing the spatial patterns and growth rates of the modes, DMD can help predict areas that are prone to decay or damage due to high energy densities, heating, or other factors. DMD can also identify unstable modes and flaws in device design or the numerical solver itself. Furthermore, understanding the dominant modes and their characteristics can help in optimize the design parameters of such devices.}
    \item A novel DMD-EMPIC algorithm (Fig. \ref{fig:DMD-EMPIC}) is presented to accelerate the EMPIC simulations showing post-transient behavior, i.e. either steady-state, equilibrium, or any type of periodic behavior. The time-domain DMD model of the current density implements rapid prediction of the current density at any time instant, and thus eliminates the need for EMPIC stages involving particles. The DMD-EMPIC strategy utilizes the on-the-fly algorithm developed in \cite{nayak2023fly} to detect the end of transience in real-time. It then replaces the computationally expensive \textit{gather, pusher} and \textit{scatter} with DMD predicted current beyond that point in time. {It is important to highlight that the key distinction between this work and previous research, as outlined in \cite{NAYAK2021110671}, lies in the application of the DMD model to current density for replacing EMPIC stages involving particles. This approach not only facilitates the prediction of future current density but also enables fast and accurate forecasting of self-field values, adhering precisely to the discrete Maxwell's equations. Moreover, this study employs a more versatile `on-the-fly' algorithm, distinguishing it from the method utilized in \cite{NAYAK2021110671}. } DMD-EMPIC has the potential to significantly expedite the EMPIC simulations for plasma systems showing long oscillations (e.g. limit-cycle behavior).
\end{enumerate}

This paper is organized as follows. Section \ref{sec:theo_back} provides a quick overview of the EMPIC algorithm used to generate the high-fidelity data, and the DMD algorithm which is trained on that high-fidelity data. Section \ref{sec:dmd-empic} introduces the DMD-EMPIC algorithm for accelerating EMPIC simulations. Section \ref{sec:results} provides a series of results showcasing the effectiveness of DMD-EMPIC for accelerating EMPIC simulations. An analysis of the computational gain for the examples considered, and of the computational complexity in general, is provided in Section \ref{sec:comp_gain}. Finally, Section \ref{sec:conclusion} summarizes the main takeaways.


\section{Theoretical Background}\label{sec:theo_back}

\subsection{EMPIC Algorithm}\label{sec:empic}
\begin{figure*} [t]
    \centering
      \includegraphics[width=0.95\linewidth]{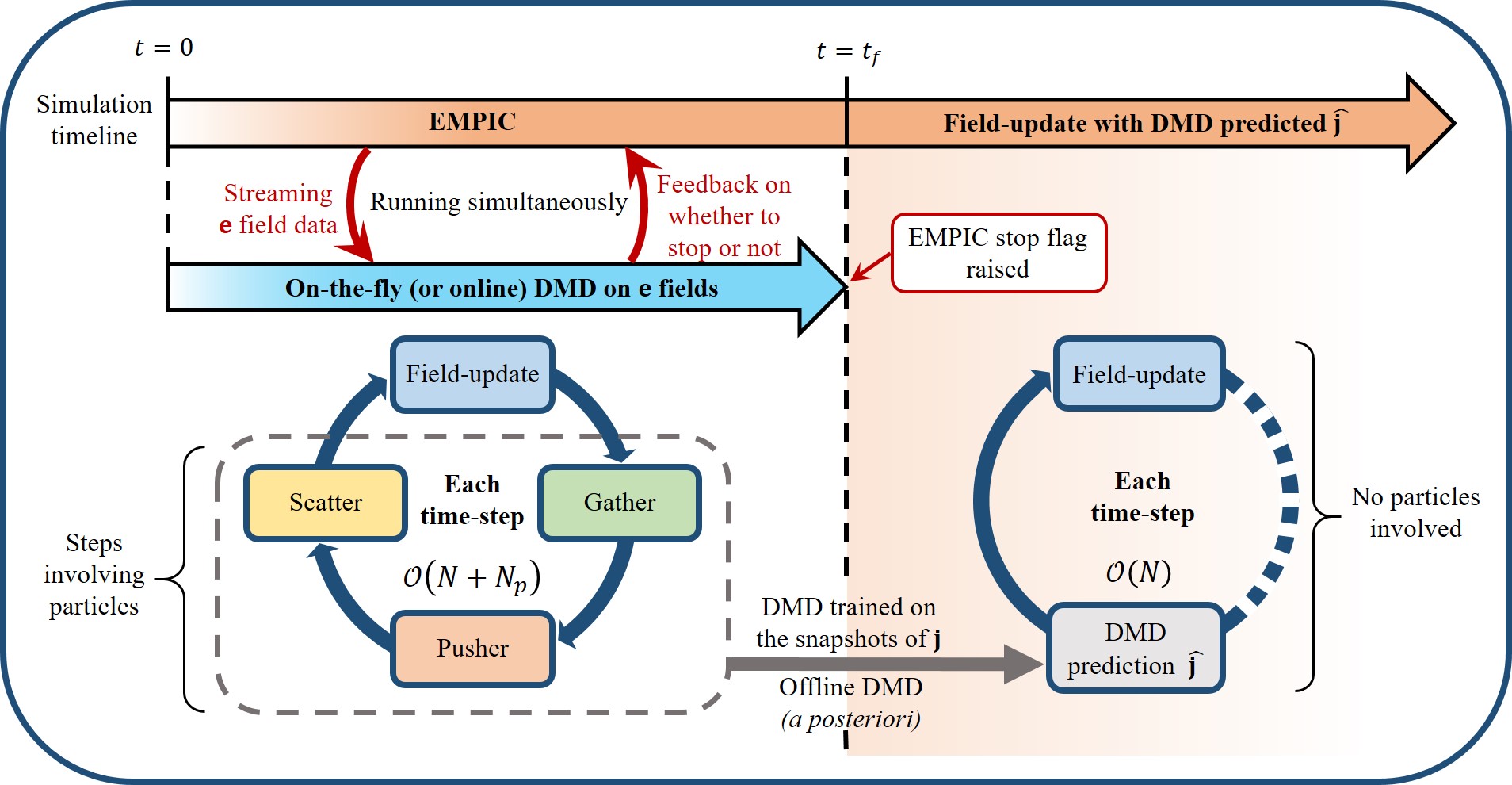}
  \caption{\small{DMD-EMPIC algorithm for accelerating EMPIC simulations. Note that for $t>t_f$, the stages are not exactly cyclic since fields no longer have any effect on the DMD predicted current density (illustrated by the broken line). }
   \label{fig:DMD-EMPIC}}
\end{figure*}

The electromagnetic particle-in-cell (EMPIC) algorithm \cite{moon2015exact,kim2011parallel,EVSTATIEV2013376,doi:10.1063/1.4976849,kraus2017gempic,jianyuan2018structure} primarily consists of four steps i) \textit{field-update}, ii) \textit{gather}, iii) \textit{pusher}, and iv) \textit{scatter} (bottom left of Fig. \ref{fig:DMD-EMPIC}) which are executed in a cyclic fashion at each time-step. For completeness, we will briefly discuss each step in the context of a fully kinetic finite-element-based charge-conserving EMPIC algorithm \cite{kim2011parallel,moon2015exact,na2016local}. We will also discuss the computational hurdle that a large number of particles poses in EMPIC simulations. \par

\textbf{\textit{Field-update}}\label{sec:field_update}: The field-update stage deals with the time-update of the electric and magnetic fields by solving Maxwell's equations. Here, the field-update is obtained on a finite element mesh where the fields are represented as (discrete) differential forms~\cite{flanders1989, teixeira1999lattice, kotiuga2004, he2007differential, deschamps1981electromagnetics, teixeira2013differential}. The electric field 1-form $\mathcal{E}\left(t, \mathbf{r}\right)$ and magnetic flux density 2-form $\mathcal{B}\left(t, \mathbf{r}\right)$ are expanded as
\begin{subequations}
    \begin{align} 
\mathcal{E}\left(t, \mathbf{r}\right) &= \sum_{i=1}^{N_{1}} e_{i}\left(t\right) w^{(1)}_{i} (\mathbf{r}), \label{eq:EDoF}\\
\mathcal{B}\left(t, \mathbf{r}\right) &= \sum_{i=1}^{N_{2}} b_{i}\left(t\right) w^{(2)}_{i} (\mathbf{r})\label{eq:BDoF}, 
    \end{align}  
\end{subequations}
where $e_i(t)$ and $b_i(t)$ are their temporal degrees of freedom (DoF), and $w^{(1)}_{i} (\mathbf{r})$ and $w^{(2)}_{j} (\mathbf{r})$ represent Whitney 1-forms (edge-based) and Whitney 2-forms (facet-based) respectively~\cite{moon2015exact, teixeira2014lattice,na2016local}. $N_1$ and $N_2$ respectively denote the number of edges and elements (faces) in the mesh. From \eqref{eq:EDoF} and \eqref{eq:BDoF} and using a Galerkin-type finite element discretization, Maxwell's curl equations in discrete form suitable for time stepping can be written as\cite{moon2015exact, na2016local}:

\begin{subequations}\label{eq:FDTD_upd}
\begin{align}
    \mathbf{b}^{(n + \frac{1}{2})} &= \mathbf{b}^{(n - \frac{1}{2})} - \Delta t \mathbf{C} \cdot \mathbf{e}^{(n)} \label{eq:bupd}\\
     \mathbf{e}^{(n + 1)} &= \mathbf{e}^{(n)} + \Delta t  \left[\star_{\epsilon}\right]^{- 1}\cdot {\mathbf{C}}^{T}\cdot\left[\star_{\mu^{- 1}}\right] \cdot \mathbf{b}^{(n+\frac{1}{2})} - \nonumber\\  &~~~~~~~~~~~~~~~~~~~~~~~~~~~~~~~~\Delta t \left[\star_{\epsilon}\right]^{- 1}\cdot\mathbf{j}^{(n+\frac{1}{2})},\label{eq:eupd}
\end{align}
\end{subequations}
where $\Delta t$ denotes the time step, $\mathbf{e}=[e_1~e_2~\ldots~e_{N_1}]^{\text{T}}$ is defined on integer time-steps, and $\mathbf{b}=[b_1~b_2~\ldots~b_{N_2}]^{\text{T}}$ and current density $\mathbf{j}=[j_1~j_2~\ldots~j_{N_1}]^{\text{T}}$ are defined on half-integer time-steps. More details on $\mathbf{j}$ are provided in the description of the scatter stage ahead.
 The incidence matrices
$\mathbf{C}$ and $\mathbf{C}^T$ encode the exterior derivative (or, equivalently, the curl operator distilled from the metric structure) on the primal and dual meshes, respectively ~\cite{na2016local,teixeira2014lattice}. The elements of $\mathbf{C}$ are in the set $\{-1,0,1\}$ as they contain information about adjacency and relative orientation among mesh elements.  The Hodge matrices $\left[\star_{\epsilon}\right]$ and $\left[\star_{\mu^{-1}}\right]$ represent generalized constitutive relations incorporating the metric information from the finite element mesh. Further details on the discrete field update equations can be found in~\citep{na2016local,moon2015exact,kim2011parallel}.\par

\begin{figure*} [t]
    \centering
      \includegraphics[width=0.95\linewidth]{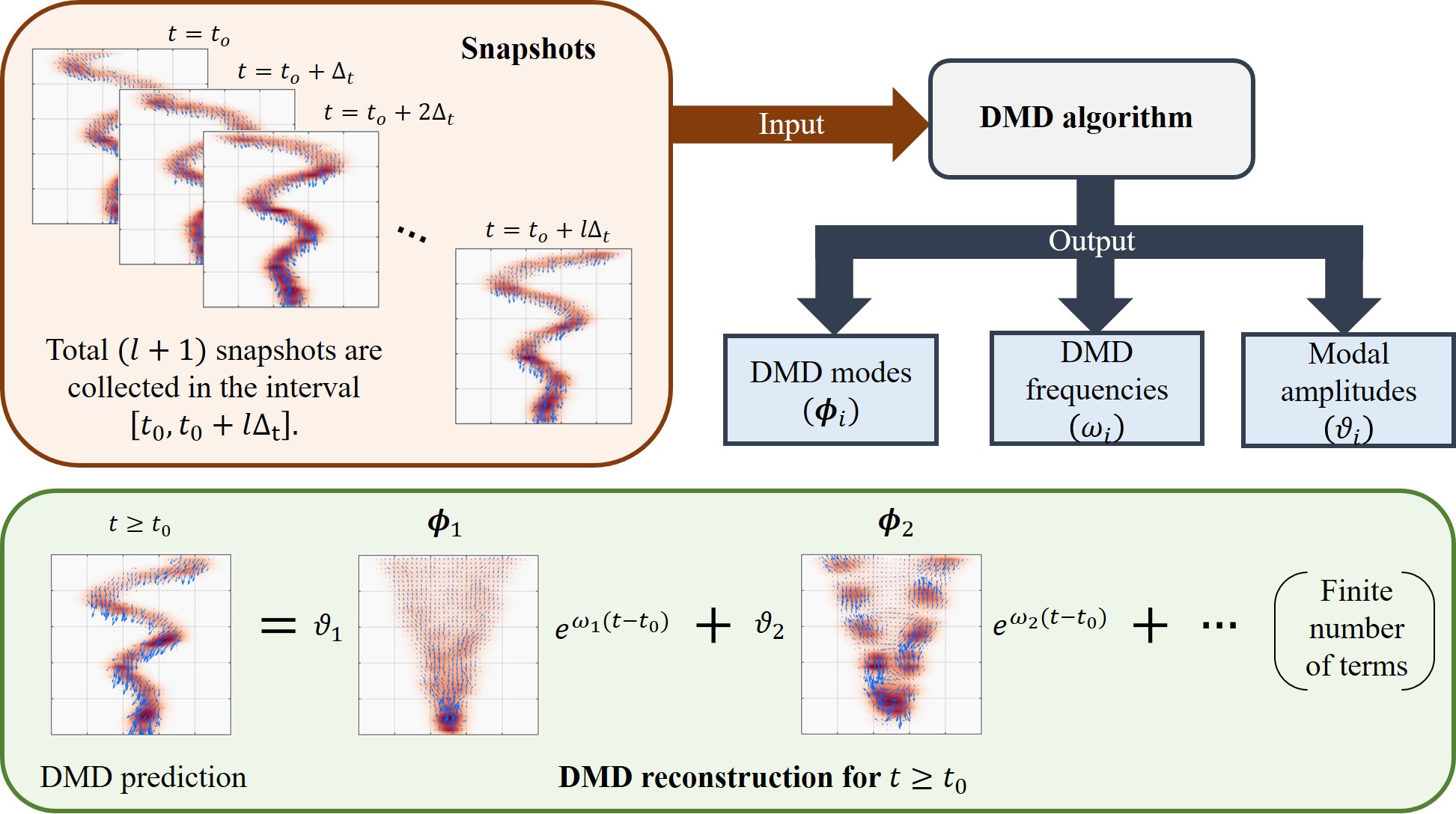}
  \caption{\small{Schematic illustration of the DMD algorithm for current density modeling of an electron beam. }
   \label{fig:DMD}}
\end{figure*}

\textbf{\textit{Gather}}: In the gather step, the fields are calculated at the particle position $\mathbf{r}_p$ using the interpolatory Whitney functions:  
\begin{subequations}
    \begin{align} 
\mathcal{E}\left(n\Delta t, \mathbf{r}_p\right) &= \mathcal{E}_p^{(n)}=\sum_{i=1}^{N_{1}} e_{i}^{(n)} w^{(1)}_{i} (\mathbf{r}_p), \label{eq:Egather}\\
\mathcal{B}((n+\frac{1}{2})\Delta t, \mathbf{r}_p) &= \mathcal{B}_p^{(n+\frac{1}{2})}=\sum_{i=1}^{N_{2}} b_{i}^{(n+\frac{1}{2})} w^{(2)}_{i} (\mathbf{r}_p)\label{eq:Bgather}. 
    \end{align}  
\end{subequations}

\textbf{\textit{Pusher}}:
Once the fields are interpolated to each charged particle location, the particle position and velocity are updated using Newton's laws of motion and Lorentz force equations. The particle position $\mathbf{r}_p$ and (non-relativistic) velocity $\mathbf{v}_p$ can be obtained as follows~\cite{moon2015exact}
\begin{subequations}
\begin{align}
	\mathbf{r}_p^{n+1} &=\mathbf{r}_p^{n} + \Delta t \mathbf{v}_p^{n+\frac{1}{2}}, \\
	\mathbf{v}_p^{n+\frac{1}{2}} &= N^{-1} \cdot N^T \cdot \mathbf{v}_p^{n-\frac{1}{2}} + \frac{q_p \Delta t}{m_p} N^{-1} \cdot \mathcal{E}_p^{(n)},
\end{align}
\end{subequations}
where
\begin{equation}
N = \begin{bmatrix}
1 & - \frac{q_p \Delta t}{2 m_p} \mathcal{B}_{p,z}^{(n)} & \frac{q_p \Delta t}{2 m_p} \mathcal{B}_{p,y}^{(n)} \\
\frac{q_p \Delta t}{2 m_p} \mathcal{B}_{p,z}^{(n)} & 1 & - \frac{q_p \Delta t}{2 m_p} \mathcal{B}_{p,x}^{(n)} \\
-\frac{q_p \Delta t}{2 m_p} \mathcal{B}_{p,y}^{(n)} & \frac{q_p \Delta t}{2 m_p} \mathcal{B}_{p,x}^{(n)} & 1
\end{bmatrix},
\end{equation}
$\mathcal{B}_{p}^{(n)} = \frac{1}{2}\left(\mathcal{B}_{p}^{(n+\frac{1}{2})} + \mathcal{B}_{p}^{(n-\frac{1}{2})}\right)$, and $m_p$ and $q_p$ are respectively the mass and charge of $p\textsuperscript{th}$ particle. Relativistic motion  can also be incorporated, if required.

\textbf{\textit{Scatter}}:
The scatter step is essentially a particle-to-mesh mapping where the motion of each charged-particle during time-step $\Delta t$ is mapped to the edges of the mesh in order to obtain each element of the current density vector $\mathbf{j}$
\begin{equation}
	j_i = \sum\limits_{p = 1}^{N_p} \frac{q_p}{\Delta t} \int\limits_{\mathbf{r}_{p1}}^{\mathbf{r}_{p2}} w_i^{(1)} (\mathbf{r}_p) \cdot d\mathbf{r}_{p},
\end{equation}
where the integration is performed along the straight line trajectory of the $p$\textsuperscript{th} particle during one time step, from $\mathbf{r}_{p1}$ to $\mathbf{r}_{p2}$\cite{Note1}. The particular mapping ensures charge conservation at the discrete level, as shown in~\cite{moon2015exact}. Mathematically, the above equation can be understood as a Galerkin projection of the 2-form $\mathcal{J}$ (resulting by the collective movement of the charged particles) onto the Whitney 1-form $w_i^{(1)}$ associated to edge $i$ of the mesh.
The edge current density vector $\mathbf{j}$ thus obtained is subsequently used to update the field values in the next time-step according to \eqref{eq:FDTD_upd}.
\par

The \textit{gather, pusher} and \textit{scatter} steps need to be performed individually for each superparticle in the solution domain. As mentioned earlier, in order to accurately capture the phase space evolution, a large number of particles is often necessary, a typical number being on the order of $10^6$ or so.

\subsection{DMD Algorithm}

Let us consider a discrete-time dynamical system consisting of the state $\mathbf{x}\in\mathcal{M}$, where $\mathcal{M}$ is an $N$-dimensional manifold, and $F$ is a sequential flow map $F:\mathcal{M}\mapsto \mathcal{M}$ such that
\begin{align}
    \mathbf{x}^{(n+1)}=F(\mathbf{x}^{(n)}).
\end{align}
The superscript `$n$'  represents the discrete time index. In practical scenarios, $\mathcal{M}$ is typically represented by $\mathbb{R}^N$, $N$ being the number of mesh elements (nodes, edges, faces etc.) over which $\mathbf{x}$ is defined. It is important to note that in our context, $\mathbf{x}^{(n)}$ may represent
 $\mathbf{e}^{(n)}$, $\mathbf{j}^{(n+\frac{1}{2})}$, or any other relevant dynamic variable as introduced in the previous section. DMD collects time-snapshots of the state $\mathbf{x}$ and generates a set of spatial DMD modes $\bm{\phi}_i(\mathbf{r})$ where $\mathbf{r}$ is the position vector, with corresponding DMD frequencies $\omega_i$ (in general, complex valued), and modal amplitudes $\vartheta_i$ ($i=1,2,\ldots$). These extracted DMD features are used to (approximately) reconstruct the spatiotemporal behavior of the original dynamics~\cite{schmid2010dynamic,tu2013dynamic,kutz2016dynamic}.\par
The first stage of the DMD process is snapshot collection, i.e. sampling the state $\mathbf{x}$ at different time instants to build the training dataset. The snapshots are typically collected with uniform sampling interval from either high-fidelity simulations (i.e. EMPIC) or experimental data. The corresponding time window is referred to as the training window/region, or DMD harvesting window/region or simply DMD window. Let the DMD harvesting window consist of $(l+1)$ snapshots, spanning from $t_0=n_0\Delta t$ to $t_l=(n_0+l\Delta_n)\Delta t$, $\Delta_n$ and $\Delta t$ being the number of time-steps between two consecutive snapshots, and the time-step interval respectively. The snapshot matrix $X$ and the shifted snapshot matrix $X'$ are formed as
\begin{subequations}\label{eq:snapshots}
\begin{gather}
 X
 =
  \begin{bmatrix}\label{eq:snapshot_eq1}
  \mathbf{x}^{(n_0)}& \mathbf{x}^{(n_0+\Delta_n)} \ldots \mathbf{x}^{(n_0+(l-1)\Delta_n)}  \\
   \end{bmatrix},
\\
X'
=
  \begin{bmatrix}\label{eq:snapshot_eq2}
  \mathbf{x}^{(n_0+\Delta_n)}& \mathbf{x}^{(n_0+2\Delta_n)} \ldots \mathbf{x}^{(n_0+l\Delta_n)}  \\
   \end{bmatrix}.
\end{gather}
   
\end{subequations}
Assuming a linear predictive relationship $X'\approx A\cdot X$, DMD proceeds to extract the eigenvalues and eigenvectors of $A$ in an efficient fashion. Singular value decomposition (SVD) of the snapshot matrix $X$, results in the $U$, $\Sigma$, and $V$ matrices such that
\begin{align}
 X=U\Sigma V^*, 
 \end{align}
 where the superscript `$*$' denotes complex conjugate transpose. The columns of $U$ are essentially the proper orthogonal decomposition (POD) modes \cite{kerschen2002physical} which captures the dominant spatial pattern. The columns of $V$ represent the corresponding temporal pattern whereas the singular values (elements of the diagonal matrix $\Sigma$) indicate the weight (importance) of corresponding modes. Typically, the singular values show an exponentially decay pattern hinting at the underlying low-dimensional structure. We only retain the first $r$ ($<l$) singular values corresponding to the first $r$ columns of $U, V$, leading to reduced SVD matrices $U_r,\Sigma_r$ and $V_r$ such that $X\approx U_r\Sigma_rV_r^*$. The value of $r$ can be chosen based on a hard energy threshold or optimal hard thresholding \cite{gavish2014optimal,opt_thr_code}. Consequently, $A$ can be approximated as 
\begin{align}
A\approx X'V_r\Sigma_r^{-1}U_r^*. 
\end{align}
Next, $A$ is projected onto the columns of $U_r$, leading to 
\begin{align}
\tilde{A} = U_r^*AU_r = U_r^*X'V_r\Sigma_r^{-1}. 
\end{align}
The lower dimensionality of $\tilde{A}$, makes the eigendecomposition operation $\tilde{A}W=W\Lambda$ computationally efficient, 
where the diagonal matrix $\Lambda$ contains the eigenvalues $\lambda_i$, $i = 1, 2, \ldots, r$ that approximate the eigenvalues of $A$. The extracted spatial DMD modes $\bm{\phi}_i(\mathbf{r})$ are given by the columns of the matrix ${\Phi} = X'V_r\Sigma_r^{-1}W$ \cite{tu2013dynamic}, resulting in the DMD reconstruction $\hat{\mathbf{x}}$ of the dynamic state for $t\geq t_0$, i.e.
\begin{align}\label{eq:DMD_recon}
   \mathbf{x}(\mathbf{r},t)\approx\hat{\mathbf{x}}(\mathbf{r},t)&=\sum_{i=1}^r \vartheta_i\bm{\phi}_i(\mathbf{r})e^{\omega_i(t-t_0)},
\end{align} 
where $\omega_i=ln(\lambda_i)/\Delta_t$, $\Delta_t$ being the time interval between two consecutive DMD snapshots i.e. $\Delta_t=\Delta_n\Delta t$. The modal amplitudes $\vartheta_i$ can be calculated by solving an optimization problem as described in \cite{jovanovic2014sparsity}. Further details can be found in \cite{schmid2010dynamic, tu2013dynamic, kutz2016dynamic,pan2020structure}. The width of DMD data harvesting window is generally taken such that it can capture the entire dynamics. For systems showing periodic behavior, the DMD window should ideally cover multiple periods. The basis for selection of $\Delta_t$ is the Nyquist criterion, and noise frequency. In practice, since we generally deal with \textit{real} data, the DMD modes $(\bm\phi_i)$, frequencies $(\omega_i)$ and modal amplitudes $(\vartheta_i)$ as in \eqref{eq:DMD_recon}  appear in complex-conjugate pairs. Eq. \eqref{eq:DMD_recon} can be written in terms of $M$  complex-conjugate (overbar) pairs as,
\begin{align}\label{eq:DMD_recon_conj}
    \hat{\mathbf{x}}(\mathbf{r},t)=\sum_{m=1}^M (\vartheta_m\bm\phi_m e^{\omega_m(t-t_0)}+\overline{\vartheta}_m\overline{\bm\phi}_m e^{\overline{\omega}_m(t-t_0)}).
\end{align}
For purely real modes (DC modes), two terms in \eqref{eq:DMD_recon_conj} can be combined into a single term with $2M\geq r$. From this point onward, by the $m^\text{th}$ DMD mode or frequency, {we will} refer to the $m^\text{th}$ complex-conjugate pair. Also, for conciseness, we will address both $\omega_m$ and $f_m$ as the $m^\text{th}$ DMD frequency where $f_m=|\frac{\Im\{\omega_m\}}{2\pi}|$ ($\Im\{\cdot\}$ represents the imaginary part). A schematic representation of the DMD process for current density modeling is provided in Fig. \ref{fig:DMD}. \par

DMD's efficacy in modeling nonlinear dynamics can be attributed to its close relation to the Koopman operator~\cite{rowley2009spectral}. The Koopman operator theory tells us that with a suitable transformation ${g}(\cdot)$, with $g:\mathcal{M}\mapsto \mathcal{C}$ residing in the infinite-dimensional Hilbert space, a dynamical system, nonlinear in the original state space $\mathbf{x}$, can be represented by a linear dynamical system in the transformed ``feature" or ``observable" space $g\mathbf{(x)}$ \cite{koopman1931hamiltonian}. If we can find a finite$(p)$-dimensional subspace invariant under the Koopman operator, the infinite-dimensional Koopman operator can be represented by a $p\times p$ matrix $\mathbf{K}$, such that the time evolution of a vector-valued observable $\mathbf{g(x)}=[g_1(\mathbf{x});g_2(\mathbf{x});\ldots;g_p(\mathbf{x})]$ belonging to that subspace is given by $\mathbf{g}(\mathbf{x}^{(n+1)})=\mathbf{K}\cdot \mathbf{g}(\mathbf{x}^{(n)})$, where $[;]$ denotes the vertical stacking. DMD assumes an identity transformation, i.e. $\mathbf{g(x)=x}$, which works well for variety of high-dimensional nonlinear dynamical systems \cite{rowley2009spectral,kaptanoglu2020characterizing,tu2013dynamic,kutz2016dynamic,PhysRevE.103.012201,PhysRevE.99.063311}. However, the assumption of $\mathbf{g(x)=x}$ might not be sufficient for highly nonlinear systems, including kinetic plasmas. In order to adequately capture nonlinearities, the Hankel variant of DMD makes use of the time-delay coordinates {where the vector valued observable at the $n^\text{th}$ timestep is given by
\begin{align}\label{eq:hankel_dmd}
    \mathbf{g(x}^{(n)})=[\mathbf{x}^{(n)};\mathbf{x}^{(n-\Delta_n)};\ldots ;\mathbf{x}^{(n-(d-1)\Delta_n)}]
\end{align} 
, where ``;" indicates vertical stacking, $d$ is the number of Hankel stacks \cite{nayak2023fly,kamb2020time,pan2020structure,brunton2017chaos}, and $\Delta_n$ is the DMD sampling interval in terms of timesteps.} The success of delay embeddings in forming a suitable vector-valued Koopman observable is rooted in the Takens' embedding theorem \cite{takens2006detecting,brunton2017chaos}, which states that under certain conditions, the attractor of a dynamical system in delayed coordinates is \textit{diffeomorphic} to the attractor of original system. Broadly speaking, incorporation of time-delayed embeddings help better model the nonlinearities through a linear model such as DMD. Here, we apply the Hankel DMD which follows the same steps as the regular DMD, but with two notable exceptions. First, instead of $\mathbf{x}$, {we use $\mathbf{g(x)}$ \eqref{eq:hankel_dmd}} to form the snapshot matrices \eqref{eq:snapshots}, and second, we retain the first $N$ elements of the DMD modes $(\bm\phi)$ to reconstruct the state as in \eqref{eq:DMD_recon}.

\section{DMD-EMPIC Algorithm}\label{sec:dmd-empic}

As described in Section \ref{sec:empic}, the primary computational bottleneck comes from the \textit{gather}, \textit{pusher} and \textit{scatter} stages as they involve each and every particle in the solution domain. However, these steps are necessary as they dictate the time evolution of the current density which subsequently helps update the electromagnetic fields. We try to address this issue by directly modeling the time evolution of current density $\mathbf{j}$ using DMD. An analytical {equation of the type \eqref{eq:DMD_recon_conj}} for time variation of current density in EMPIC serves the following purpose,
\begin{itemize}
    \item DMD helps in rapid prediction of $\mathbf{j}$ with negligible computation cost compared to the particle operations. Also, a linear time-evolution model for current density can facilitate control theory applications.
    \item The \textit{gather}, \textit{pusher} and \textit{scatter} stages can be replaced by the DMD predicted $\mathbf{j}$ which is then used to update the electric and magnetic fields in the \textit{field-update} stage (Section \ref{sec:field_update}) within the finite-element time-domain (FETD) setting. This can drastically reduce the computation cost of EMPIC.
\end{itemize}

The DMD-EMPIC algorithm is illustrated in Fig. \ref{fig:DMD-EMPIC}. It consists of two main phases, the transient ($t \leq t_f$) and post-transient phase ($t > t_f$). 
\begin{itemize}
    \item \textbf{Transient phase ($t \leq t_f$):} The high-fidelity EMPIC simulation is run until the transient phase ends, denoted by the final time $t_{f}$. The on-the-fly DMD algorithm~\cite{nayak2023fly} is run simultaneously with the ongoing EMPIC simulation to identify $t_f$ on the fly. The electric field data $(\mathbf{e})$ is fed to the Hankel DMD algorithm and it provides feedback in real-time regarding whether to stop the EMPIC simulation or not. In order to maximize computational gains, it is desirable to terminate the time-consuming EMPIC simulation at the earliest opportunity. At the same time, if the simulation is terminated too early (i.e. before the transient ends), DMD will not be able to make accurate time-extrapolation due to the lack of quality training data \cite{NAYAK2021110671}. Thus, a real-time algorithm for timely termination of high-fidelity EMPIC simulation is necessary. In the DMD-EMPIC algorithm, this is achieved by the sliding-window on-the-fly DMD algorithm developed in {\cite{nayak2023fly}} which analyzes the time evolution of $\mathbf{e}$ in order to detect end of transience at $t=t_f$. {However, in order to handle the repetitive execution of DMD for large datasets, we make  modifications to the algorithm described in \cite{nayak2023fly}. Instead of using the standard version of  DMD, we perform randomized DMD to reduce the computational load. The on-the-fly algorithm is described in Appendix \ref{sec:onthefly_algo}. }

    \item \textbf{Post-transient phase ($t > t_f$):} Following the detection  of end-of-transience $(t_f)$, Hankel DMD is performed in offline or \textit{a posteriori} fashion on the snapshots of $\mathbf{j}$ collected inside the final DMD window. The purpose of the offline DMD is to predict the current density beyond $t_f$, denoted by $\hat{\mathbf{j}}$. Typically, the \textit{gather}, \textit{pusher} and \textit{scatter} steps are required for the time-update of $\mathbf{j}$. Here, since an analytical expression (similar to \eqref{eq:DMD_recon}) for the time evolution of $\mathbf{j}$ is available from the offline DMD, we avoid these steps by using the DMD predicted $\hat{\mathbf{j}}$. The predicted current $\hat{\mathbf{j}}$ is then used in consecutive time-steps to update the self electric and magnetic fields in the \textit{field-update} stage (Section \ref{sec:field_update}). Note that for $t > t_f$, the relation between the self-fields and current density is not exactly cyclic. Beyond $t=t_f$, the fields do not have any effect on the time-evolution of the current density $\hat{\mathbf{j}}$.
    
\end{itemize} \par

As shown in Fig. \ref{fig:DMD-EMPIC}, before the end of transience is detected $(t<t_{f})$, the computation cost for each time-step of the simulation is same as the cost of typical EMPIC $(\mathcal{O}(N+N_p))$ with the added cost of on-the-fly DMD, where $N$ is aggregate mesh dimension and $N_p$ is the total number of charged particles. However, beyond $t_{f}$ the computation cost per time step reduces to $\mathcal{O}(N)$, which is a significant reduction in computation time given that $N_p\gg N$ in typical EMPIC settings \cite{Note3}. The computational gain of the proposed DMD-EMPIC is discussed in details in Section \ref{sec:comp_gain}.

\begin{figure*} [t]

    \centering
  \subfloat[Electron beam snapshot at $t=64$ ns.\label{fig:beam_snap_wavy} ]{%
       \includegraphics[width=0.29\linewidth]{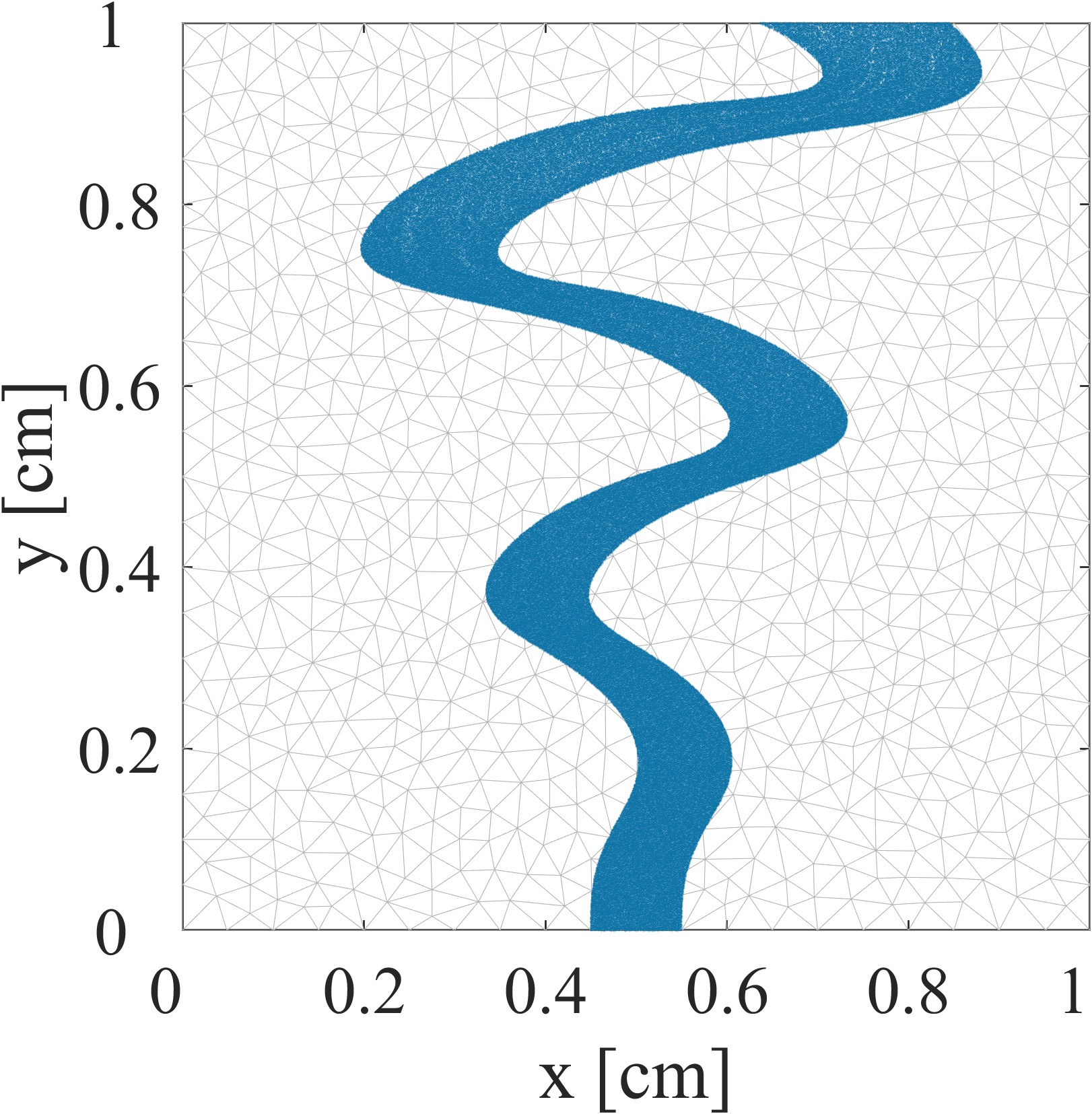}}
       \hfill
  \subfloat[Current density snapshot at $t=64$ ns.\label{fig:crnt_snap_wavy} ]{%
        \includegraphics[width=0.35\linewidth]{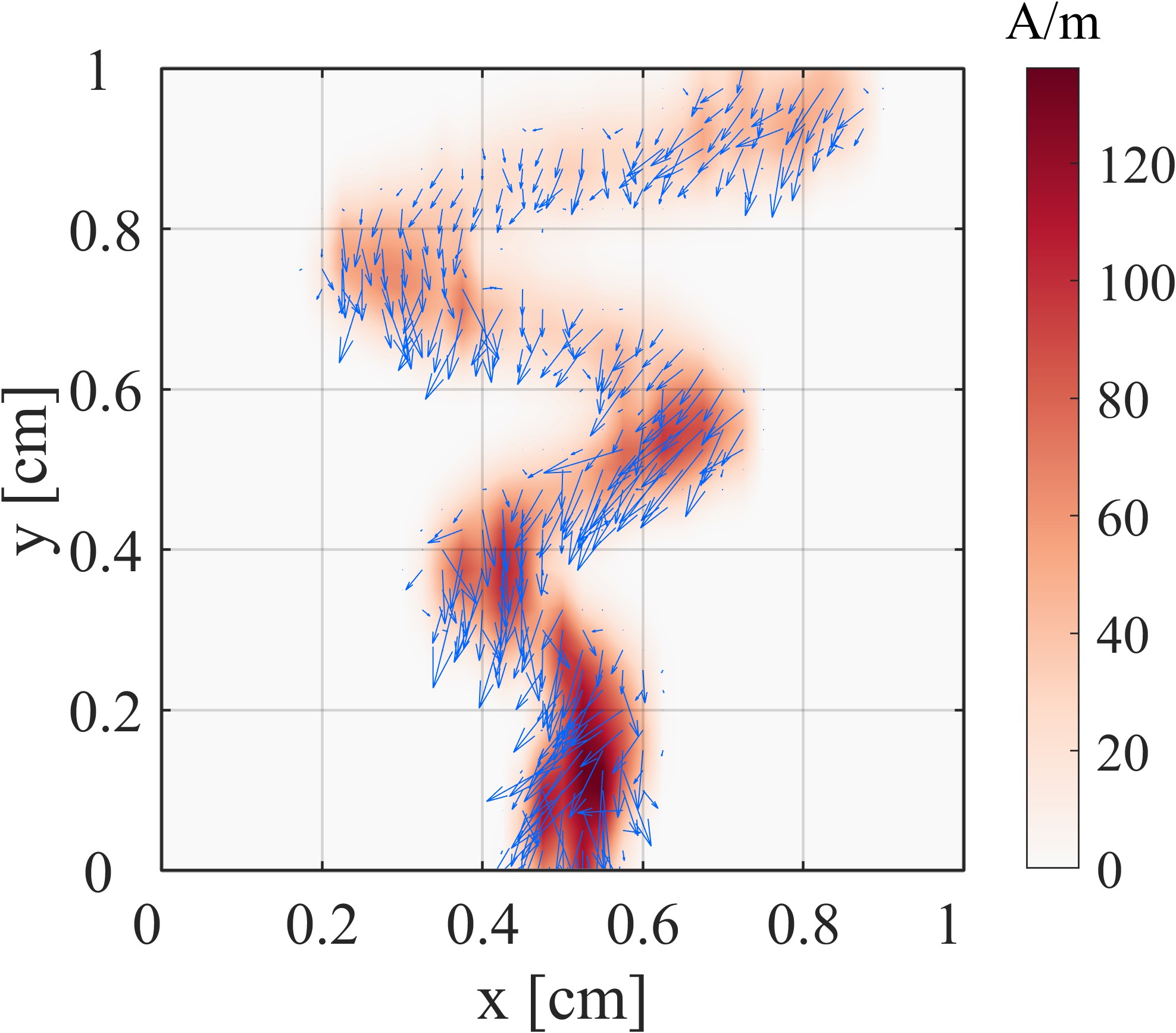}}
        \hfill
  \subfloat[Singular values for oscillating electron beam.\label{fig:singvals_wavy} ]{%
    \includegraphics[width=0.312\linewidth]{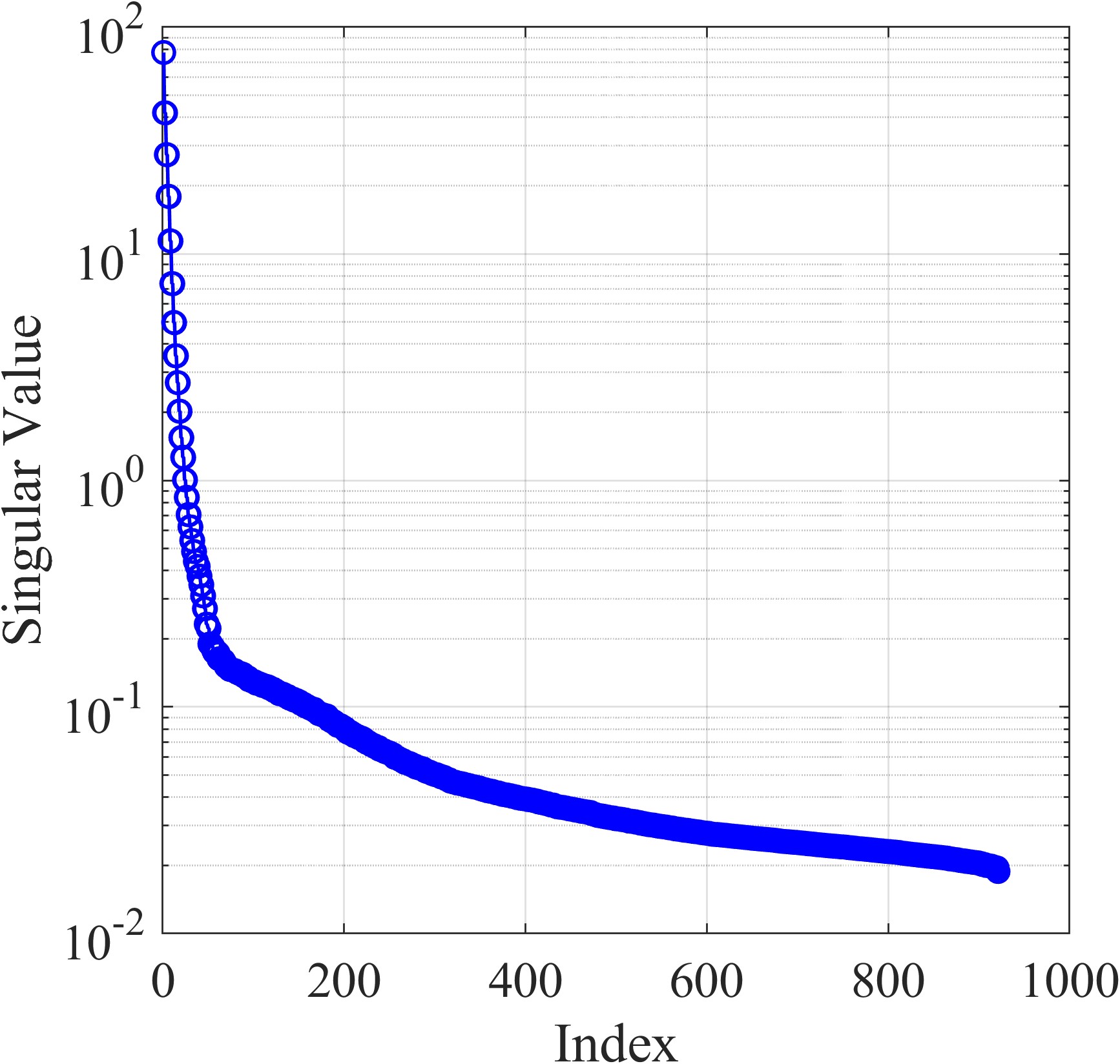}}
  
  \caption{\small{ (a) Snapshot of the electron beam at {$t=64$ ns.} The beam is propagating along the $+ve$ $y$ direction, and oscillating under the influence of a $z$-directed transverse magnetic flux. The blue dots represent superparticles and grey lines show the finite element mesh edges. (b) Snapshot of the current density at $t=16$ ns. The magnitude is shown by a colormap \cite{colormap}, whereas the direction is denoted by the blue arrows. Note that in all the current density plots, the magnitude colormap is smoothed for visualization purpose. (c) Singular values after performing SVD on the snapshots of current density.}  }\label{fig:setup_sing_wavy}
\end{figure*}

\begin{figure*} [t]

    \centering
  \subfloat[DMD mode 1 with $f_1=0$ (DC). \label{fig:wavy_mode01} ]{%
       \includegraphics[width=0.32\linewidth]{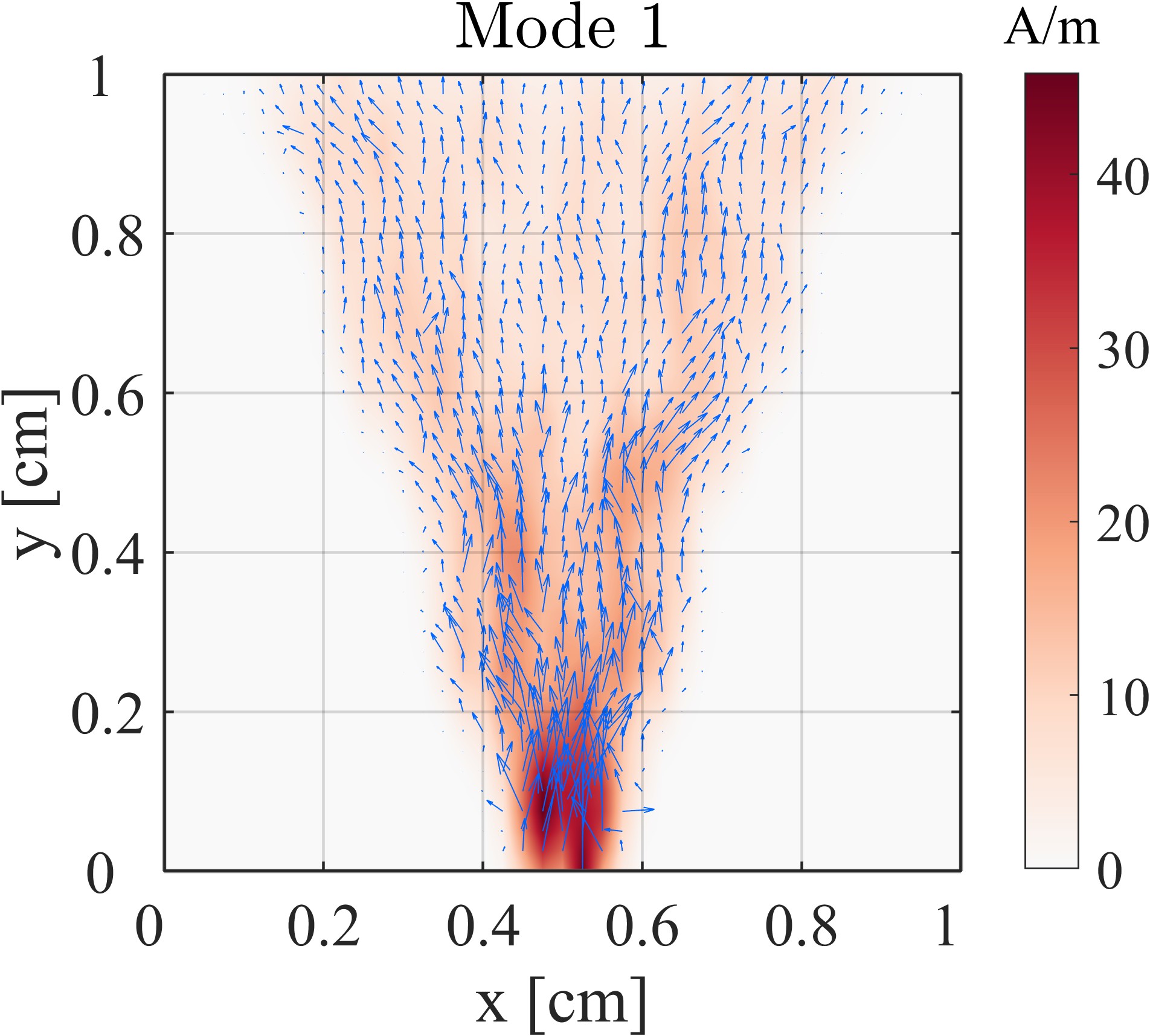}}
    \hfill
  \subfloat[DMD Mode 2 with $f_2=1.25$ GHz. \label{fig:wavy_mode02} ]{%
        \includegraphics[width=0.32\linewidth]{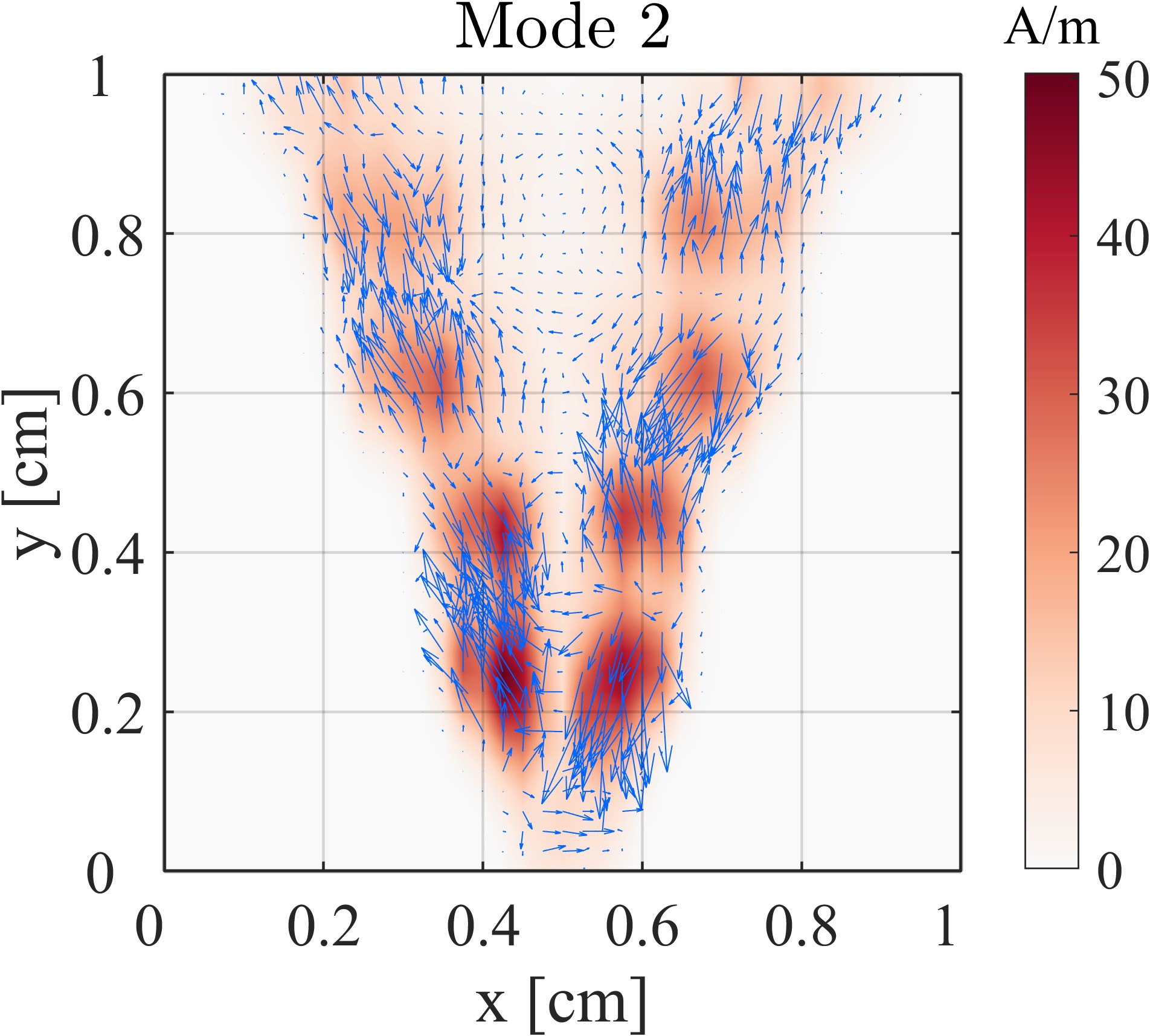}}
    \hfill 
   \subfloat[DMD Mode 3 with $f_3=2.50$ GHz.\label{fig:wavy_mode03} ]{%
        \includegraphics[width=0.32\linewidth]{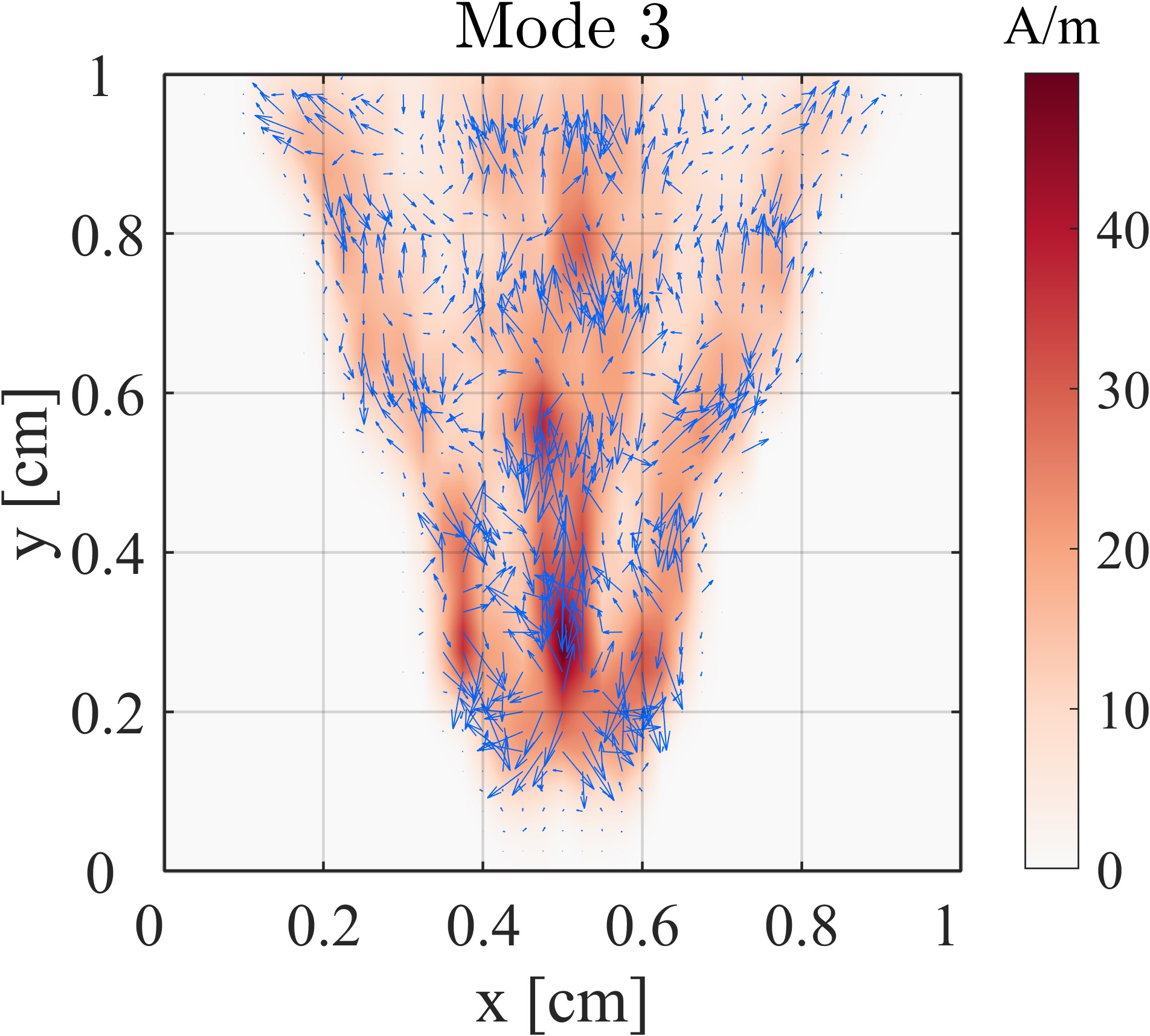}}\\ 
   \subfloat[DMD mode 4 with $f_4=3.75$ GHz.\label{fig:wavy_mode04} ]{%
        \includegraphics[width=0.32\linewidth]{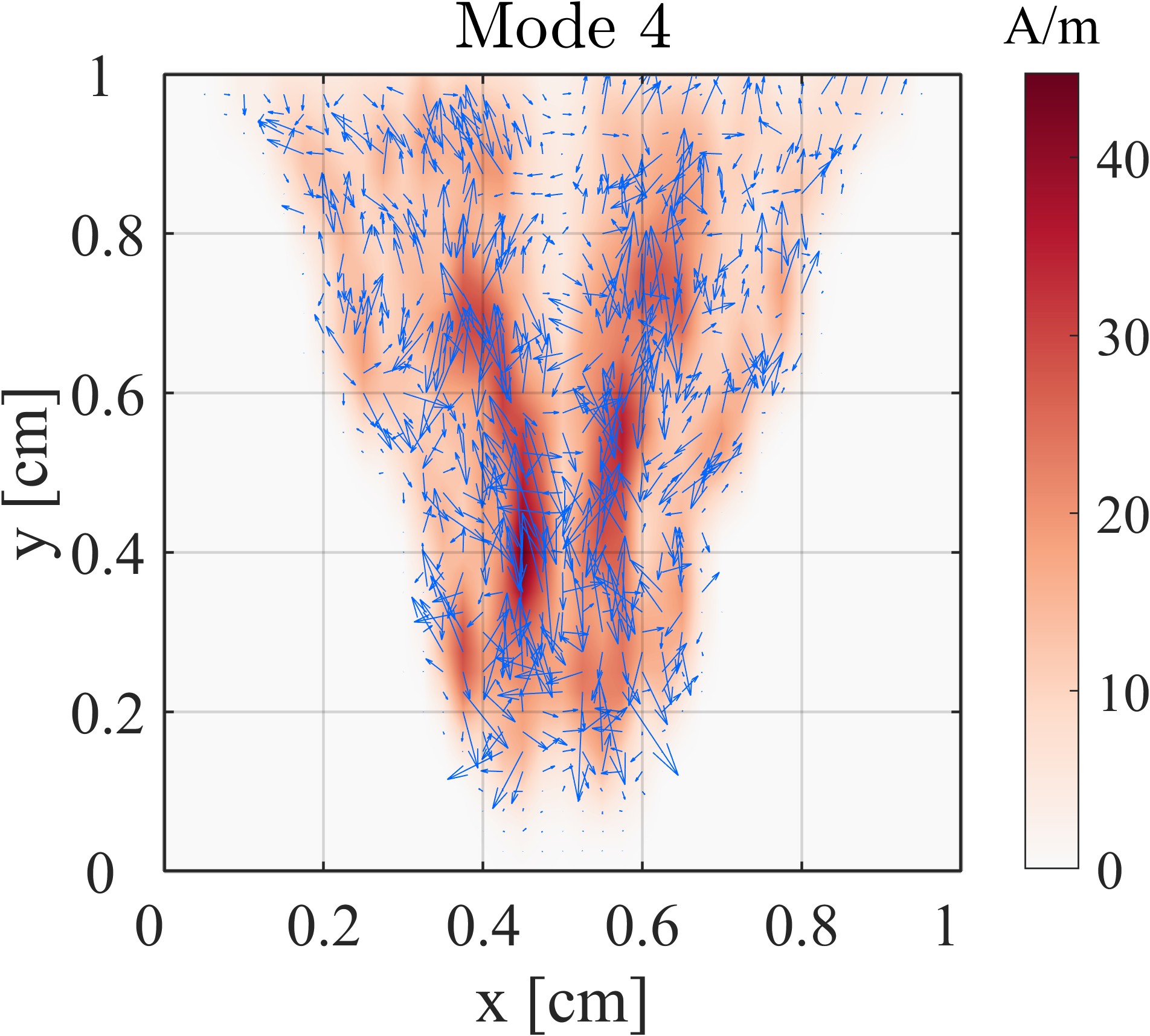}}
    \hfill 
    \subfloat[DMD eigenvalues.\label{fig:eigs_wavy} ]{%
        \includegraphics[width=0.34\linewidth]{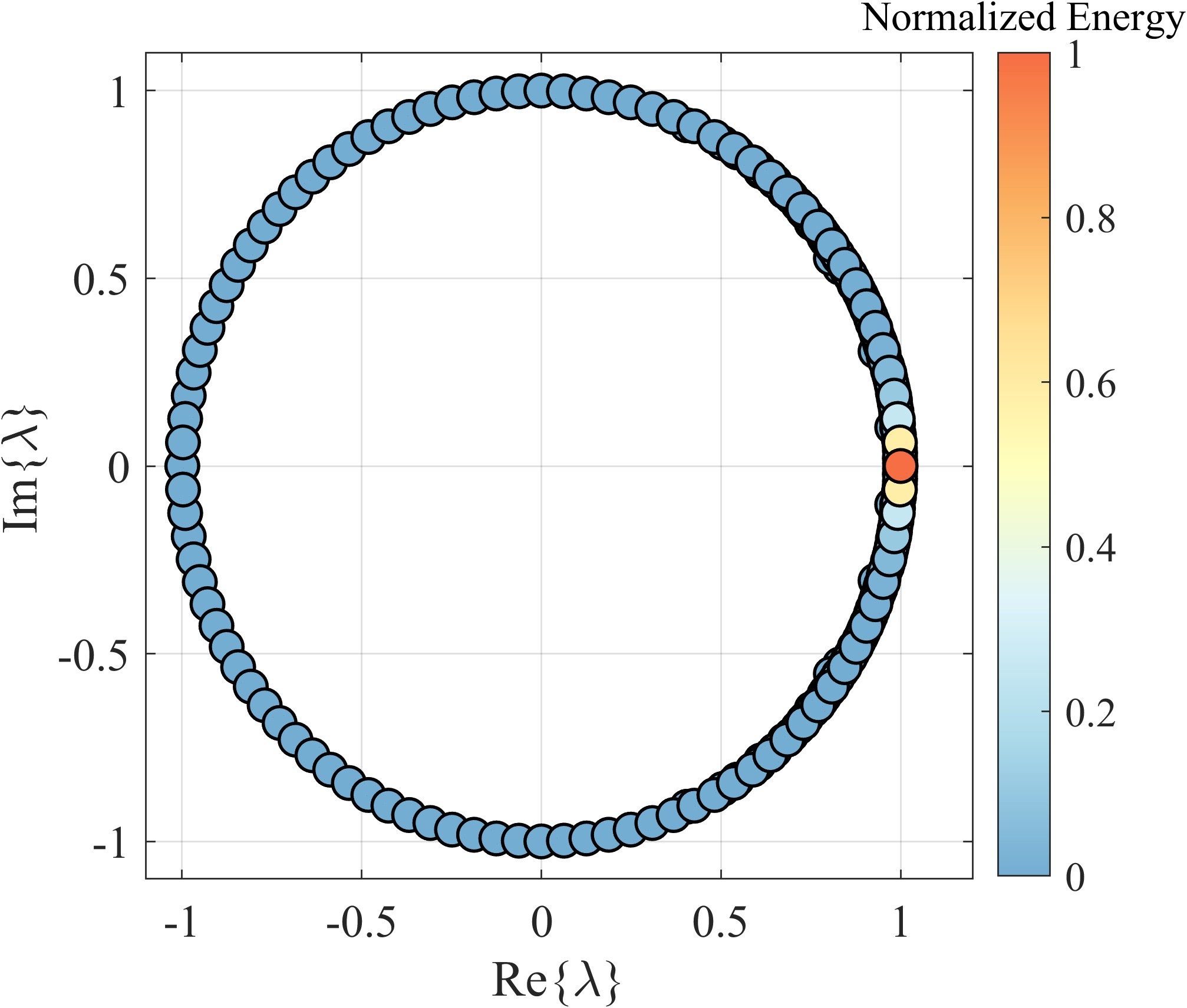}}
    \hfill
      \subfloat[Spatial correlation among DMD modes.\label{fig:corr_wavy} ]{%
        \includegraphics[width=0.316\linewidth]{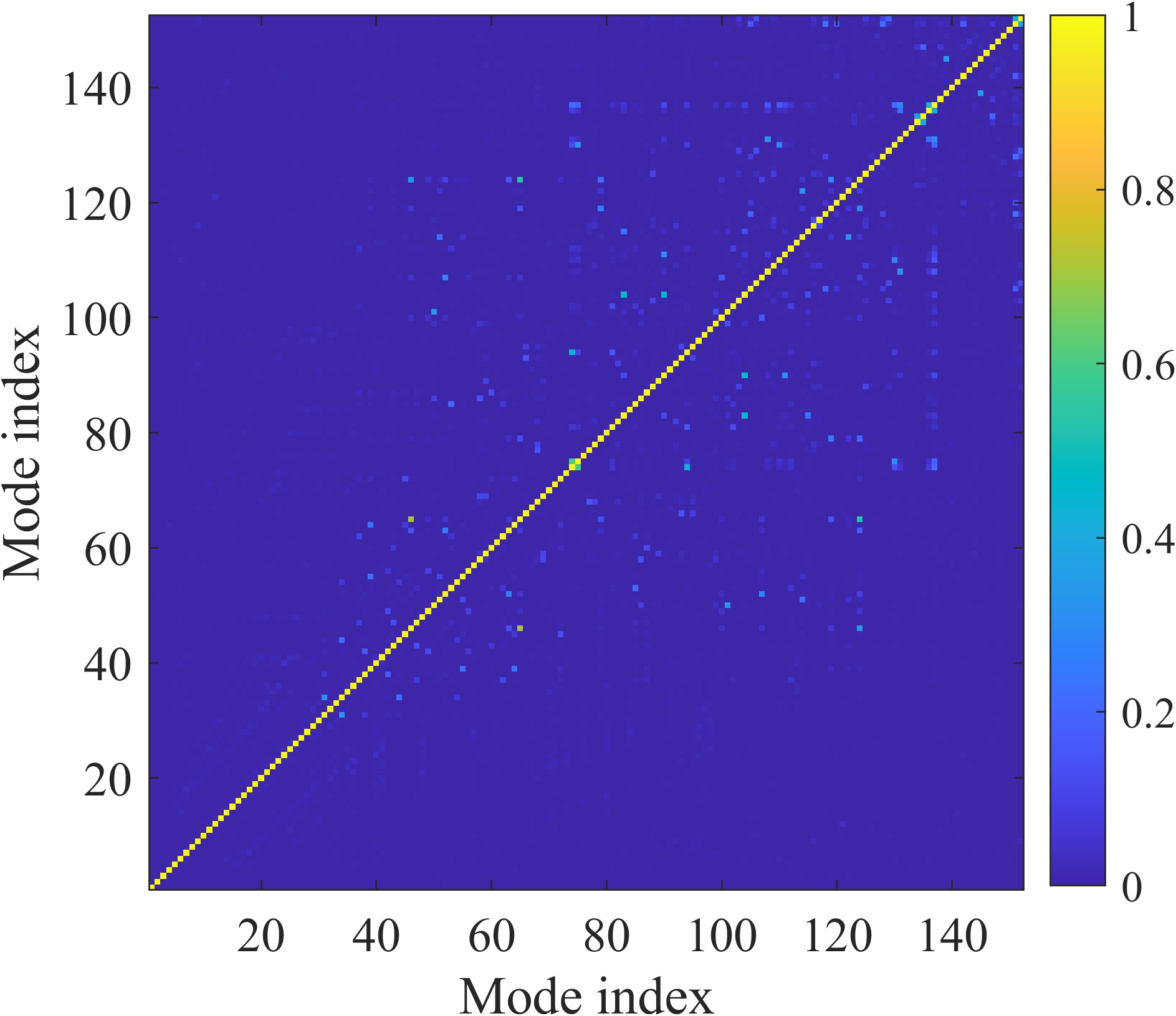}}
    \hfill   
  \caption{\small{ Extracted DMD features for oscillating electron beam. (a,b,c,d) The DMD modes $(\bm\phi_m+\overline{\bm\phi}_m)$ and their corresponding frequencies $(f_m)$ for the current density modeling. (e) The DMD eigenvalues on the complex plane wrt. the unit circle. The DMD eigenvalues are color-mapped according to their normalized energy. (f) Spatial correlation $(\rho)$ among different DMD modes. }  }\label{fig:DMD_wavy}
\end{figure*}

\begin{figure*} [t]

    \centering
  \subfloat[EMPIC simulation of current density at $t=64$ ns.\label{fig:crnt_wavy_EMPIC} ]{%
       \includegraphics[width=0.3\linewidth]{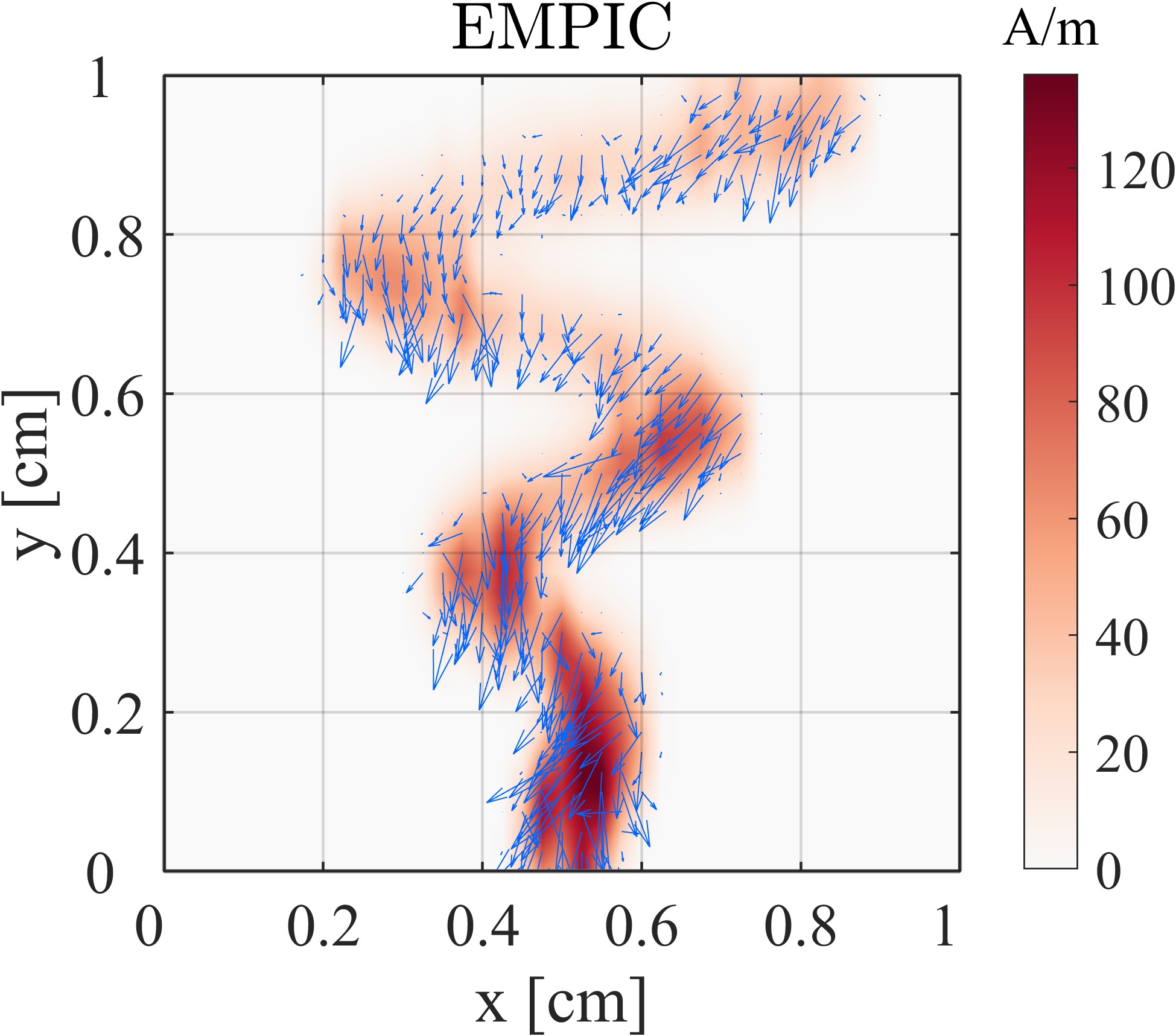}}
       \hfill
  \subfloat[DMD prediction of current density at $t=64$ ns.\label{fig:crnt_wavy_DMD} ]{%
        \includegraphics[width=0.3\linewidth]{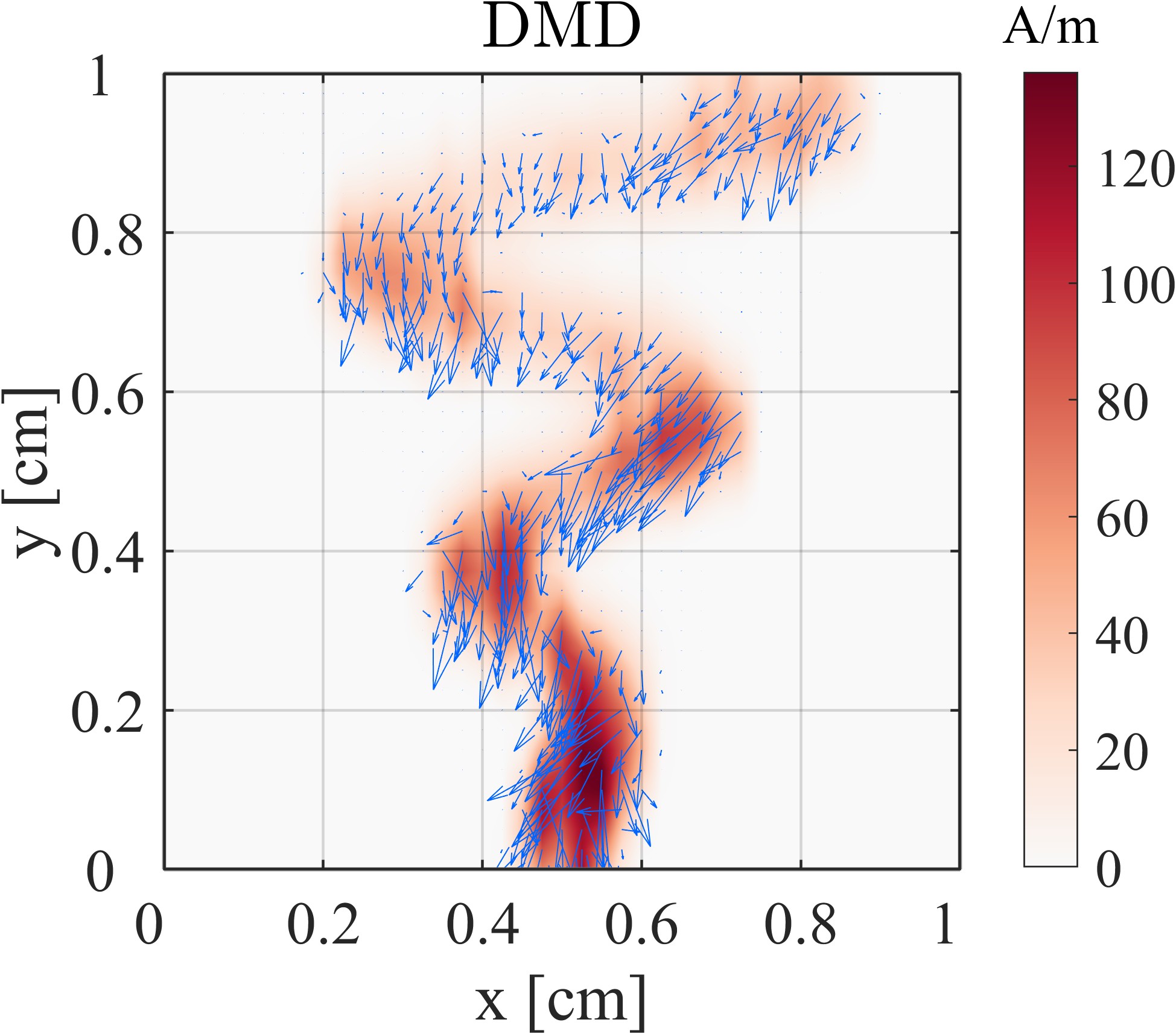}}
        \hfill
  \subfloat[Relative error between EMPIC and DMD predicted current density according to \eqref{eq:rel_error}. \label{fig:err_crnt_wavy} ]{%
    \includegraphics[width=0.34\linewidth]{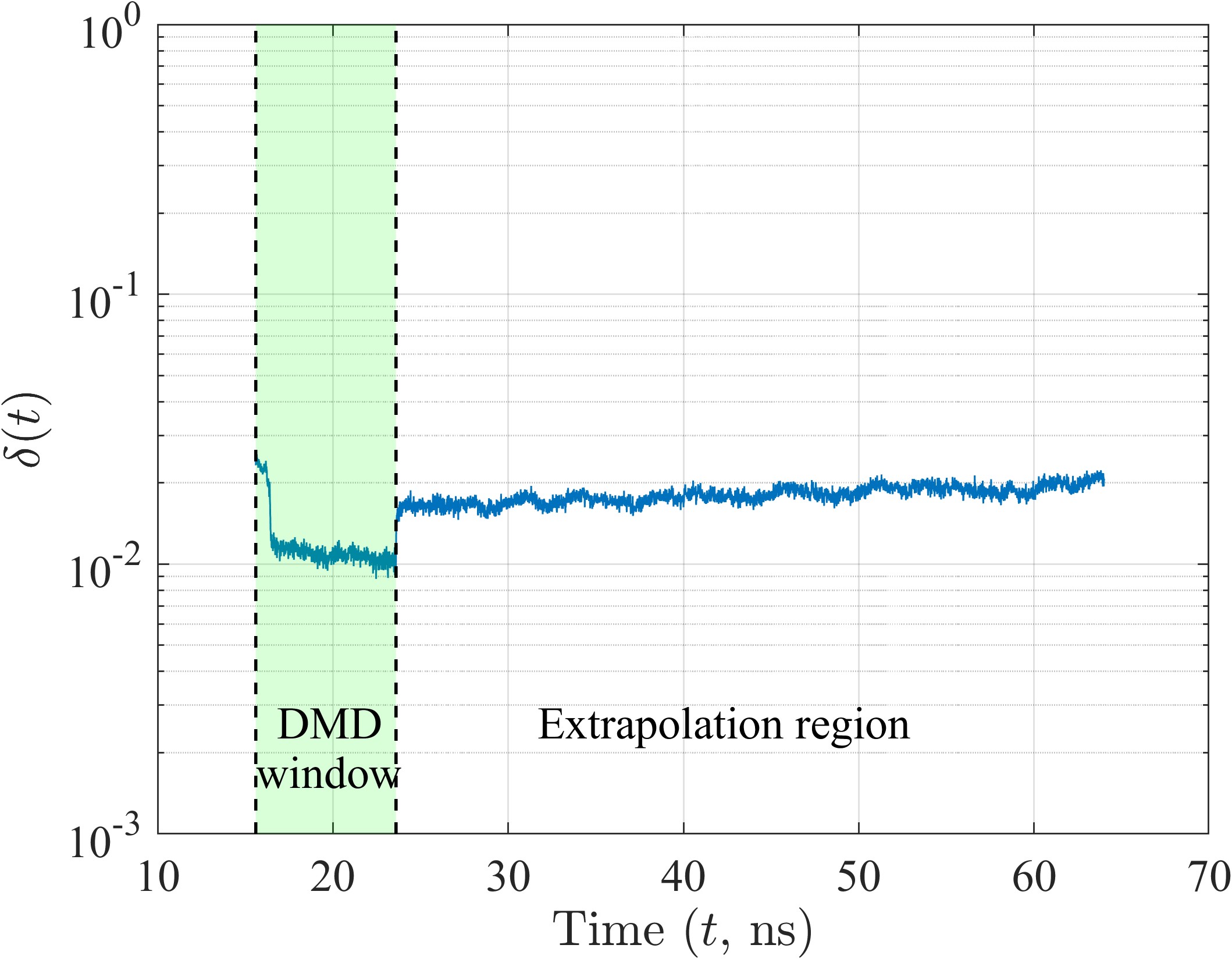}}
  
  \caption{\small{ Comparison between EMPIC and DMD predicted current density. The shaded green region in (c) denotes the DMD training window. Note that the gap at the end of DMD window is due to the time-delayed stacking.}  }\label{fig:DMD_comp_wavy}
\end{figure*}

\begin{figure*} [t]

    \centering
  \subfloat[EMPIC simulation of electric field at $t=64$ ns.\label{fig:e_wavy_EMPIC} ]{%
       \includegraphics[width=0.31\linewidth]{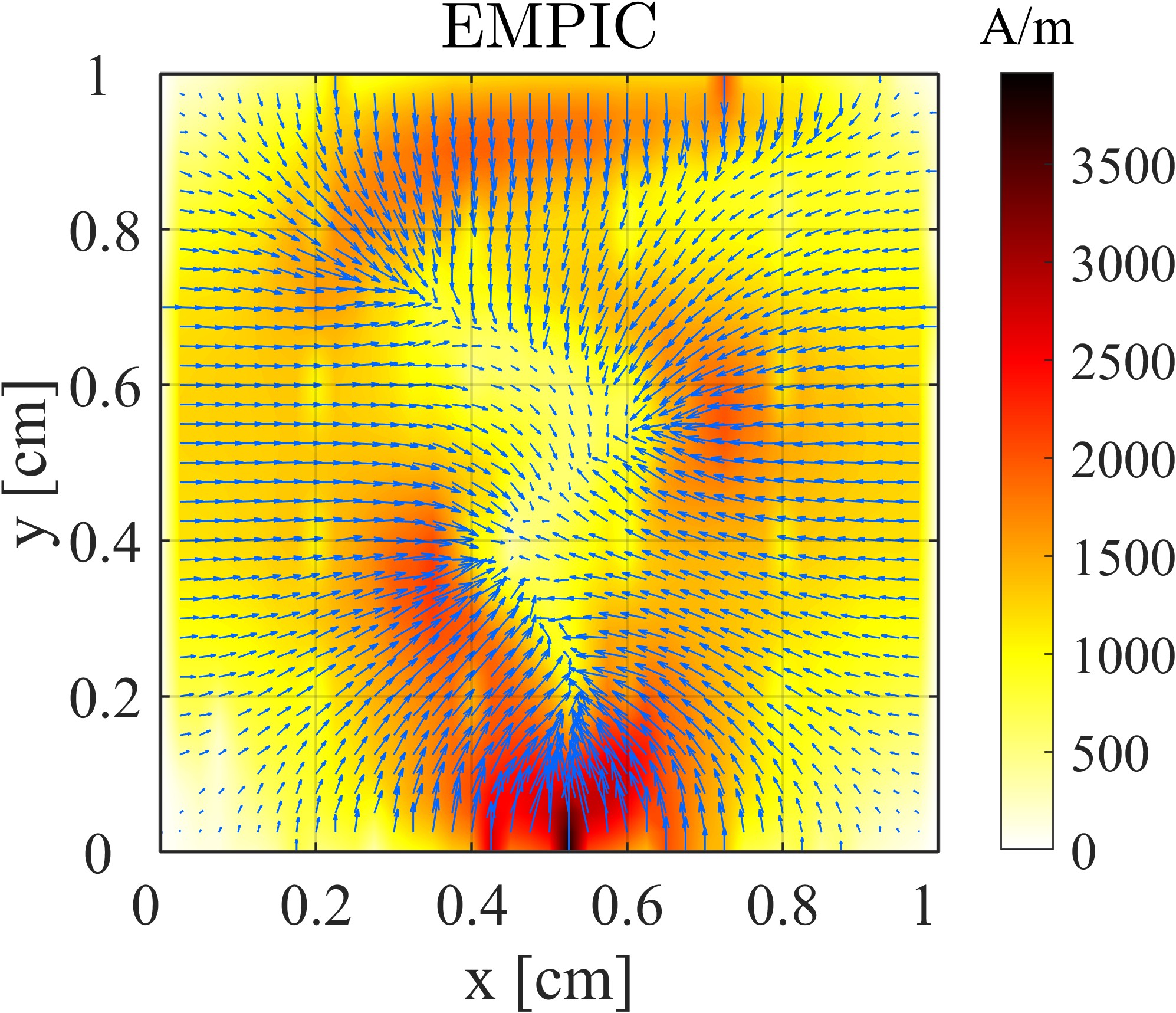}}
       \hfill
  \subfloat[DMD-EMPIC simulation of electric field at $t=64$ ns.\label{fig:e_wavy_DMD} ]{%
        \includegraphics[width=0.31\linewidth]{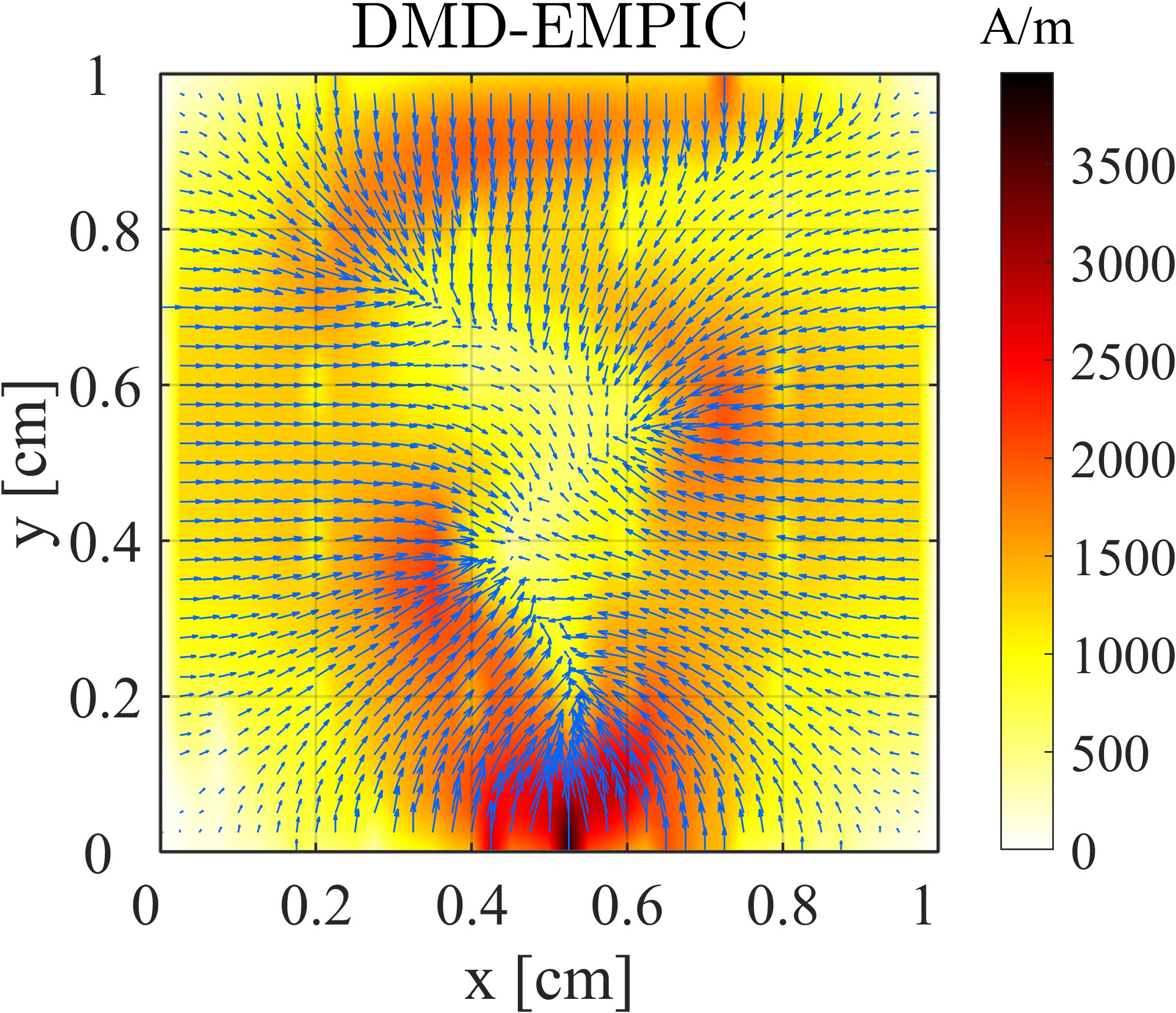}}
        \hfill
  \subfloat[Relative errors between EMPIC and DMD-EMPIC simulated self electric $(\mathbf{e})$ and magnetic $(\mathbf{b})$ fields.\label{fig:err_eb_wavy} ]{%
    \includegraphics[width=0.32\linewidth]{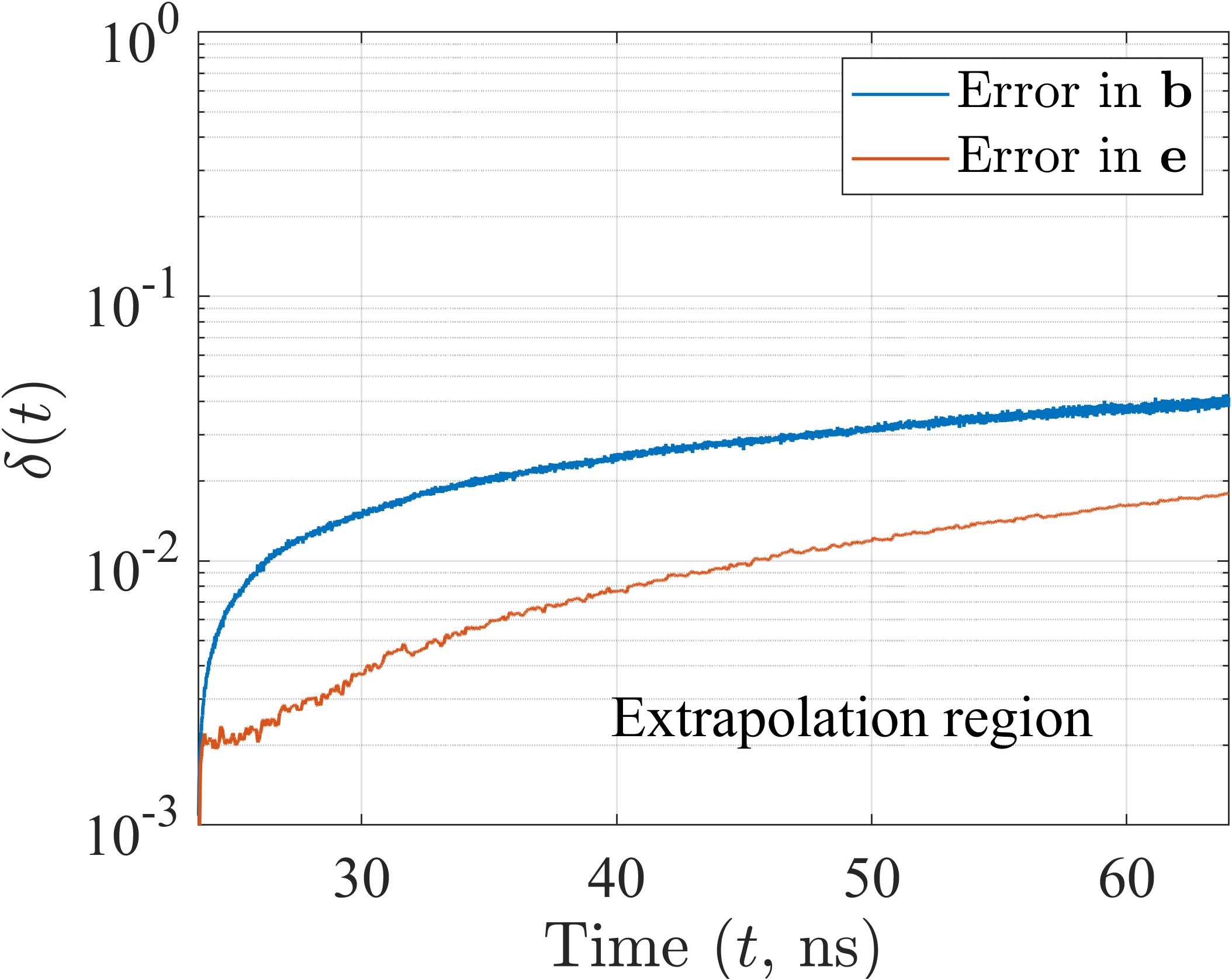}}
  
  \caption{\small{ Comparison between EMPIC and DMD-EMPIC simulated self-fields in the extrapolation region.}  }\label{fig:DMD_comp_fields_wavy}
\end{figure*}
\section{Results}\label{sec:results}

\subsection{Oscillating Electron Beam}\label{sec:eg_wavy_beam}
A 2-D electron beam propagating along the positive $y$ direction and oscillating under the influence of an external transverse magnetic flux is shown in Fig. \ref{fig:beam_snap_wavy}. The solution domain ($xy$ plane) is a square cavity of dimension $1~\text{cm}\times 1~\text{cm}$, which is discretized via an unstructured mesh composed of triangular elements. The mesh consists of $N_0=844$ nodes, $N_1=2447$ edges and $N_2=1604$ elements (triangles). Superparticles (blue dots in Fig. \ref{fig:beam_snap_wavy}), are injected at the bottom of the cavity in the $+ve$ $y$ direction with a velocity of $5\times10^6$ m/s. The superparticles are injected at the rate of $50$ per time-step in a random fashion uniformly in the range $[0.45~\text{cm},~0.55~\text{cm}]$. The superparticles discretize the phase-space of the electrons assuming a delta distribution in both position and velocity space. Superparticles are treated as point charges with mass $m_{sp}=r_{sp}m_e$ and charge $q_{sp}=r_{sp}q_e$, where $m_e$ and $q_e$ are respectively the mass and charge of an electron, and $r_{sp}=5000$ is the number of actual electrons represented by each superparticle (superparticle ratio). An external oscillating magnetic flux $\mathcal{B}_{ext}=\mathcal{B}_0~\text{sin}(2\pi/T_{osc})~\hat{z}$ is applied in the $z$-direction, where $\mathcal{B}_0=2.5\times10^{-2}$ T, and $T_{osc}=0.8$ ns. The simulation is run until $n=320000$ time-steps or $t=64$ ns with the time-step interval $\Delta t=0.2$ ps.\par

The post-transient snapshot of the current density $\mathbf{j}$ is shown in Fig. \ref{fig:crnt_snap_wavy} at $t=16$ ns $(n=80000)$. The goal is to model the time evolution of such snapshots inside the cavity using DMD. Unlike electromagnetic fields, $\mathbf{j}$ is restricted to only the mesh elements interacting with the particles. In other words, the number of active edges $N_{1a}$ over which $\mathbf{j}$ is nonzero (within the DMD window span) is less than total number of mesh edges $N_1$. We only consider those active edges for DMD modeling with $\mathbf{x}=\mathbf{j}_a$, where $\mathbf{j}_a$ is the current density vector with active edges. After performing the DMD, we revert back to the original state space with zero padding. \par

\textbf{On-the-fly DMD on fields:} The on-the-fly DMD is carried out on the electric field data to detect the end of transience or onset of the periodic behavior as {described in Appendix \ref{sec:onthefly_algo} \cite{nayak2023fly}. Approximate prior knowledge about the time-scale is required to choose the DMD {window} width $\Delta t_w$ accordingly, ensuring that it covers multiple oscillation cycles. We select $\Delta t_w = 8$ ns. Once the EMPIC simulation reaches $t=\Delta t_w$, fast Fourier transform (FFT) is performed on a randomly chosen set of 20 points in space. Averaged FFT allows us to select the DMD sampling interval $\Delta_t=8$ ps. We select the target rank $r=200$, and number of Hankel stacks $d=10$. The shift in consecutive sliding DMD windows is $\delta t_w = 0.4$ ns. The on-the-fly DMD parameters are mentioned in details in the Table \ref{table:onthefly_DMD}. on The onset of equilibrium (end of transience) is detected at $t=t_f=23.06$ ns. }

\begingroup
\setlength{\tabcolsep}{10pt}
\renewcommand{\arraystretch}{1.5}
\begin{table}

\begin{center}
\caption{DMD parameters for modeling current density.}
\label{table:dmd_params}
\begin{ruledtabular}
\begin{tabular}{ c c c c } 
\textbf{\makecell{Parameters}} & \textbf{\makecell{Osc. Beam}} & \textbf{\makecell{Vircator}} & \textbf{\makecell{BWO}}\\
\hline
$t_{st}$ & $15.60$ ns & $22.40$ ns & $91.20$ ns\\ 
$t_{en}$ & $23.60$ ns & $30.40$ ns & $107.20$ ns\\
$\Delta t_w$ & $8$ ns & $8$ ns & $16$ ns\\
$\Delta_t$ & $8$ ps & $40$ ps & $2$ ps\\
\makecell{$d$} & $80$ & $50$ & $20$\\
$r$ & $302$ & $42$ & $1999$\\
\makecell{$M$} & $152$ & $22$ & $1006$\\
\end{tabular}
\end{ruledtabular}
\end{center} 
\end{table} 
\endgroup

\textbf{Offline DMD on current density:} {The primary contribution of this work is to accelerate EMPIC simulations by DMD modeling of the plasma current density.} EMPIC stops at $t=t_f$ (detected end of transience), and offline DMD on current density $\mathbf{j}$ is performed in the {window $t\in [t_f-\Delta t_w,t_f]$} for time-extrapolation. We first identify the active edges to construct $\mathbf{j}_a$, perform DMD on the snapshots of $\mathbf{j}_a$ to get the predictions $\hat{\mathbf{j}}_a$, and then revert back to $\hat{\mathbf{j}}$. {The DMD parameters are summarized in Table \ref{table:dmd_params}. The $t_{st},t_{en}$, i.e. the location of the DMD window for current density is already determined from the on-the-fly DMD with $t_{st}=t_f-\Delta t_w$ and $t_{en}=t_f$. FFT is performed in $[t_{st},t_{en}]$ to decide the DMD sampling interval $\Delta_t$. As a rule of thumb, we choose DMD sampling frequency to be four times the Nyquist frequency.} \par

The sharp decay in the singular values (Fig. \ref{fig:singvals_wavy}) indicates the existence of a low-dimensional structure in the plasma current dynamics. Fig. \ref{fig:DMD_wavy} shows the first four most energetic DMD modes and corresponding DMD eigenvalues, as well as the spatial cross-correlation matrix. The DMD modes are indexed according to their energy \cite{Note2}, with mode 1 (Fig. \ref{fig:wavy_mode01}) being the most energetic one which is essentially a DC mode. Mode 2 (Fig. \ref{fig:wavy_mode02}) captures the oscillation frequency of the external magnetic flux with DMD frequency of $f_2=1.25$ GHz $(f_m=|\frac{\Im\{\omega_m\}}{2\pi}|)$. {Mode 3 (Fig. \ref{fig:wavy_mode03}) indicates the first harmonic. Together these two modes capture $>90\%$ of the total energy (Fig. \ref{fig:eigs_wavy}).} As the mode index increases, the spatial pattern becomes less structured due to the effect of numerical noise. The frequencies associated with mode 3 (Fig. \ref{fig:wavy_mode03}) and mode 4 (Fig. \ref{fig:wavy_mode04}) indicate that those are essentially the harmonics of mode 2, generated due to the nonlinear wave-particle interaction. The correlation $\rho$ between different spatial patterns of DMD modes indicates their extent of orthogonality. We use the absolute value of modal assurance criterion (MAC) \cite{beaverstock2015automatic,NAYAK2021110671} to compute the spatial correlation among DMD modes,
\begin{align}
    \rho(\bm\psi_i,\bm\psi_j)=|\text{MAC}(\boldsymbol{\psi}_i,\boldsymbol{\psi}_j)|=|\frac{\big| \boldsymbol{\psi}_i^\text{T}  \ {\boldsymbol{\overline{\psi}}}_j\big|^2 }{(\boldsymbol{\psi}_i^\text{T} {\boldsymbol{\overline{\psi}}}_i)\cdot(\boldsymbol{\psi}_j^\text{T} {\boldsymbol{\overline{\psi}}}_j) }|.
\end{align}

\begin{figure*} [t]

    \centering
  \subfloat[Virtual cathode snapshot at {$t=64$ ns}.\label{fig:beam_snap_virt} ]{%
       \includegraphics[width=0.285\linewidth]{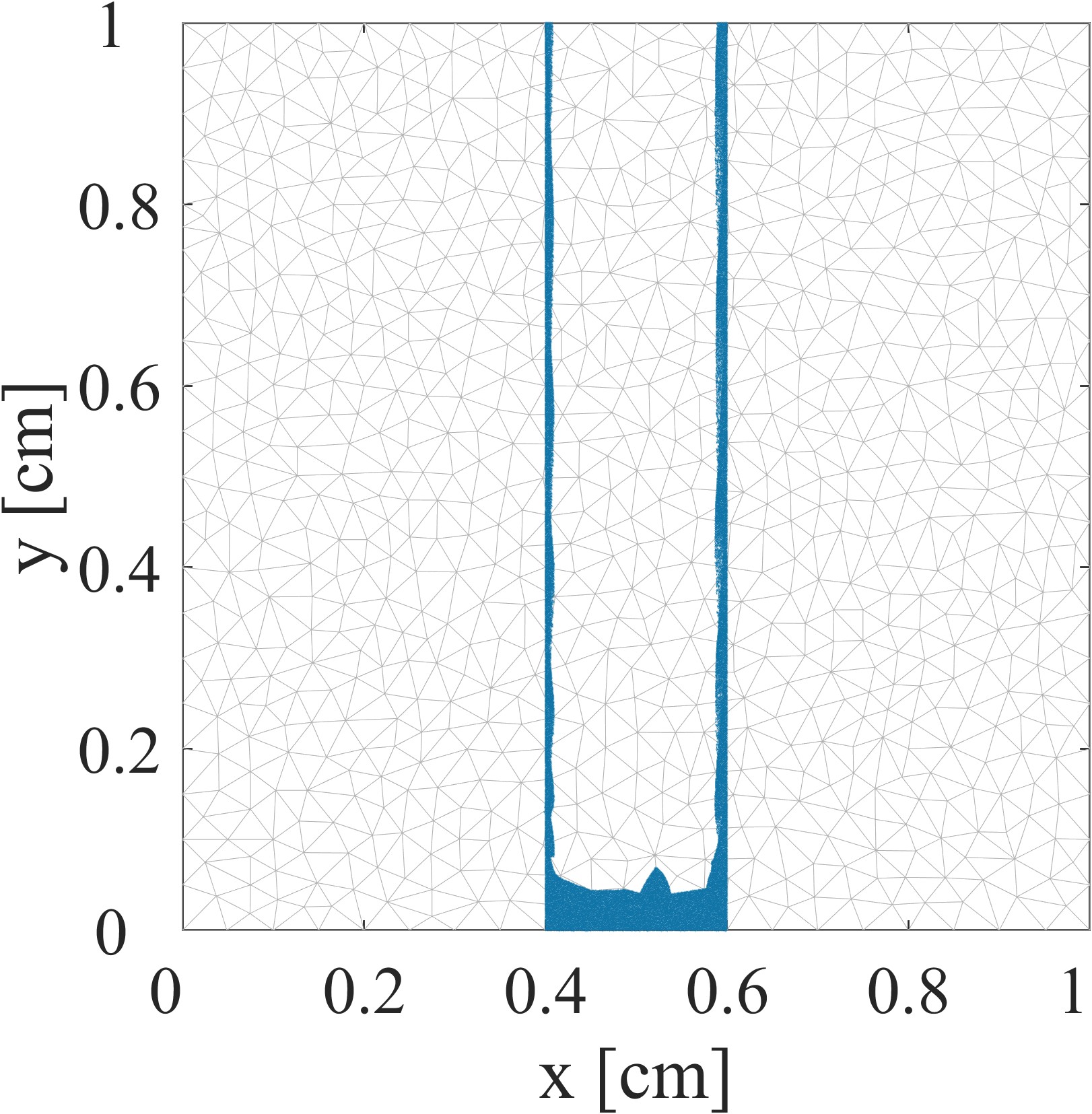}}
       \hfill
  \subfloat[Current density snapshot at {$t=64$ ns}.\label{fig:crnt_snap_virt} ]{%
        \includegraphics[width=0.345\linewidth]{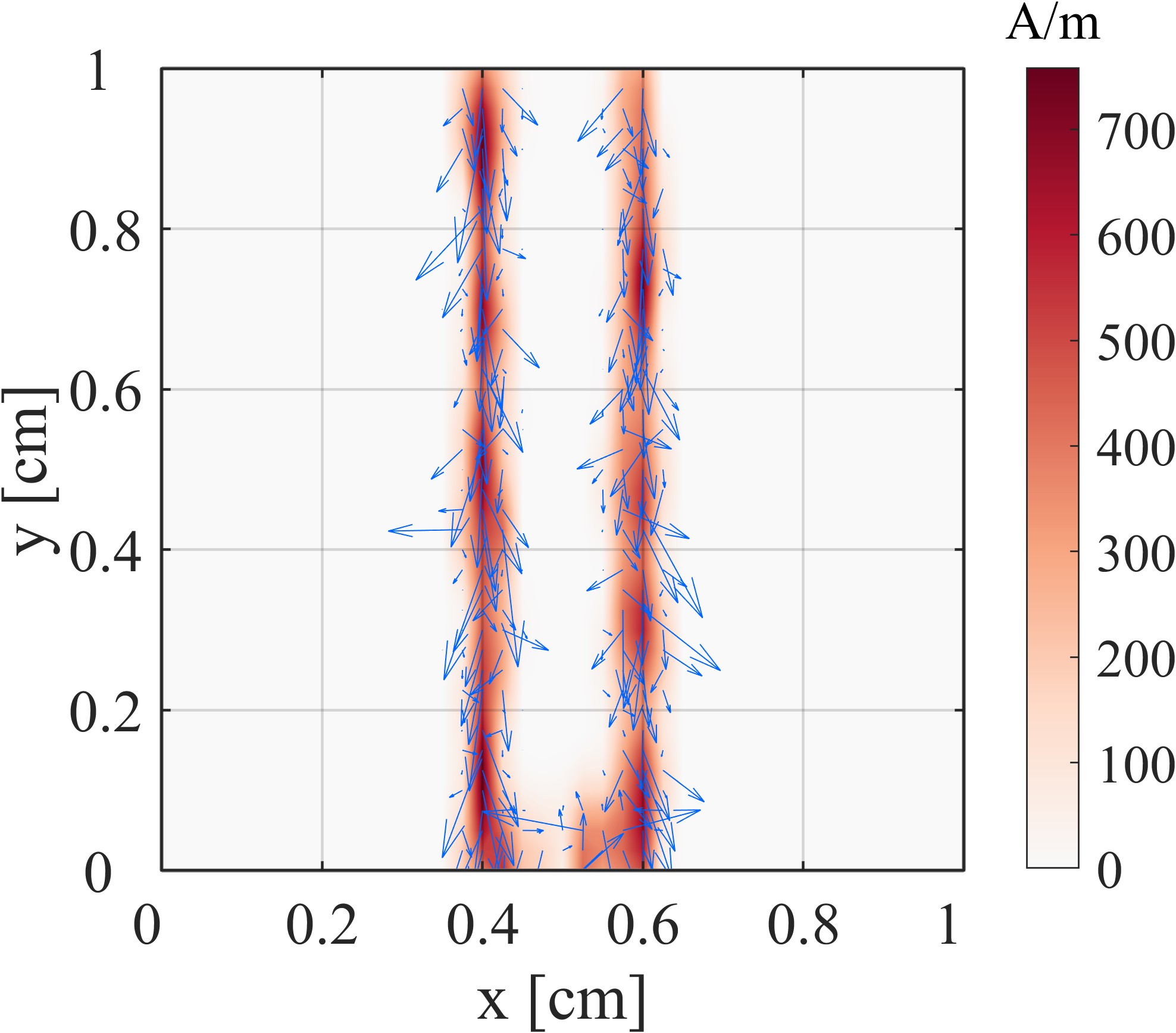}}
        \hfill
  \subfloat[Singular values for virtual cathode oscillations.\label{fig:singvals_virt} ]{%
    \includegraphics[width=0.30\linewidth]{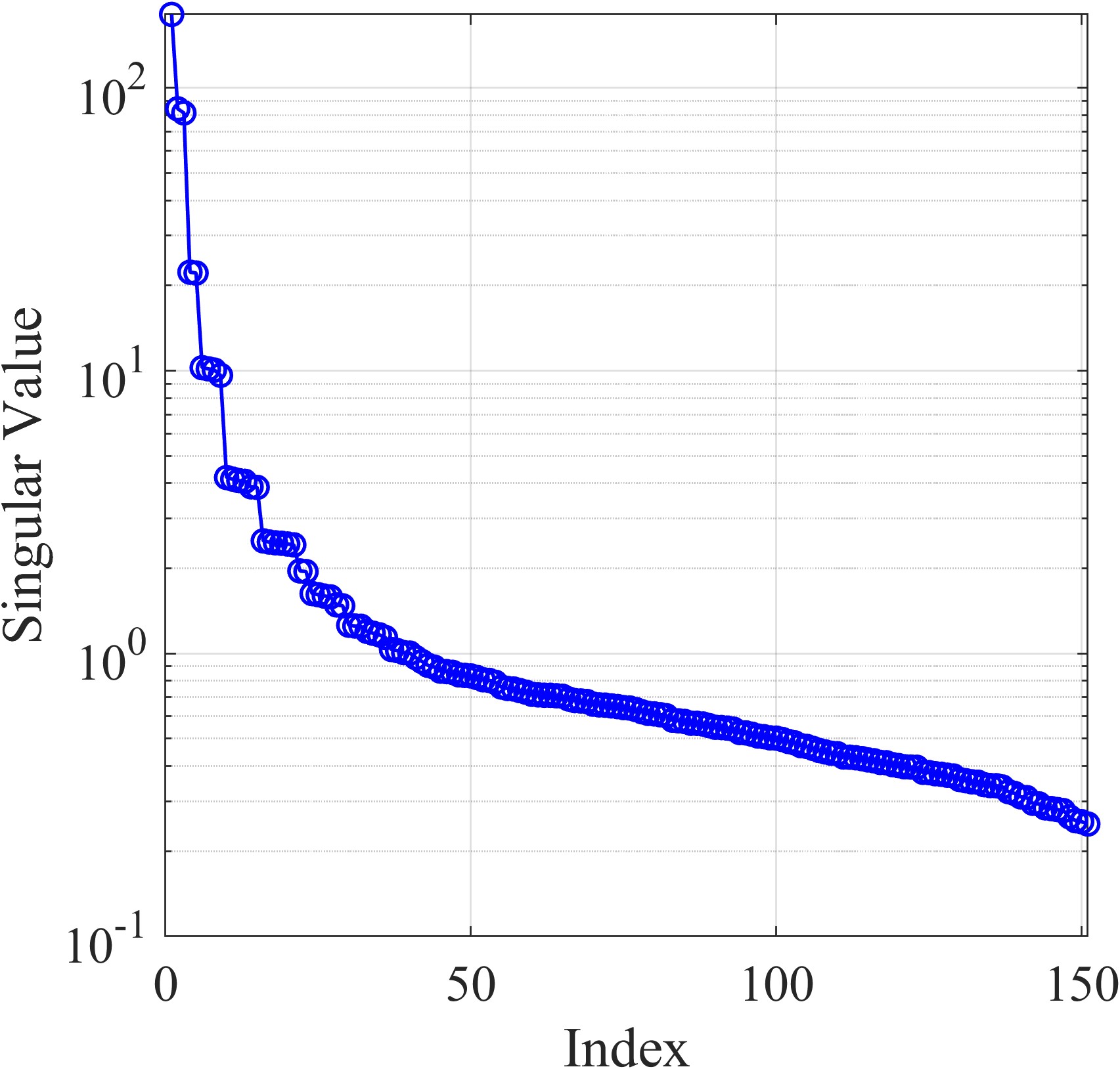}}
  
  \caption{\small{ (a) Snapshot of the virtual cathode formation at $t=64$ ns. (b) Snapshot of the current density at $t=64$ ns. (c) The singular values after performing SVD on the snapshots of current density.}  }\label{fig:setup_sing_virt}
\end{figure*}

We reiterate that by $m^\text{th}$ DMD mode, we refer to the $\{\bm\phi_m,\overline{\bm\phi}_m\}$ pair in \eqref{eq:DMD_recon_conj}. While plotting the modes (Fig. \ref{fig:DMD_wavy}) as well as calculating $\rho$, we use $\bm\psi_m=(\bm\phi_m+\overline{\bm\phi}_m)=2\Re\{\bm\phi_m\}$, where $\Re\{\cdot\}$ represents the real part. The spatial correlation matrix shows that the dominant DMD modes are orthogonal to each other with off-diagonal elements close to zero. However, it is important to note that unlike POD, DMD does not ensure orthogonality in space, but guarantee orthogonality in time. \par

The DMD predicted current density (Fig. \ref{fig:crnt_wavy_DMD}) deep into the prediction (extrapolation) region is plotted against the current density from high-fidelity EMPIC simulation (Fig. \ref{fig:crnt_wavy_EMPIC}) for side-to-side comparison. The relative error given by \eqref{eq:rel_error}, is also plotted in Fig. \ref{fig:err_crnt_wavy}.
{
\begin{align}\label{eq:rel_error}
    \delta(t)=\frac{||\hat{\mathbf{j}}(t)-\mathbf{j}(t)||_2}{ ||\mathbf{j}(t)||_2},
\end{align}
where $||\cdot||_2$ indicates the 2-norm.} The average relative error in the extrapolation region is {$1.80\%$.} The \textit{gather, pusher} and \textit{scatter} stages are replaced by the DMD prediction $\hat{\mathbf{j}}$ for $t > t_f$ $(\equiv n > n_f)$ in the EMPIC simulation, as illustrated in Fig. \ref{fig:DMD-EMPIC}. The self-fields $\mathbf{e}$ and $\mathbf{b}$ generated beyond $t_f$ is compared against the self-fields generated from EMPIC simulation. The relative error is calculated in a similar manner as in \eqref{eq:rel_error}. The electric field patterns from EMPIC and DMD-EMPIC in the extrapolation region at $t=64$ ns are shown in Fig. \ref{fig:e_wavy_EMPIC}, and Fig. \ref{fig:e_wavy_DMD} showing good agreement. {The relative errors in the self electric and magnetic field are $0.94 \%$ and $2.60 \%$ respectively (Fig. \ref{fig:err_eb_wavy}). The gain in runtime is discussed in Section }\ref{sec:comp_gain}.

\begin{figure*} [t]

    \centering
  \subfloat[DMD mode 1 with $f_1=0$ (DC). \label{fig:virt_mode01} ]{%
       \includegraphics[width=0.315\linewidth]{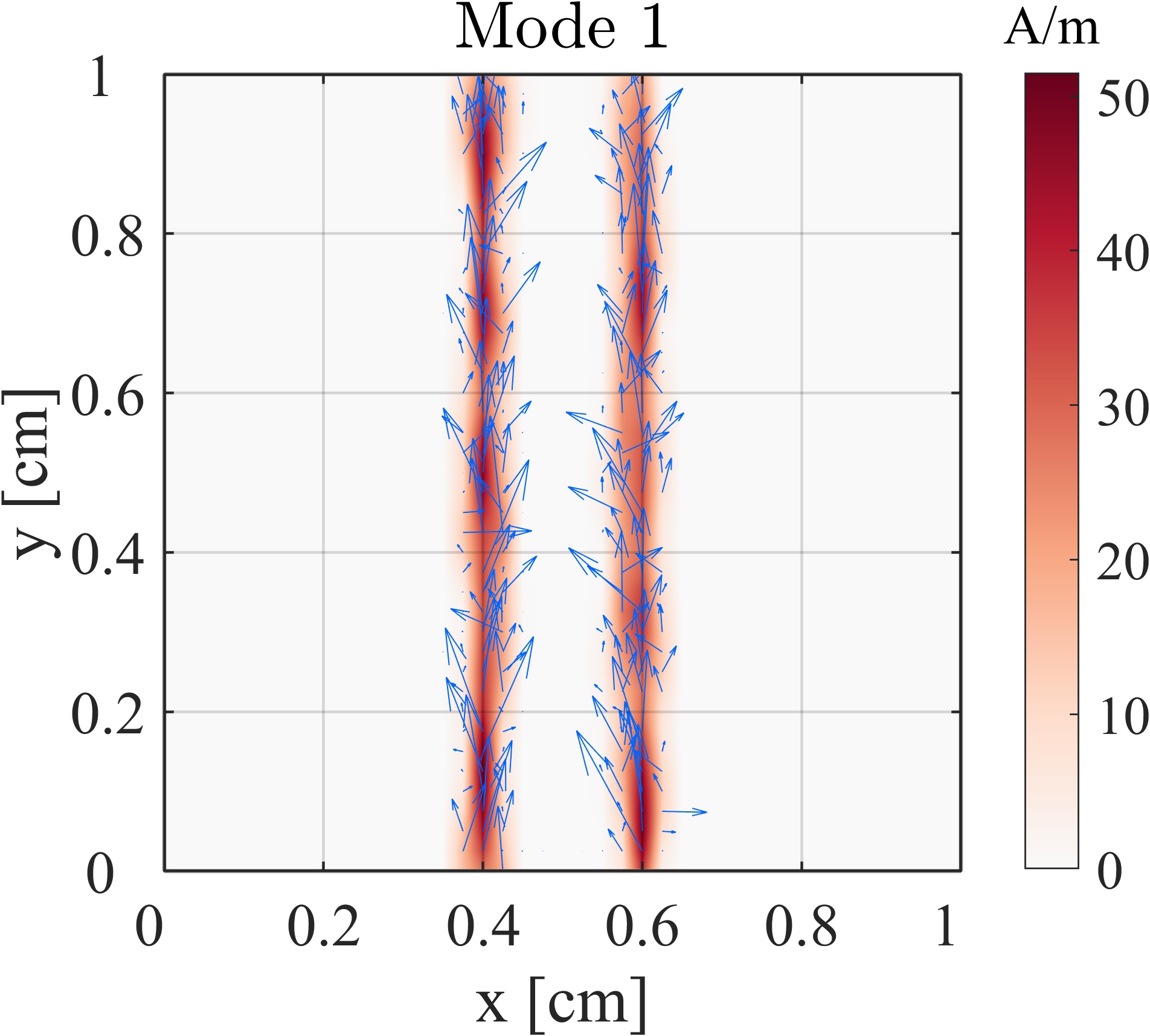}}
    \hfill
  \subfloat[DMD Mode 2 with $f_2=1.66$ GHz. \label{fig:virt_mode02} ]{%
        \includegraphics[width=0.325\linewidth]{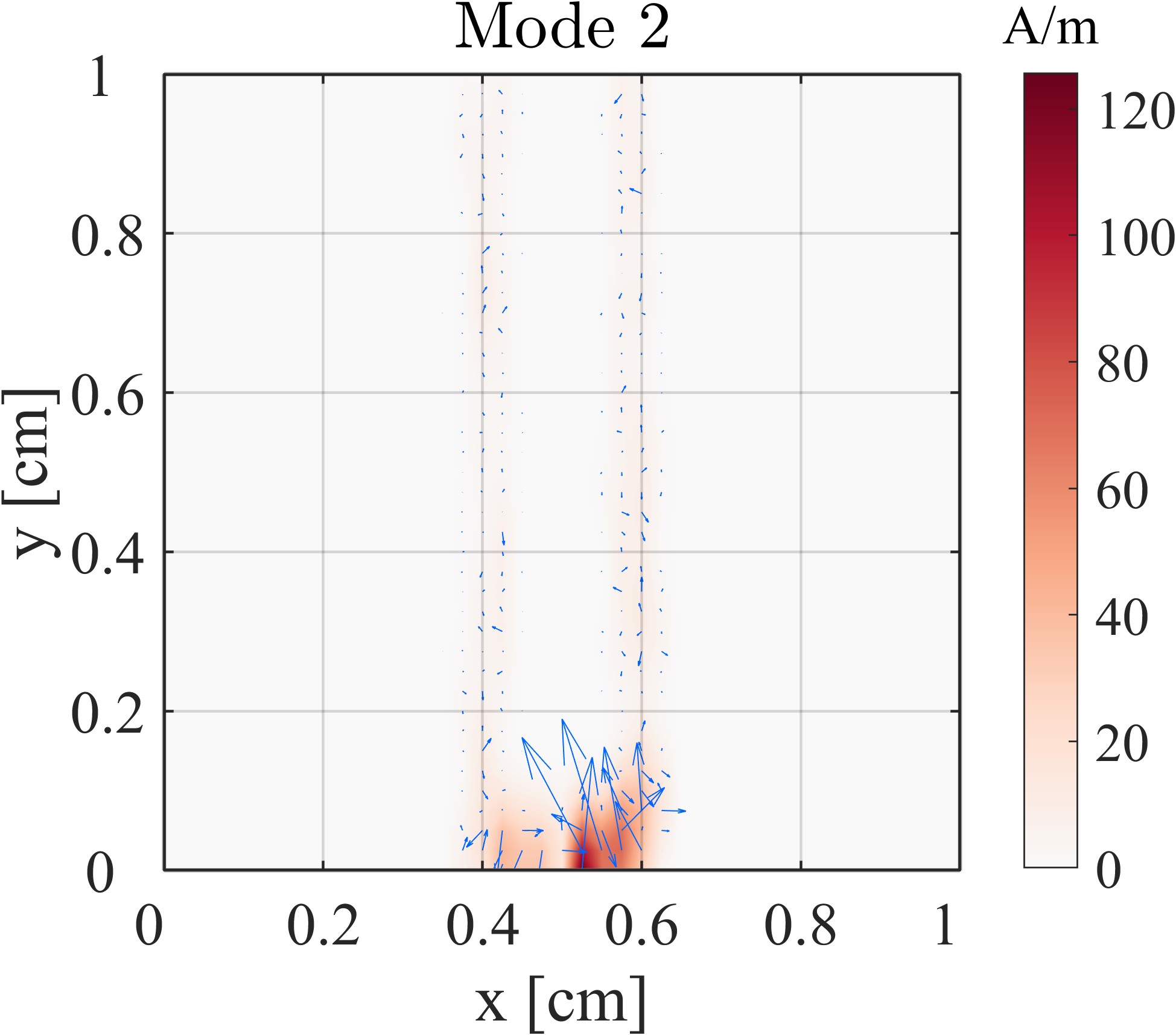}}
     \hfill       
    \subfloat[DMD eigenvalues.\label{fig:eigs_virt} ]{%
        \includegraphics[width=0.34\linewidth]{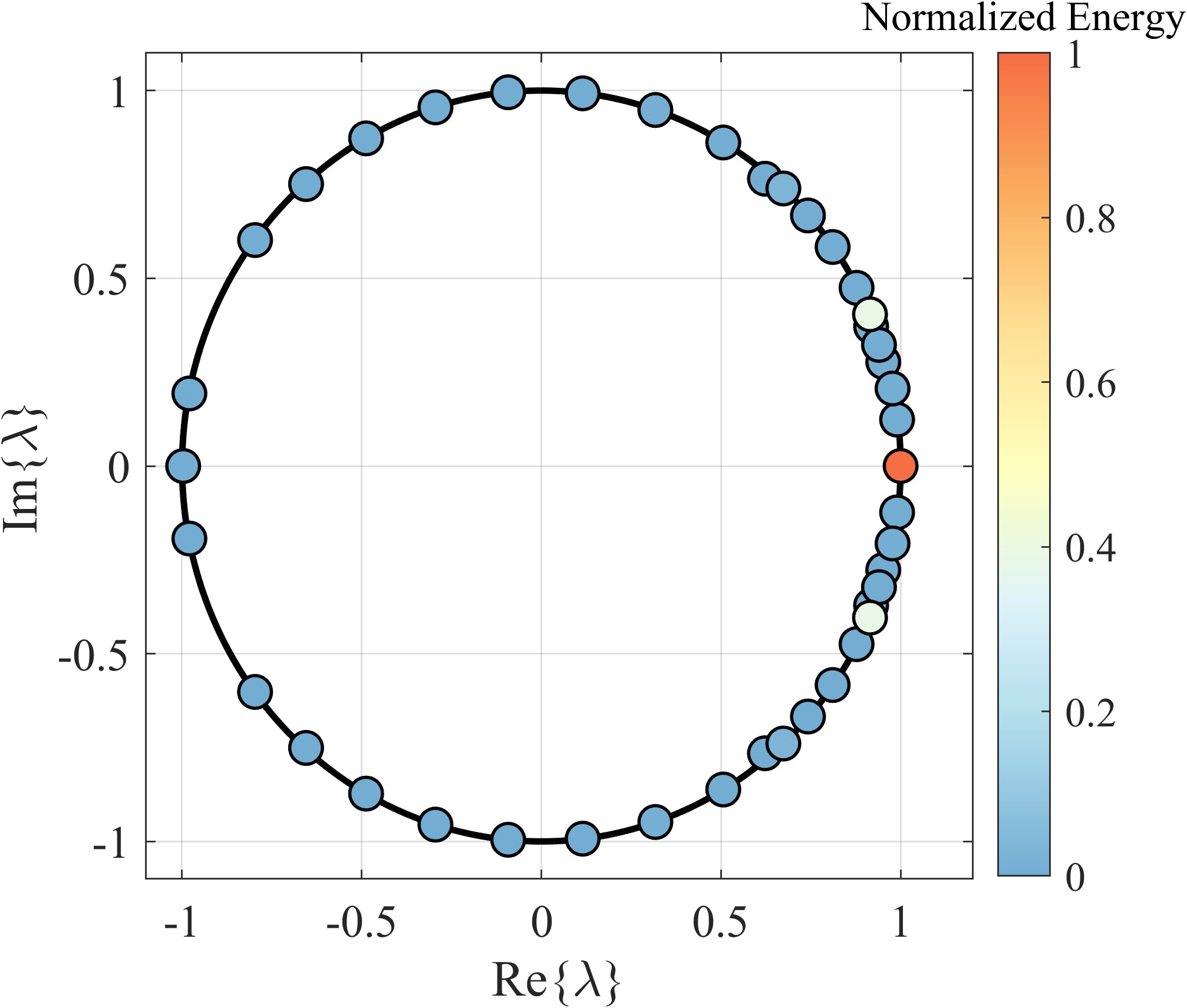}}
    \hfill 
  \caption{\small{ Extracted DMD modes and eigenvalues for current density modeling in virtual cathode oscillations. }  }\label{fig:DMD_virt}
\end{figure*}

\subsection{Virtual Cathode Oscillations}
Next we consider a more challenging example of virtual cathode oscillation. We use the same setup as in Section \ref{sec:eg_wavy_beam}, however with the following modifications. We increase the current injection by both increasing the superparticle ratio to $r_{sp}=75000$ and the superparticle injection rate to $200$ per time-step. The superparticles are injected at the bottom in the region $[0.4~\text{cm},~0.6~\text{cm}]$. Instead of a transverse oscillating magnetic flux, we apply a strong confining magnetic field, $\mathbf{B} = B_y\hat{y}$ along the $y$ direction, with $B_y=100$ T. The simulation is run until $n=320000$ time-steps or $t=64$ ns with time-step $\Delta t=0.2$ ps. The snapshot of the beam after virtual cathode formation at $t=16$ ns, and the corresponding current density plot are shown in Fig. \ref{fig:beam_snap_virt} and Fig. \ref{fig:crnt_snap_virt} respectively. Modeling the current density for virtual cathode oscillations is particularly challenging because there are no external forces dictating a clear oscillation pattern of the electrons. The majority of oscillations are limited to a small region (near the bottom) causing possible rank deficiency, and the leakage from the sides makes variation of $\mathbf{j}$ more prone to the particle noise. \par

\textbf{On-the-fly DMD on fields:} Similar to the oscillating beam case, the on-the-fly DMD is carried out to detect the end of transience indicating the onset of virtual cathode oscillations. {We select $\Delta t_w = 8$ ns, $\Delta_t=8$ ps, $\delta t_w=0.4$ ns, $r=200$, and number of Hankel stacks $d=10$. The onset of the virtual cathode oscillations is detected at $t_f=30.40$ ns. The on-the-fly DMD parameters are provided in details in Table \ref{table:onthefly_DMD}.}

\textbf{Offline DMD on current density:} The training parameters for the Offline DMD is summarized in Table \ref{table:dmd_params}. Fast decay of the singular values in Fig. \ref{fig:singvals_virt} indicates that most of the energy is concentrated in small number of modes. First two most energetic DMD modes carry more than $95\%$ of the total energy, and are plotted in Fig. \ref{fig:DMD_virt}. Mode 1 (Fig. \ref{fig:virt_mode01}) represent the DC leakage current from sides of the virtual cathode, whereas mode 2 (Fig. \ref{fig:virt_mode02}) captures the oscillations near the root of the beam, having frequency {of $1.65$ GHz}. The DMD eigenvalues (Fig. \ref{fig:eigs_virt}) also indicate the dominance of the DC mode in terms of energy. \par

\begin{figure} [t]

    \centering
  \subfloat[Relative error in DMD predicted current density.\label{fig:err_c_virt} ]{%
       \includegraphics[width=0.95\linewidth]{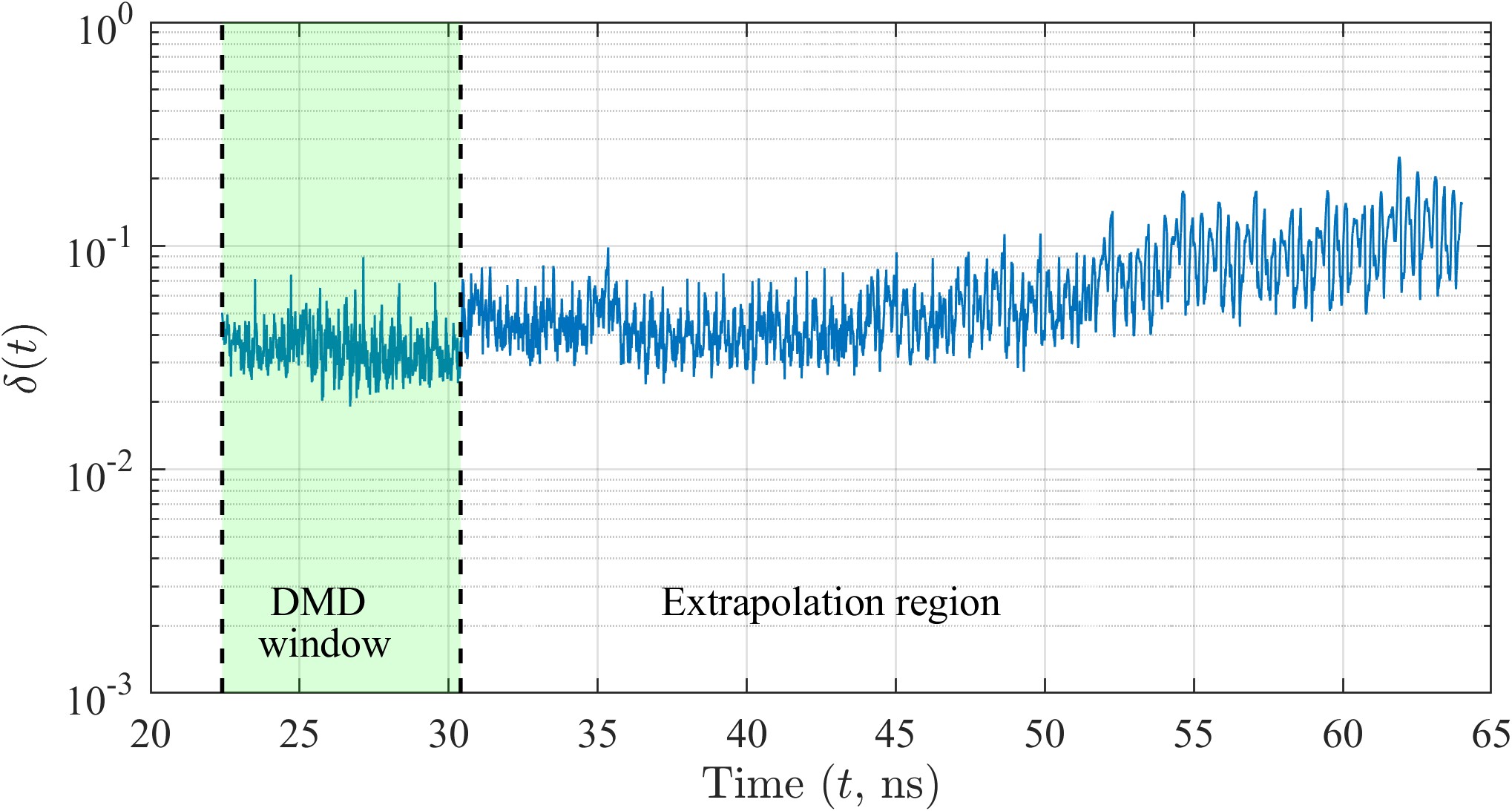}}\\
  \subfloat[Relative errors in the DMD-EMPIC simulated self-fields. \label{fig:virt_eb_err} ]{%
        \includegraphics[width=0.95\linewidth]{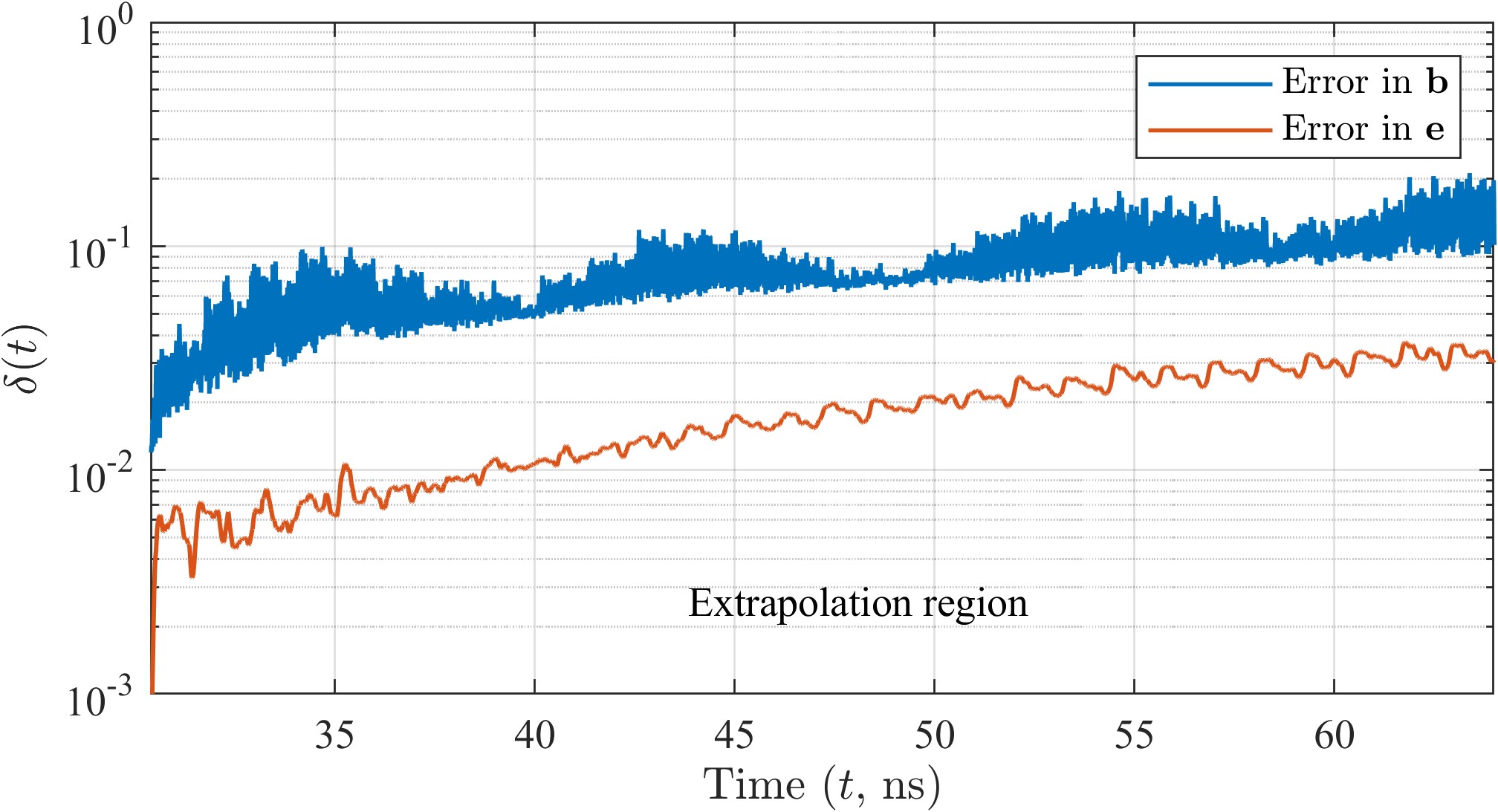}}
  \caption{\small{ Relative errors for the virtual cathode oscillations. }  }\label{fig:err_virt}
\end{figure}

With the help of extracted DMD modes, frequencies and modal amplitudes, current density is extrapolated for $t>t_f$ using DMD. The relative error in predicted current is shown in Fig. \ref{fig:err_c_virt}. The average error in extrapolation region around {$6.65\%$, which is higher compared to the oscillating beam case $(1.80\%)$.} This is expected because the leakage from both sides of the virtual cathode results in non-smooth current variation due to high-particle noise. Also, the localized nature of the oscillation contributes to a possible rank-deficiency resulting in higher error. The relative error in self electric field $\mathbf{e}$ and magnetic flux $\mathbf{b}$ is shown in Fig. \ref{fig:virt_eb_err}. {The average relative error in $\mathbf{e}$ is around $1.81\%$ whereas in $\mathbf{b}$ is around $8.10\%$.} Error in $\mathbf{b}$ is typically higher for both the test-cases since self magnetic field is generally very low in magnitude, and more susceptible to particle and numerical noise.

\begin{figure*}[t]
    \centering
    \subfloat[Snapshot of electron beam inside the BWO at $t=140$ ns.]{\includegraphics[width=1\linewidth] {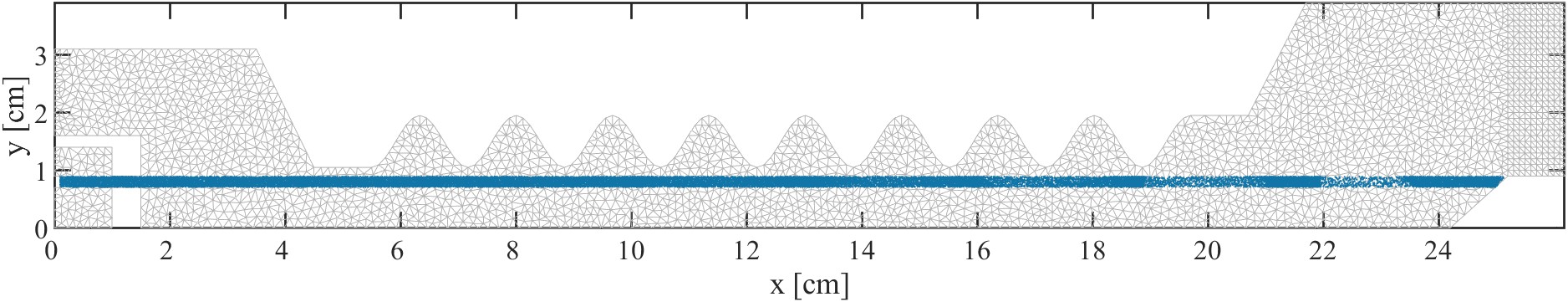}\label{fig:bwo_snap}}\\
    \subfloat[DMD mode 1 (DC mode)]{\includegraphics[width=1\linewidth]{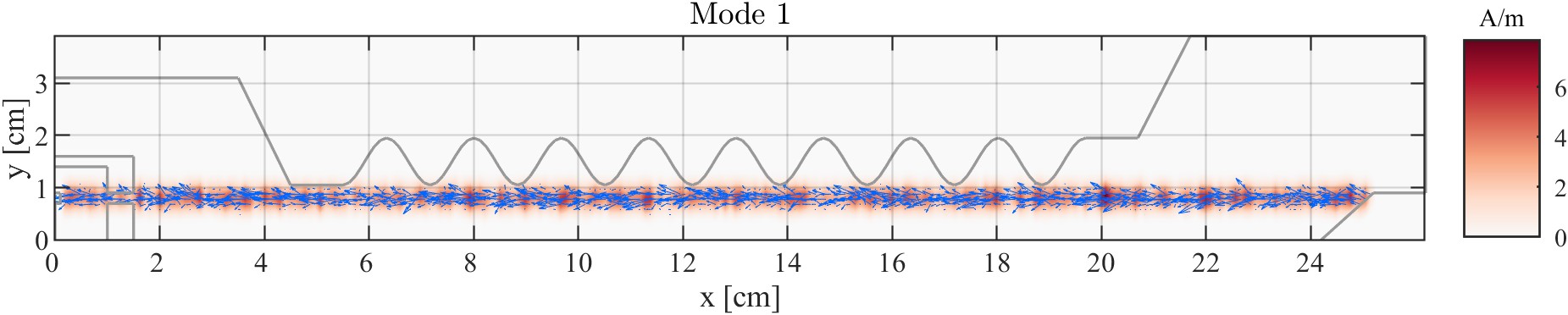}\label{fig:bwo_mode01}}\\
    \subfloat[DMD mode 2 with $f_2=8.32$ GHz.]{\includegraphics[width=1\linewidth]{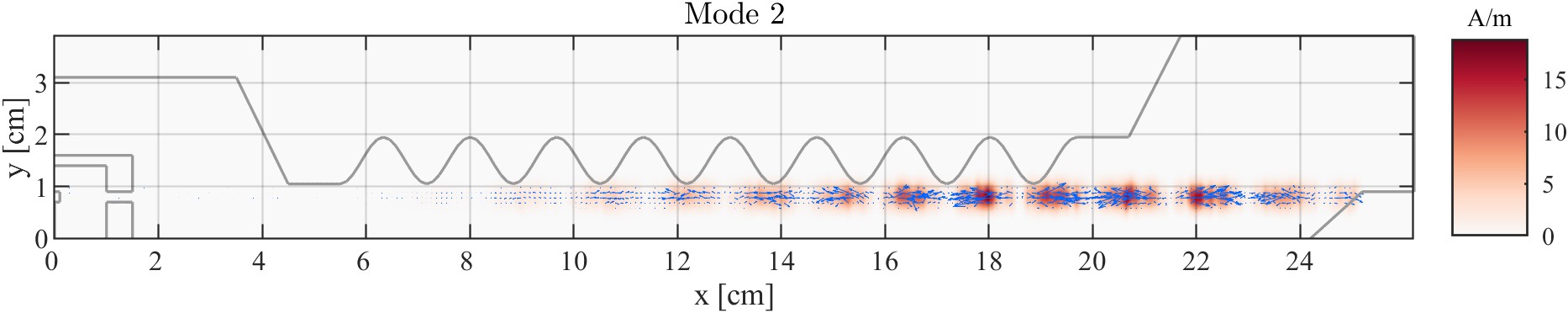}\label{fig:bwo_mode02}}
    \caption{Offline DMD on current density of a backward wave oscillator (BWO).}
    \label{fig:bwo_dmd}
\end{figure*}

{
\subsection{Backward Wave Oscillator}
Now we consider a more challenging and practical 2.5D case of a backward wave oscillator (BWO) \cite{na2017axisymmetric}. The finite element discretization of the longitudinl cross-section of a sinusoidally corrugated slow-wave structure (SCSWS) is depicted Fig. \ref{fig:bwo_snap}. The SCWS has boundary profile $R(z) = \frac{1}{2}(A + B) + \frac{1}{2}(A - B) \cos(\frac{2\pi}{C}z)$. Based on an eigenmode analysis, the SCSWS is designed to have $A = 19.5$ mm, $B = 10.5$ mm, $C = 16.7$ mm, and $N_{\text{crg}} = 8.5$. Each superparticle in the EMPIC model represents $r_{sp}=1.25 \times 10^8$ electrons, with injection rate of $20$ superparticles per time-step ($\Delta_t=0.5$ ps). The superparticles are injected in a random fashion with uniform distribution within the region centered around $y=0.008$ m with beam width $0.0018$ m. Superparticles are injected with a velocity of $2.5\times 10^8$ m/s in the $x$-direction. The simulation is run for $440,001$ time-steps or $t=220.005$ ns. The fundamental frequency of the BWO is $8.31$ GHz. \par

The on-the-fly algorithm detects  end of transience at $t_f=107.20$ ns with $\Delta t_w = 16$ ns, $\Delta_t= 2$ ps and $\delta t_w=1.6$ ns (see Table \ref{table:onthefly_DMD} for details). The training parameters for offline DMD on current density are provided in Table \ref{table:dmd_params}. The average extrapolation error in the current density is $13.52\%$ (Fig. \ref{fig:bwo_err}). The higher error can be attributed to particle noise which plays a significant role in the absence of a strong external force dictating the oscillations (oscillating beam). The most energetic mode is the DC mode (Fig. \ref{fig:bwo_mode01}) followed by the mode (Fig. \ref{fig:bwo_mode02}) oscillating with fundamental frequency at $f_2=8.32$ GHz. As expected, the oscillations are concentrated towards the end of BWO where the bunching of electrons (superparticles) occurs. The error in the self electric field from the DMD-EMPIC framework is $5.58\%$ (Fig. \ref{fig:bwo_err}).

\begin{figure} [t]

    \centering
      \includegraphics[width=0.95\linewidth]{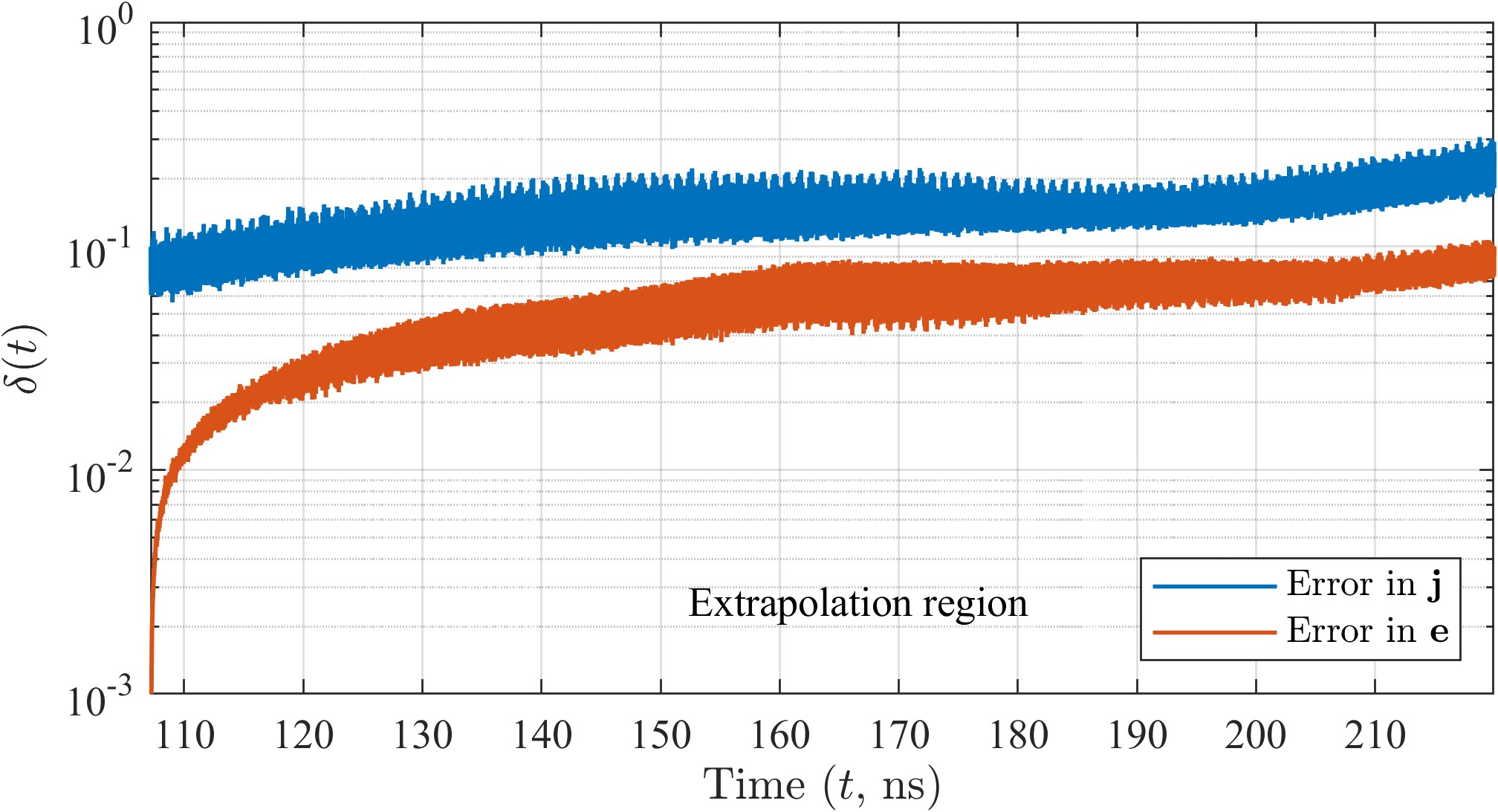}
  \caption{\small{Relative 2-norm error in current density $\mathbf{j}$ (DMD extrapolation) and self electric field $\mathbf{e}$ (DMD-EMPIC framework) for the backward wave oscillator. }
   \label{fig:bwo_err}}
\end{figure}

} 

\section{Computational gain}\label{sec:comp_gain}
\subsection{Computational complexity}
The time complexity of typical EMPIC algorithm for each time-step {with \textit{explicit} field solver} is given by $\mathcal{O}(N+N_p)$, where $N$ is the aggregate mesh dimension and $N_p$ is the total number of (super)particles. In a traditional EMPIC setting $N_p\gg N$, thus the number of particles remains the primary bottleneck for expediting EMPIC simulations. If the simulation is run until $n_q$ time-steps, the total time-complexity is given by $\mathcal{O}(Nn_q+N_pn_q)$. \par

The computation cost of repeated calculation of the DMD features for the on-the-fly application also adds to the typical EMPIC cost for $n \leq n_f~(\equiv t \leq t_f)$ in the DMD-EMPIC algorithm. The computation complexity of DMD is dominated by the SVD step, {resulting in time-complexity of $\mathcal{O}(l_d^2Nd)$} \cite{Note4}, where $l_d$ is the number of DMD snapshots after $d$ number of Hankel stacking. {Time complexity for randomized DMD with target rank $r(<l_d)$, is typically $\mathcal{O}(rl_dNd)$.}Note that as mentioned in \cite{hemati2014dynamic}, SVD need not be recalculated every time we shift the DMD window. The resulting DMD features can be calculated in an incremental manner. In the worst case scenario, let us consider that the SVD is performed for each sliding window. Also, in the worst case scenario, the DMD window is shifted by only one snapshot, i.e. the amount of shift in terms of time-steps $\Delta_{ns}=\Delta_n$. With these considerations, {the cost of the sliding-window on-the-fly DMD (randomized) is approximately $\mathcal{O}(\frac{rl_d Nd n_f}{\Delta_n})$}, where $n_f$ is the time-step at which EMPIC is stopped, and $\Delta_n$ is the number of time-steps between two consecutive DMD snapshots.\par

The time-complexity of the DMD-EMPIC algorithm for $n<n_f~(\equiv t<t_f)$ is {$\mathcal{O}(Nn_f+N_pn_f+\frac{rl_dNdn_f}{\Delta_n})$, whereas for $n > n_f~(\equiv t > t_f)$ is $\mathcal{O}(N(n_q-n_f)+l_d^2Nd)$ considering one-time cost of offline DMD with $d$ Hankel stacks. Since there are no particles involved in case of DMD-EMPIC for $n > n_f$, $N_p$ does not appear in the time complexity.} For a simulation run until $n=n_q$, the time-complexity of the DMD-EMPIC is {$\mathcal{O}(Nn_q+N_pn_f+\frac{rl_dNdn_f}{\Delta_n}+l_d^2Nd)$.} The time-complexities for EMPIC and DMD-EMPIC are summarized in Table \ref{table:time_complexity}.

\begingroup
\setlength{\tabcolsep}{10pt} 
\renewcommand{\arraystretch}{1.8} 
\begin{table}
\begin{center}
\caption{EMPIC and DMD-EMPIC complexities {with explicit field solver.}}
\label{table:time_complexity}
\begin{ruledtabular}
\begin{tabular}{ cc } 
\textbf{EMPIC} & $\mathcal{O}(Nn_q+N_pn_q)$ \\
\textbf{DMD-EMPIC} &  {$\mathcal{O}(Nn_q+N_pn_f+\frac{rl_dNdn_f}{\Delta_n}+l_d^2Nd)$}\\
\end{tabular}
\end{ruledtabular}
\end{center} 
\end{table} 
\endgroup

\subsection{Runtime comparison }
As mentioned earlier, in typical EMPIC setting $N_p\gg N$. As long as $n_q$ is moderately larger than $n_f$, there will be significant gain in the runtime. For late-time queries $(n_q\gg n_f)$, the ratio of runtimes for EMPIC $(T_q)$ and DMD-EMPIC $(\widehat{T}_q)$ can be roughly given by, {$\widehat{T}_q/T_q\approx n_f/n_q$, given field solver takes negligible time compared to the entire simulation. Note that for large scale problems such as BWO, this approximation does not necessarily hold as evident from Table \ref{table:runtime}.}\par

{The} simulation runtime depends on several factors including the specific computational platform and hardware, and code optimization. In this work, the numerical experiments are run on Intel Xeon E5-2680 v4 (Broadwell) compute CPU with 2.4 GHz and 14 cores per processor. Each node has $128$ GB of memory. The interconnect used is Mellanox EDR Infiniband Networking (100Gbps). Each simulation job was allocated 1 node and 5 cores. {The node runtimes for all the test cases are listed in Table \ref{table:runtime}. Note that the CPU runtime is approximately 5 times the node runtime, exhibiting good shared-memory parallelization across cores. }

\begingroup
\setlength{\tabcolsep}{10pt} 
\renewcommand{\arraystretch}{1.5} 
\begin{table}
\begin{center}
\caption{Node runtime in days.} 
\label{table:runtime} 
\begin{ruledtabular}
\begin{tabular}{ c c c } 
\textbf{\makecell{}} & \textbf{\makecell{EMPIC}} & \textbf{\makecell{DMD-EMPIC}}\\
\hline
Osc. Beam & 4.43 & 1.66\\ 
Vircator & 5.96 & 2.93\\
BWO & 7.04 & 4.08\\
\end{tabular}
\end{ruledtabular}
\end{center} 
\end{table} 
\endgroup

{
\subsection{Effect of Parallelization}
It is important to note that particle-in-cell (PIC) algorithms are highly parallelizable and an appropriate parallel computing architecture can be employed to accelerate EMPIC simulations \cite{decyk2014particle}. Single nodes with multiple CPUs (shared memory) or multiple nodes (each with one or more CPUs and distributed memory) can reduce the simulation time from days to several hours. {Fortunately, the DMD-EMPIC algorithm should also achieve comparable acceleration from parallelization for the following reason:} let the computation time for performing DMD (online + offline) be $T_{DMD}$. Let the runtime for the original EMPIC simulation up to the desired query time be $T_q$. Suppose the on-the-fly (online) DMD raises the flag to stop the EMPIC simulation at $T_f (<T_q)$. The relative gain in computation time $G_T$ is,
\begin{align}\label{eq:comp_gain}
G_T = \frac{T_q}{T_f+T_{DMD}+T_{FS}} \approx \frac{T_q}{T_f}
\end{align}

, where $T_{FS}$ is the time taken by the field solver {beyond $T_f$}. This approximation holds if $T_{DMD}$ and $T_{FS}$ are negligible compared to $T_f$. Depending on the scale of the problem, type of the field solver and parallelization, $T_{FS}$ can be negligible compared to $T_f$. $T_{DMD}$ is usually much less than $T_f$ even with parallelization, because parallelization not only helps accelerate EMPIC but also the DMD computation. For example, authors in \cite{sayadi2016parallel} utilize a parallel Tall and Skinny QR (TSQR) algorithm for parallelizing the SVD computation. The construction of the Koopman operator, eigendecomposition, and the construction of DMD modes are also done in an embarrassingly parallel fashion. Recently, authors in \cite{maryada2022reduced} used a communication-optimal parallel TSQR algorithm for reduced-communication parallel DMD. These parallel DMD methods are reported to scale well. A detailed study of how parallel EMPIC scales compared to parallel DMD in the presence of parallel computing architecture is beyond the scope of this paper. Since both DMD and EMPIC can benefit from parallel computation, the overall acceleration by DMD-EMPIC depends on how early the system reaches equilibrium or ``pseudo'' equilibrium, i.e., how small $T_f$ is compared to $T_q$. Consequently, Eq. \eqref{eq:comp_gain} is independent of the hardware platform or the type of PIC algorithms used, as both $T_f$ and $T_q$ are similarly affected. 
}
\section{Conclusion and Look Ahead}\label{sec:conclusion}

In this work we have demonstrated a reduced-order framework for modeling space-charge dynamics from EMPIC simulations by performing dynamic mode decomposition (DMD) on the current density. Extracted features such as DMD spatial modes and frequencies help to discover and analyze the dominant physics. DMD-EMPIC shows great promise in reducing overall runtime of EMPIC simulations via time-extrapolation of the current density data to future time. However, similar to any other data-driven method, the extrapolation accuracy depends on the quality of the training data. {While our ultimate goal is to extrapolate from very early (transient) time-series data, current time-extrapolation methods like DMD and recurrent neural networks (RNNs) require further development to realize this fully. In the meantime, a more viable and effective approach is to interpolate time-series data of current density across various simulation parameters. This technique, rooted in reduced-order modeling of current density, offers a promising solution to the longstanding challenge of accelerating EMPIC plasma simulations.} Another important aspect is ensuring conservation laws in the extrapolated evolution. Currently the DMD-EMPIC algorithm does not enforce energy conservation explicitly but only approximately as the predicted current density and fields are close to the original EMPIC quantities. As a future work, physical constraints such as conservation laws may be explicitly included in the training and extrapolation process. Another possible venue of future work is to make the DMD parameter selection fully automatic.

\section{Acknowledgments}
This work was funded in part by the Department of Energy Grant DE-SC0022982 through the NSF/DOE Partnership in Basic Plasma Science and Engineering and in part by the Ohio Supercomputer Center Grant PAS-0061.

\appendix

{
\section{\label{sec:onthefly_algo}On-the-Fly DMD Algorithm}

The on-the-fly DMD algorithm \cite{nayak2023fly} for identifying the end of transient is based on the invariance of the DMD extracted features irrespective of the location of DMD window in equilibrium, as long as the DMD window is wide enough to capture the dynamics. Invariance in DMD extracted features lead to invariance in the DMD predicted solution at a fixed time window. However, when the system is in transience, no such claim regarding the invariance of extracted DMD features can be made. It is important to note that the on-the-fly DMD necessitates frequent recalculations of DMD algorithm, which can become computationally intensive for large datasets. To address this challenge, we employ the randomized DMD algorithm as proposed in \cite{erichson2019randomized}. While this approach offers greater efficiency, it does so at the cost of some accuracy. However, this trade-off is acceptable in our context, as the primary focus during the on-the-fly stage is to capture the system's overall behavior rather than precise reconstruction.\par

The hyperparameters in Algorithm \ref{algo:onthefly_DMD} include $\delta n_w$, $\eta$, $q$, and $\delta_0$. The shift between consecutive DMD windows, denoted by $\delta n_w$ (equivalently $\delta t_w=\delta n_w\Delta t$), balances precision and computational cost. Specifically, a smaller $\delta n_w$ enhances the accuracy in identifying $t_f$ but increases the number of required DMD computations. Conversely, a larger $\delta n_w$ reduces computation at the expense of precision. For large-scale problems, a larger $\delta n_w$ is advisable. The factor $\eta$ determines the extent of time for comparing DMD predictions, reflecting our prediction goals for equilibrium behavior. The decision to halt the EMPIC simulation is based on whether the relative error between DMD predictions remains below $\delta_0$ across $q$ consecutive windows, indicating the end of the transient state. The choice of $\delta_0$ reflects the desired precision in DMD extrapolation. Opting for a higher value of $q$ increases robustness against noise in data, but may delay the detection of $t_f$. The hyperparameters for Algorithm \ref{algo:onthefly_DMD}, including parameters for randomized DMD \cite{erichson2019randomized} $(r,\Delta_t,d)$, are provided in Table \ref{table:onthefly_DMD}. Note that $r$ for randomized DMD corresponds to the target rank of randomized SVD operation.

\begingroup
\setlength{\tabcolsep}{10pt}
\renewcommand{\arraystretch}{1.5}
\begin{table}
\begin{center}

\caption{On-the-fly DMD parameters}
\label{table:onthefly_DMD}
\begin{ruledtabular}
\begin{tabular}{ c c c c } 
\textbf{\makecell{Parameters}} & \textbf{\makecell{Osc. Beam}} & \textbf{\makecell{Vircator}} & \textbf{\makecell{BWO}}\\
\hline
$\Delta t_w$ & $8$ ns & $8$ ns & $16$ ns \\ 
$\delta t_w$ & $0.4$ ns & $0.4$ ns & $1.6$ ns \\ 
$\eta$ & $10$ & $10$ & $10$ \\
$q$ & $5$ & $5$ & $5$ \\
$\delta_0$ & $0.05$ & $0.05$ & $0.05$ \\
$r$ & $200$ & $200$ & $200$ \\
$\Delta_t$ & $8$ ps & $8$ ps & $2$ ps \\
$d$ & $10$ & $10$ & $5$ \\
\end{tabular}
\end{ruledtabular}
\end{center} 
\end{table}
\endgroup

\begin{algorithm}[H] 
 \caption{On-the-fly sliding-window DMD algorithm for detecting $t_f$ for real-time termination of EMPIC.}
   
 \begin{algorithmic}[1]\label{algo:onthefly_DMD}
 \renewcommand{\algorithmicrequire}{\textbf{Input: }}
 \renewcommand{\algorithmicensure}{\textbf{Output:}}
 \REQUIRE 
      i) DMD window width $\Delta t_w$ or equivalently $\Delta n_w$ based on the approximate idea of the timescale of the problem.\\
     ii) Electric field data from the ongoing EMPIC simulation.\\

 \ENSURE EMPIC termination flag.
 \STATE {After the EMPIC simulation reaches $n=\Delta n_w$, shift the DMD window by $\delta n_{w}$ time-steps (represented by increasing window index $k$) as the simulation progresses. Let us denote the starting and end of the $k^{\text{th}}$ window by $t_{st,k}(\equiv n_{st,k})$ and $t_{en,k}(\equiv n_{en,k})$ respectively.} 
 \FOR {Current ($k^\text{th}$) DMD window}
 \STATE {Perform randomized DMD for $k^{\text{th}}$ window and obtain the DMD prediction from $n_k=(n_{en,k}+\eta \Delta n_w)$ to $(n_k + \Delta n_w)$. Let us denote this DMD prediction corresponding to $k^{\text{th}}$ DMD window as $\widehat{\mathbf{x}}_{k}^{(n)}$.}
 \STATE {Get the average relative 2-norm error between the overlapping region of $\widehat{\mathbf{x}}_k^{(n)}$ and $\widehat{\mathbf{x}}_{k-1}^{(n)}$ denoted by $\langle \delta^{(n)} \rangle_{w}$, where }
 \begin{align}
     \langle \delta^{(n)} \rangle_{k} = \frac{1}{(n_{k}-\delta n_w+1)}\sum_{n=n_k}^{n_k+\Delta n_w -\delta n_w}\frac{||\widehat{\mathbf{x}}_w^{(n)}-\widehat{\mathbf{x}}_{w-1}^{(n)}||_2}{||\widehat{\mathbf{x}}_{w-1}^{(n)}||_2}
 \end{align}
 \IF{(\(\langle \delta^{(n)} \rangle_{k}\)+ \(\langle \delta^{(n)} \rangle_{k-1}\)+ \(\ldots\)+ \(\langle \delta^{(n)} \rangle_{k-q+1}\))/q $<$ \(\delta_0\)}
 \STATE{Raise the EMPIC termination flag.}
 \RETURN
 \ELSE
 \STATE{
 Move to the next DMD window, $w=w+1$.}
 \ENDIF
 \ENDFOR
 \end{algorithmic}
 \end{algorithm}

} 


\begin{thebibliography}{59}%
\makeatletter
\providecommand \@ifxundefined [1]{%
 \@ifx{#1\undefined}
}%
\providecommand \@ifnum [1]{%
 \ifnum #1\expandafter \@firstoftwo
 \else \expandafter \@secondoftwo
 \fi
}%
\providecommand \@ifx [1]{%
 \ifx #1\expandafter \@firstoftwo
 \else \expandafter \@secondoftwo
 \fi
}%
\providecommand \natexlab [1]{#1}%
\providecommand \enquote  [1]{``#1''}%
\providecommand \bibnamefont  [1]{#1}%
\providecommand \bibfnamefont [1]{#1}%
\providecommand \citenamefont [1]{#1}%
\providecommand \href@noop [0]{\@secondoftwo}%
\providecommand \href [0]{\begingroup \@sanitize@url \@href}%
\providecommand \@href[1]{\@@startlink{#1}\@@href}%
\providecommand \@@href[1]{\endgroup#1\@@endlink}%
\providecommand \@sanitize@url [0]{\catcode `\\12\catcode `\$12\catcode
  `\&12\catcode `\#12\catcode `\^12\catcode `\_12\catcode `\%12\relax}%
\providecommand \@@startlink[1]{}%
\providecommand \@@endlink[0]{}%
\providecommand \url  [0]{\begingroup\@sanitize@url \@url }%
\providecommand \@url [1]{\endgroup\@href {#1}{\urlprefix }}%
\providecommand \urlprefix  [0]{URL }%
\providecommand \Eprint [0]{\href }%
\providecommand \doibase [0]{https://doi.org/}%
\providecommand \selectlanguage [0]{\@gobble}%
\providecommand \bibinfo  [0]{\@secondoftwo}%
\providecommand \bibfield  [0]{\@secondoftwo}%
\providecommand \translation [1]{[#1]}%
\providecommand \BibitemOpen [0]{}%
\providecommand \bibitemStop [0]{}%
\providecommand \bibitemNoStop [0]{.\EOS\space}%
\providecommand \EOS [0]{\spacefactor3000\relax}%
\providecommand \BibitemShut  [1]{\csname bibitem#1\endcsname}%
\let\auto@bib@innerbib\@empty
\bibitem [{\citenamefont {Gold}\ and\ \citenamefont
  {Nusinovich}(1997)}]{gold1997review}%
  \BibitemOpen
  \bibfield  {author} {\bibinfo {author} {\bibfnamefont {S.~H.}\ \bibnamefont
  {Gold}}\ and\ \bibinfo {author} {\bibfnamefont {G.~S.}\ \bibnamefont
  {Nusinovich}},\ }\bibfield  {title} {\bibinfo {title} {Review of high-power
  microwave source research},\ }\href@noop {} {\bibfield  {journal} {\bibinfo
  {journal} {Rev. Sci. Instrum.}\ }\textbf {\bibinfo {volume} {68}},\ \bibinfo
  {pages} {3945} (\bibinfo {year} {1997})}\BibitemShut {NoStop}%
\bibitem [{\citenamefont {Booske}(2008)}]{booske2008plasma}%
  \BibitemOpen
  \bibfield  {author} {\bibinfo {author} {\bibfnamefont {J.~H.}\ \bibnamefont
  {Booske}},\ }\bibfield  {title} {\bibinfo {title} {Plasma physics and related
  challenges of millimeter-wave-to-terahertz and high power microwave
  generation},\ }\href@noop {} {\bibfield  {journal} {\bibinfo  {journal}
  {Phys. Plasmas}\ }\textbf {\bibinfo {volume} {15}},\ \bibinfo {pages}
  {055502} (\bibinfo {year} {2008})}\BibitemShut {NoStop}%
\bibitem [{\citenamefont {Benford}\ \emph {et~al.}(2015)\citenamefont
  {Benford}, \citenamefont {Swegle},\ and\ \citenamefont
  {Schamiloglu}}]{benford2015high}%
  \BibitemOpen
  \bibfield  {author} {\bibinfo {author} {\bibfnamefont {J.}~\bibnamefont
  {Benford}}, \bibinfo {author} {\bibfnamefont {J.~A.}\ \bibnamefont
  {Swegle}},\ and\ \bibinfo {author} {\bibfnamefont {E.}~\bibnamefont
  {Schamiloglu}},\ }\href@noop {} {\emph {\bibinfo {title} {High Power
  Microwaves}}}\ (\bibinfo  {publisher} {CRC Press},\ \bibinfo {address} {New
  York},\ \bibinfo {year} {2015})\BibitemShut {NoStop}%
\bibitem [{\citenamefont {Wan}\ \emph {et~al.}(2015)\citenamefont {Wan},
  \citenamefont {Matthaeus}, \citenamefont {Roytershteyn}, \citenamefont
  {Karimabadi}, \citenamefont {Parashar}, \citenamefont {Wu},\ and\
  \citenamefont {Shay}}]{PhysRevLett.114.175002}%
  \BibitemOpen
  \bibfield  {author} {\bibinfo {author} {\bibfnamefont {M.}~\bibnamefont
  {Wan}}, \bibinfo {author} {\bibfnamefont {W.~H.}\ \bibnamefont {Matthaeus}},
  \bibinfo {author} {\bibfnamefont {V.}~\bibnamefont {Roytershteyn}}, \bibinfo
  {author} {\bibfnamefont {H.}~\bibnamefont {Karimabadi}}, \bibinfo {author}
  {\bibfnamefont {T.}~\bibnamefont {Parashar}}, \bibinfo {author}
  {\bibfnamefont {P.}~\bibnamefont {Wu}},\ and\ \bibinfo {author}
  {\bibfnamefont {M.}~\bibnamefont {Shay}},\ }\bibfield  {title} {\bibinfo
  {title} {Intermittent dissipation and heating in {3D} kinetic plasma
  turbulence},\ }\href {https://doi.org/10.1103/PhysRevLett.114.175002}
  {\bibfield  {journal} {\bibinfo  {journal} {Phys. Rev. Lett.}\ }\textbf
  {\bibinfo {volume} {114}},\ \bibinfo {pages} {175002} (\bibinfo {year}
  {2015})}\BibitemShut {NoStop}%
\bibitem [{\citenamefont {Birdsall}\ and\ \citenamefont
  {Langdon}(2004)}]{birdsall2004plasma}%
  \BibitemOpen
  \bibfield  {author} {\bibinfo {author} {\bibfnamefont {C.~K.}\ \bibnamefont
  {Birdsall}}\ and\ \bibinfo {author} {\bibfnamefont {A.~B.}\ \bibnamefont
  {Langdon}},\ }\href@noop {} {\emph {\bibinfo {title} {Plasma Physics via
  Computer Simulation}}}\ (\bibinfo  {publisher} {CRC Press},\ \bibinfo
  {address} {New York},\ \bibinfo {year} {2004})\BibitemShut {NoStop}%
\bibitem [{\citenamefont {Eppley}(1988)}]{eppley1988use}%
  \BibitemOpen
  \bibfield  {author} {\bibinfo {author} {\bibfnamefont {K.}~\bibnamefont
  {Eppley}},\ }\href@noop {} {\emph {\bibinfo {title} {The use of
  electromagnetic particle-in-cell codes in accelerator applications}}},\
  \bibinfo {type} {Tech. Rep.}\ (\bibinfo  {institution} {Stanford Linear
  Accelerator Center, Menlo Park, CA (USA)},\ \bibinfo {year}
  {1988})\BibitemShut {NoStop}%
\bibitem [{\citenamefont {Lehe}\ \emph {et~al.}(2016)\citenamefont {Lehe},
  \citenamefont {Kirchen}, \citenamefont {Godfrey}, \citenamefont {Maier},\
  and\ \citenamefont {Vay}}]{PhysRevE.94.053305}%
  \BibitemOpen
  \bibfield  {author} {\bibinfo {author} {\bibfnamefont {R.}~\bibnamefont
  {Lehe}}, \bibinfo {author} {\bibfnamefont {M.}~\bibnamefont {Kirchen}},
  \bibinfo {author} {\bibfnamefont {B.~B.}\ \bibnamefont {Godfrey}}, \bibinfo
  {author} {\bibfnamefont {A.~R.}\ \bibnamefont {Maier}},\ and\ \bibinfo
  {author} {\bibfnamefont {J.-L.}\ \bibnamefont {Vay}},\ }\bibfield  {title}
  {\bibinfo {title} {Elimination of numerical {C}herenkov instability in
  flowing-plasma particle-in-cell simulations by using {G}alilean
  coordinates},\ }\href {https://doi.org/10.1103/PhysRevE.94.053305} {\bibfield
   {journal} {\bibinfo  {journal} {Phys. Rev. E}\ }\textbf {\bibinfo {volume}
  {94}},\ \bibinfo {pages} {053305} (\bibinfo {year} {2016})}\BibitemShut
  {NoStop}%
\bibitem [{\citenamefont {Cruz}\ \emph {et~al.}(2021)\citenamefont {Cruz},
  \citenamefont {Grismayer},\ and\ \citenamefont
  {Silva}}]{PhysRevE.103.L051201}%
  \BibitemOpen
  \bibfield  {author} {\bibinfo {author} {\bibfnamefont {F.}~\bibnamefont
  {Cruz}}, \bibinfo {author} {\bibfnamefont {T.}~\bibnamefont {Grismayer}},\
  and\ \bibinfo {author} {\bibfnamefont {L.~O.}\ \bibnamefont {Silva}},\
  }\bibfield  {title} {\bibinfo {title} {Kinetic instability in inductively
  oscillatory plasma equilibrium},\ }\href
  {https://doi.org/10.1103/PhysRevE.103.L051201} {\bibfield  {journal}
  {\bibinfo  {journal} {Phys. Rev. E}\ }\textbf {\bibinfo {volume} {103}},\
  \bibinfo {pages} {L051201} (\bibinfo {year} {2021})}\BibitemShut {NoStop}%

\bibitem [{\citenamefont {Nayak}\ \emph {et~al.}(2023)\citenamefont {Nayak},
  \citenamefont {Teixeira},\ and\ \citenamefont
  {Burkholder}}]{nayak2023fly}
  \BibitemOpen
  \bibfield  {author} {\bibinfo {author} {\bibfnamefont {Indranil}~\bibnamefont
  {Nayak}}, \bibinfo {author} {\bibfnamefont {Fernando L.}~\bibnamefont {Teixeira}},\
  and\ \bibinfo {author} {\bibfnamefont {Robert J.}\ \bibnamefont {Burkholder}},\
  }\bibfield  {title} {\bibinfo {title} {{On-the-Fly Dynamic Mode Decomposition for Rapid Time-Extrapolation and Analysis of Cavity Resonances}},\ }\href
  {https://doi.org/10.1103/IEEETransAntennasPropag} {\bibfield  {journal}
  {\bibinfo  {journal} {IEEE Transactions on Antennas and Propagation}\ }\textbf {\bibinfo {volume} {2023}},\
  \bibinfo {pages} {} (\bibinfo {year} {2023})}\BibitemShut {NoStop}%
  
\bibitem [{\citenamefont {Davidson}\ \emph {et~al.}(2015)\citenamefont
  {Davidson}, \citenamefont {Tableman}, \citenamefont {An}, \citenamefont
  {Tsung}, \citenamefont {Lu}, \citenamefont {Vieira}, \citenamefont {Fonseca},
  \citenamefont {Silva},\ and\ \citenamefont {Mori}}]{DAVIDSON20151063}%
  \BibitemOpen
  \bibfield  {author} {\bibinfo {author} {\bibfnamefont {A.}~\bibnamefont
  {Davidson}}, \bibinfo {author} {\bibfnamefont {A.}~\bibnamefont {Tableman}},
  \bibinfo {author} {\bibfnamefont {W.}~\bibnamefont {An}}, \bibinfo {author}
  {\bibfnamefont {F.}~\bibnamefont {Tsung}}, \bibinfo {author} {\bibfnamefont
  {W.}~\bibnamefont {Lu}}, \bibinfo {author} {\bibfnamefont {J.}~\bibnamefont
  {Vieira}}, \bibinfo {author} {\bibfnamefont {R.}~\bibnamefont {Fonseca}},
  \bibinfo {author} {\bibfnamefont {L.}~\bibnamefont {Silva}},\ and\ \bibinfo
  {author} {\bibfnamefont {W.}~\bibnamefont {Mori}},\ }\bibfield  {title}
  {\bibinfo {title} {Implementation of a hybrid particle code with a {PIC}
  description in $r–z$ and a gridless description in $\phi$ into {OSIRIS}},\
  }\href@noop {} {\bibfield  {journal} {\bibinfo  {journal} {Journal of
  Computational Physics}\ }\textbf {\bibinfo {volume} {281}},\ \bibinfo {pages}
  {1063} (\bibinfo {year} {2015})}\BibitemShut {NoStop}%
\bibitem [{\citenamefont {Shapoval}\ \emph {et~al.}(2021)\citenamefont
  {Shapoval}, \citenamefont {Lehe}, \citenamefont {Th\'evenet}, \citenamefont
  {Zoni}, \citenamefont {Zhao},\ and\ \citenamefont
  {Vay}}]{PhysRevE.104.055311}%
  \BibitemOpen
  \bibfield  {author} {\bibinfo {author} {\bibfnamefont {O.}~\bibnamefont
  {Shapoval}}, \bibinfo {author} {\bibfnamefont {R.}~\bibnamefont {Lehe}},
  \bibinfo {author} {\bibfnamefont {M.}~\bibnamefont {Th\'evenet}}, \bibinfo
  {author} {\bibfnamefont {E.}~\bibnamefont {Zoni}}, \bibinfo {author}
  {\bibfnamefont {Y.}~\bibnamefont {Zhao}},\ and\ \bibinfo {author}
  {\bibfnamefont {J.-L.}\ \bibnamefont {Vay}},\ }\bibfield  {title} {\bibinfo
  {title} {Overcoming timestep limitations in boosted-frame particle-in-cell
  simulations of plasma-based acceleration},\ }\href
  {https://doi.org/10.1103/PhysRevE.104.055311} {\bibfield  {journal} {\bibinfo
   {journal} {Phys. Rev. E}\ }\textbf {\bibinfo {volume} {104}},\ \bibinfo
  {pages} {055311} (\bibinfo {year} {2021})}\BibitemShut {NoStop}%
\bibitem [{\citenamefont {Pandya}(2016)}]{pandya2016low}%
  \BibitemOpen
  \bibfield  {author} {\bibinfo {author} {\bibfnamefont {M.}~\bibnamefont
  {Pandya}},\ }\emph {\bibinfo {title} {Low edge safety factor disruptions in
  the Compact Toroidal Hybrid: Operation in the low-{Q} regime, passive
  disruption avoidance and the nature of MHD precursors}},\ \href@noop {}
  {Ph.D. thesis},\ \bibinfo  {school} {Auburn University} (\bibinfo {year}
  {2016})\BibitemShut {NoStop}%
\bibitem [{\citenamefont {Van~Milligen}\ \emph {et~al.}(2014)\citenamefont
  {Van~Milligen}, \citenamefont {S{\'a}nchez}, \citenamefont {Alonso},
  \citenamefont {Pedrosa}, \citenamefont {Hidalgo}, \citenamefont
  {De~Aguilera},\ and\ \citenamefont {Fraguas}}]{van2014use}%
  \BibitemOpen
  \bibfield  {author} {\bibinfo {author} {\bibfnamefont {B.~P.}\ \bibnamefont
  {Van~Milligen}}, \bibinfo {author} {\bibfnamefont {E.}~\bibnamefont
  {S{\'a}nchez}}, \bibinfo {author} {\bibfnamefont {A.}~\bibnamefont {Alonso}},
  \bibinfo {author} {\bibfnamefont {M.}~\bibnamefont {Pedrosa}}, \bibinfo
  {author} {\bibfnamefont {C.}~\bibnamefont {Hidalgo}}, \bibinfo {author}
  {\bibfnamefont {A.~M.}\ \bibnamefont {De~Aguilera}},\ and\ \bibinfo {author}
  {\bibfnamefont {A.~L.}\ \bibnamefont {Fraguas}},\ }\bibfield  {title}
  {\bibinfo {title} {The use of the biorthogonal decomposition for the
  identification of zonal flows at {TJ-II}},\ }\href@noop {} {\bibfield
  {journal} {\bibinfo  {journal} {Plasma Physics and Controlled Fusion}\
  }\textbf {\bibinfo {volume} {57}},\ \bibinfo {pages} {025005} (\bibinfo
  {year} {2014})}\BibitemShut {NoStop}%
\bibitem [{\citenamefont {Byrne}(2017)}]{byrne2017study}%
  \BibitemOpen
  \bibfield  {author} {\bibinfo {author} {\bibfnamefont {P.~J.}\ \bibnamefont
  {Byrne}},\ }\emph {\bibinfo {title} {Study of External Kink Modes in Shaped
  {HBT-EP} Plasmas}},\ \href@noop {} {Ph.D. thesis},\ \bibinfo  {school}
  {Columbia University} (\bibinfo {year} {2017})\BibitemShut {NoStop}%
\bibitem [{\citenamefont {Kaptanoglu}\ \emph {et~al.}(2021)\citenamefont
  {Kaptanoglu}, \citenamefont {Morgan}, \citenamefont {Hansen},\ and\
  \citenamefont {Brunton}}]{kaptanoglu2021physics}%
  \BibitemOpen
  \bibfield  {author} {\bibinfo {author} {\bibfnamefont {A.~A.}\ \bibnamefont
  {Kaptanoglu}}, \bibinfo {author} {\bibfnamefont {K.~D.}\ \bibnamefont
  {Morgan}}, \bibinfo {author} {\bibfnamefont {C.~J.}\ \bibnamefont {Hansen}},\
  and\ \bibinfo {author} {\bibfnamefont {S.~L.}\ \bibnamefont {Brunton}},\
  }\bibfield  {title} {\bibinfo {title} {Physics-constrained, low-dimensional
  models for magnetohydrodynamics: First-principles and data-driven
  approaches},\ }\href@noop {} {\bibfield  {journal} {\bibinfo  {journal}
  {Physical Review E}\ }\textbf {\bibinfo {volume} {104}},\ \bibinfo {pages}
  {015206} (\bibinfo {year} {2021})}\BibitemShut {NoStop}%
\bibitem [{\citenamefont {Sasaki}\ \emph {et~al.}(2019)\citenamefont {Sasaki},
  \citenamefont {Kawachi}, \citenamefont {Dendy}, \citenamefont {Arakawa},
  \citenamefont {Kasuya}, \citenamefont {Kin}, \citenamefont {Yamasaki},\ and\
  \citenamefont {Inagaki}}]{sasaki2019using}%
  \BibitemOpen
  \bibfield  {author} {\bibinfo {author} {\bibfnamefont {M.}~\bibnamefont
  {Sasaki}}, \bibinfo {author} {\bibfnamefont {Y.}~\bibnamefont {Kawachi}},
  \bibinfo {author} {\bibfnamefont {R.}~\bibnamefont {Dendy}}, \bibinfo
  {author} {\bibfnamefont {H.}~\bibnamefont {Arakawa}}, \bibinfo {author}
  {\bibfnamefont {N.}~\bibnamefont {Kasuya}}, \bibinfo {author} {\bibfnamefont
  {F.}~\bibnamefont {Kin}}, \bibinfo {author} {\bibfnamefont {K.}~\bibnamefont
  {Yamasaki}},\ and\ \bibinfo {author} {\bibfnamefont {S.}~\bibnamefont
  {Inagaki}},\ }\bibfield  {title} {\bibinfo {title} {Using dynamical mode
  decomposition to extract the limit cycle dynamics of modulated turbulence in
  a plasma simulation},\ }\href@noop {} {\bibfield  {journal} {\bibinfo
  {journal} {Plasma Physics and Controlled Fusion}\ }\textbf {\bibinfo {volume}
  {61}},\ \bibinfo {pages} {112001} (\bibinfo {year} {2019})}\BibitemShut
  {NoStop}%
\bibitem [{\citenamefont {Nayak}\ \emph
  {et~al.}(2021{\natexlab{a}})\citenamefont {Nayak}, \citenamefont {Kumar},\
  and\ \citenamefont {Teixeira}}]{NAYAK2021110671}%
  \BibitemOpen
  \bibfield  {author} {\bibinfo {author} {\bibfnamefont {I.}~\bibnamefont
  {Nayak}}, \bibinfo {author} {\bibfnamefont {M.}~\bibnamefont {Kumar}},\ and\
  \bibinfo {author} {\bibfnamefont {F.~L.}\ \bibnamefont {Teixeira}},\
  }\bibfield  {title} {\bibinfo {title} {Detection and prediction of
  equilibrium states in kinetic plasma simulations via mode tracking using
  reduced-order dynamic mode decomposition},\ }\href@noop {} {\bibfield
  {journal} {\bibinfo  {journal} {Journal of Computational Physics}\ }\textbf
  {\bibinfo {volume} {447}},\ \bibinfo {pages} {110671} (\bibinfo {year}
  {2021}{\natexlab{a}})}\BibitemShut {NoStop}%
\bibitem [{\citenamefont {{Nicolini}}\ \emph {et~al.}(2019)\citenamefont
  {{Nicolini}}, \citenamefont {{Na}},\ and\ \citenamefont
  {{Teixeira}}}]{julio2019}%
  \BibitemOpen
  \bibfield  {author} {\bibinfo {author} {\bibfnamefont {J.~L.}\ \bibnamefont
  {{Nicolini}}}, \bibinfo {author} {\bibfnamefont {D.}~\bibnamefont {{Na}}},\
  and\ \bibinfo {author} {\bibfnamefont {F.~L.}\ \bibnamefont {{Teixeira}}},\
  }\bibfield  {title} {\bibinfo {title} {Model order reduction of
  electromagnetic particle-in-cell kinetic plasma simulations via proper
  orthogonal decomposition},\ }\href {https://doi.org/10.1109/TPS.2019.2950377}
  {\bibfield  {journal} {\bibinfo  {journal} {IEEE Transactions on Plasma
  Science}\ }\textbf {\bibinfo {volume} {47}},\ \bibinfo {pages} {5239}
  (\bibinfo {year} {2019})}\BibitemShut {NoStop}%
\bibitem [{\citenamefont {Nayak}\ and\ \citenamefont
  {Teixeira}(2021)}]{nayak2021dynamic}%
  \BibitemOpen
  \bibfield  {author} {\bibinfo {author} {\bibfnamefont {I.}~\bibnamefont
  {Nayak}}\ and\ \bibinfo {author} {\bibfnamefont {F.~L.}\ \bibnamefont
  {Teixeira}},\ }\bibfield  {title} {\bibinfo {title} {Dynamic mode
  decomposition reduced-order models for multiscale kinetic plasma analysis},\
  }in\ \href@noop {} {\emph {\bibinfo {booktitle} {2021 United States National
  Committee of URSI National Radio Science Meeting (USNC-URSI NRSM)}}}\
  (\bibinfo {organization} {IEEE},\ \bibinfo {year} {2021})\ pp.\ \bibinfo
  {pages} {261--262}\BibitemShut {NoStop}%
\bibitem [{\citenamefont {Nayak}\ and\ \citenamefont
  {Teixeira}(2020)}]{nayak2020dynamic}%
  \BibitemOpen
  \bibfield  {author} {\bibinfo {author} {\bibfnamefont {I.}~\bibnamefont
  {Nayak}}\ and\ \bibinfo {author} {\bibfnamefont {F.~L.}\ \bibnamefont
  {Teixeira}},\ }\bibfield  {title} {\bibinfo {title} {Dynamic mode
  decomposition for prediction of kinetic plasma behavior},\ }in\ \href@noop {}
  {\emph {\bibinfo {booktitle} {2020 International Applied Computational
  Electromagnetics Society Symposium (ACES)}}}\ (\bibinfo {organization}
  {IEEE},\ \bibinfo {year} {2020})\ pp.\ \bibinfo {pages} {1--2}\BibitemShut
  {NoStop}%
  
  
\bibitem [{\citenamefont {Kraus}\ \emph
  {et~al.}(2017{\natexlab{a}})\citenamefont {Kraus}, \citenamefont {Kormann},
  \citenamefont {Morrison},\ and\ \citenamefont
  {Sonnendr{\"u}cker}}]{kraus2017gempic}%
  \BibitemOpen
  \bibfield  {author} {\bibinfo {author} {\bibfnamefont {M.}~\bibnamefont
  {Kraus}}, \bibinfo {author} {\bibfnamefont {K.}~\bibnamefont {Kormann}},
  \bibinfo {author} {\bibfnamefont {P.~J.}\ \bibnamefont {Morrison}},\ and\
  \bibinfo {author} {\bibfnamefont {E.}~\bibnamefont {Sonnendr{\"u}cker}},\
  }\bibfield  {title} {\bibinfo {title} {{GEMPIC}: geometric electromagnetic
  particle-in-cell methods},\ }\href@noop {} {\bibfield  {journal} {\bibinfo
  {journal} {Journal of Plasma Physics}\ }\textbf {\bibinfo {volume} {83}},\
  \bibinfo {pages} {905830401} (\bibinfo {year}
  {2017}{\natexlab{a}})}\BibitemShut {NoStop}%
\bibitem [{\citenamefont {Alves}\ and\ \citenamefont
  {Fiuza}(2022)}]{alves2022data}%
  \BibitemOpen
  \bibfield  {author} {\bibinfo {author} {\bibfnamefont {E.~P.}\ \bibnamefont
  {Alves}}\ and\ \bibinfo {author} {\bibfnamefont {F.}~\bibnamefont {Fiuza}},\
  }\bibfield  {title} {\bibinfo {title} {Data-driven discovery of reduced
  plasma physics models from fully kinetic simulations},\ }\href@noop {}
  {\bibfield  {journal} {\bibinfo  {journal} {Physical Review Research}\
  }\textbf {\bibinfo {volume} {4}},\ \bibinfo {pages} {033192} (\bibinfo {year}
  {2022})}\BibitemShut {NoStop}%
\bibitem [{\citenamefont {Schmid}(2010)}]{schmid2010dynamic}%
  \BibitemOpen
  \bibfield  {author} {\bibinfo {author} {\bibfnamefont {P.~J.}\ \bibnamefont
  {Schmid}},\ }\bibfield  {title} {\bibinfo {title} {Dynamic mode decomposition
  of numerical and experimental data},\ }\href@noop {} {\bibfield  {journal}
  {\bibinfo  {journal} {{Journal of Fluid Mechanics}}\ }\textbf {\bibinfo
  {volume} {656}},\ \bibinfo {pages} {5} (\bibinfo {year} {2010})}\BibitemShut
  {NoStop}%
\bibitem [{\citenamefont {Rowley}\ \emph {et~al.}(2009)\citenamefont {Rowley},
  \citenamefont {Mezi{\'c}}, \citenamefont {Bagheri}, \citenamefont
  {Schlatter},\ and\ \citenamefont {Henningson}}]{rowley2009spectral}%
  \BibitemOpen
  \bibfield  {author} {\bibinfo {author} {\bibfnamefont {C.~W.}\ \bibnamefont
  {Rowley}}, \bibinfo {author} {\bibfnamefont {I.}~\bibnamefont {Mezi{\'c}}},
  \bibinfo {author} {\bibfnamefont {S.}~\bibnamefont {Bagheri}}, \bibinfo
  {author} {\bibfnamefont {P.}~\bibnamefont {Schlatter}},\ and\ \bibinfo
  {author} {\bibfnamefont {D.~S.}\ \bibnamefont {Henningson}},\ }\bibfield
  {title} {\bibinfo {title} {Spectral analysis of nonlinear flows},\
  }\href@noop {} {\bibfield  {journal} {\bibinfo  {journal} {Journal of Fluid
  Mechanics}\ }\textbf {\bibinfo {volume} {641}},\ \bibinfo {pages} {115}
  (\bibinfo {year} {2009})}\BibitemShut {NoStop}%
\bibitem [{\citenamefont {Taylor}\ \emph {et~al.}(2018)\citenamefont {Taylor},
  \citenamefont {Kutz}, \citenamefont {Morgan},\ and\ \citenamefont
  {Nelson}}]{taylor2018dynamic}%
  \BibitemOpen
  \bibfield  {author} {\bibinfo {author} {\bibfnamefont {R.}~\bibnamefont
  {Taylor}}, \bibinfo {author} {\bibfnamefont {J.~N.}\ \bibnamefont {Kutz}},
  \bibinfo {author} {\bibfnamefont {K.}~\bibnamefont {Morgan}},\ and\ \bibinfo
  {author} {\bibfnamefont {B.~A.}\ \bibnamefont {Nelson}},\ }\bibfield  {title}
  {\bibinfo {title} {Dynamic mode decomposition for plasma diagnostics and
  validation},\ }\href@noop {} {\bibfield  {journal} {\bibinfo  {journal} {Rev.
  Sci. Instr.}\ }\textbf {\bibinfo {volume} {89}},\ \bibinfo {pages} {053501}
  (\bibinfo {year} {2018})}\BibitemShut {NoStop}%
\bibitem [{\citenamefont {Kaptanoglu}\ \emph {et~al.}(2020)\citenamefont
  {Kaptanoglu}, \citenamefont {Morgan}, \citenamefont {Hansen},\ and\
  \citenamefont {Brunton}}]{kaptanoglu2020characterizing}%
  \BibitemOpen
  \bibfield  {author} {\bibinfo {author} {\bibfnamefont {A.~A.}\ \bibnamefont
  {Kaptanoglu}}, \bibinfo {author} {\bibfnamefont {K.~D.}\ \bibnamefont
  {Morgan}}, \bibinfo {author} {\bibfnamefont {C.~J.}\ \bibnamefont {Hansen}},\
  and\ \bibinfo {author} {\bibfnamefont {S.~L.}\ \bibnamefont {Brunton}},\
  }\bibfield  {title} {\bibinfo {title} {Characterizing magnetized plasmas with
  dynamic mode decomposition},\ }\href@noop {} {\bibfield  {journal} {\bibinfo
  {journal} {Physics of Plasmas}\ }\textbf {\bibinfo {volume} {27}},\ \bibinfo
  {pages} {032108} (\bibinfo {year} {2020})}\BibitemShut {NoStop}%
\bibitem [{\citenamefont {Nayak}\ \emph
  {et~al.}(2021{\natexlab{b}})\citenamefont {Nayak}, \citenamefont {Teixeira},\
  and\ \citenamefont {Kumar}}]{nayak2021koopman}%
  \BibitemOpen
  \bibfield  {author} {\bibinfo {author} {\bibfnamefont {I.}~\bibnamefont
  {Nayak}}, \bibinfo {author} {\bibfnamefont {F.~L.}\ \bibnamefont
  {Teixeira}},\ and\ \bibinfo {author} {\bibfnamefont {M.}~\bibnamefont
  {Kumar}},\ }\bibfield  {title} {\bibinfo {title} {Koopman autoencoder
  architecture for current density modeling in kinetic plasma simulations},\
  }in\ \href@noop {} {\emph {\bibinfo {booktitle} {2021 International Applied
  Computational Electromagnetics Society Symposium (ACES)}}}\ (\bibinfo
  {organization} {IEEE},\ \bibinfo {year} {2021})\ pp.\ \bibinfo {pages}
  {1--3}\BibitemShut {NoStop}%

\bibitem [{\citenamefont {Nayak}\ \emph
  {et~al.}(2023)\citenamefont {Nayak}, \citenamefont {Kumar},\
  and\ \citenamefont {Teixeira}}]{nayak2023koopman}
  \BibitemOpen
  \bibfield  {author} {\bibinfo {author} {\bibfnamefont {I.}\ \bibnamefont
  {Nayak}}, \bibinfo {author} {\bibfnamefont {M.}\ \bibnamefont
  {Kumar}},\ and\ \bibinfo {author} {\bibfnamefont {F. L.}\ \bibnamefont
  {Teixeira}},\ }\bibfield  {title} {\bibinfo {title} {{Koopman Autoencoders for Reduced-Order Modeling of Kinetic Plasmas}},\ }\href
  {Add-DOI-URL-Here} {\bibfield  {journal}
  {\bibinfo  {journal} {Advances in Electromagnetics Empowered by Artificial Intelligence and Deep Learning}\ },\
  \bibinfo {pages} {515--542} (\bibinfo {year} {2023}),\ \bibinfo {publisher} {Wiley Online Library}}\BibitemShut {NoStop}



\bibitem [{\citenamefont {Moon}\ \emph {et~al.}(2015)\citenamefont {Moon},
  \citenamefont {Teixeira},\ and\ \citenamefont {Omelchenko}}]{moon2015exact}%
  \BibitemOpen
  \bibfield  {author} {\bibinfo {author} {\bibfnamefont {H.}~\bibnamefont
  {Moon}}, \bibinfo {author} {\bibfnamefont {F.~L.}\ \bibnamefont {Teixeira}},\
  and\ \bibinfo {author} {\bibfnamefont {Y.~A.}\ \bibnamefont {Omelchenko}},\
  }\bibfield  {title} {\bibinfo {title} {Exact charge-conserving
  scatter–gather algorithm for particle-in-cell simulations on unstructured
  grids: {A} geometric perspective},\ }\href@noop {} {\bibfield  {journal}
  {\bibinfo  {journal} {Comput. Phys. Commun.}\ }\textbf {\bibinfo {volume}
  {194}},\ \bibinfo {pages} {43} (\bibinfo {year} {2015})}\BibitemShut
  {NoStop}%
\bibitem [{\citenamefont {Kim}\ and\ \citenamefont
  {Teixeira}(2011)}]{kim2011parallel}%
  \BibitemOpen
  \bibfield  {author} {\bibinfo {author} {\bibfnamefont {J.}~\bibnamefont
  {Kim}}\ and\ \bibinfo {author} {\bibfnamefont {F.~L.}\ \bibnamefont
  {Teixeira}},\ }\bibfield  {title} {\bibinfo {title} {Parallel and explicit
  finite-element time-domain method for {M}axwell's equations},\ }\href
  {https://doi.org/10.1109/TAP.2011.2143682} {\bibfield  {journal} {\bibinfo
  {journal} {IEEE Trans. Antennas Propag.}\ }\textbf {\bibinfo {volume} {59}},\
  \bibinfo {pages} {2350} (\bibinfo {year} {2011})}\BibitemShut {NoStop}%
\bibitem [{\citenamefont {Evstatiev}\ and\ \citenamefont
  {Shadwick}(2013)}]{EVSTATIEV2013376}%
  \BibitemOpen
  \bibfield  {author} {\bibinfo {author} {\bibfnamefont {E.}~\bibnamefont
  {Evstatiev}}\ and\ \bibinfo {author} {\bibfnamefont {B.}~\bibnamefont
  {Shadwick}},\ }\bibfield  {title} {\bibinfo {title} {Variational formulation
  of particle algorithms for kinetic plasma simulations},\ }\href@noop {}
  {\bibfield  {journal} {\bibinfo  {journal} {Journal of Computational
  Physics}\ }\textbf {\bibinfo {volume} {245}},\ \bibinfo {pages} {376}
  (\bibinfo {year} {2013})}\BibitemShut {NoStop}%
\bibitem [{\citenamefont {Burby}(2017)}]{doi:10.1063/1.4976849}%
  \BibitemOpen
  \bibfield  {author} {\bibinfo {author} {\bibfnamefont {J.~W.}\ \bibnamefont
  {Burby}},\ }\bibfield  {title} {\bibinfo {title} {Finite-dimensional
  collisionless kinetic theory},\ }\href@noop {} {\bibfield  {journal}
  {\bibinfo  {journal} {Physics of Plasmas}\ }\textbf {\bibinfo {volume}
  {24}},\ \bibinfo {pages} {032101} (\bibinfo {year} {2017})}\BibitemShut
  {NoStop}%
\bibitem [{\citenamefont {Jianyuan}\ \emph {et~al.}(2018)\citenamefont
  {Jianyuan}, \citenamefont {Hong},\ and\ \citenamefont
  {Jian}}]{jianyuan2018structure}%
  \BibitemOpen
  \bibfield  {author} {\bibinfo {author} {\bibfnamefont {X.}~\bibnamefont
  {Jianyuan}}, \bibinfo {author} {\bibfnamefont {Q.}~\bibnamefont {Hong}},\
  and\ \bibinfo {author} {\bibfnamefont {L.}~\bibnamefont {Jian}},\ }\bibfield
  {title} {\bibinfo {title} {Structure-preserving geometric particle-in-cell
  methods for {Vlasov-Maxwell} systems},\ }\href@noop {} {\bibfield  {journal}
  {\bibinfo  {journal} {Plasma Science and Technology}\ }\textbf {\bibinfo
  {volume} {20}},\ \bibinfo {pages} {110501} (\bibinfo {year}
  {2018})}\BibitemShut {NoStop}%
\bibitem [{\citenamefont {Na}\ \emph {et~al.}(2016)\citenamefont {Na},
  \citenamefont {Moon}, \citenamefont {Omelchenko},\ and\ \citenamefont
  {Teixeira}}]{na2016local}%
  \BibitemOpen
  \bibfield  {author} {\bibinfo {author} {\bibfnamefont {D.-Y.}\ \bibnamefont
  {Na}}, \bibinfo {author} {\bibfnamefont {H.}~\bibnamefont {Moon}}, \bibinfo
  {author} {\bibfnamefont {Y.~A.}\ \bibnamefont {Omelchenko}},\ and\ \bibinfo
  {author} {\bibfnamefont {F.~L.}\ \bibnamefont {Teixeira}},\ }\bibfield
  {title} {\bibinfo {title} {Local, explicit, and charge-conserving
  electromagnetic particle-in-cell algorithm on unstructured grids},\
  }\href@noop {} {\bibfield  {journal} {\bibinfo  {journal} {IEEE Trans. Plasma
  Sci.}\ }\textbf {\bibinfo {volume} {44}},\ \bibinfo {pages} {1353} (\bibinfo
  {year} {2016})}\BibitemShut {NoStop}%
\bibitem [{\citenamefont {Flanders}(1989)}]{flanders1989}%
  \BibitemOpen
  \bibfield  {author} {\bibinfo {author} {\bibfnamefont {H.}~\bibnamefont
  {Flanders}},\ }\href@noop {} {\emph {\bibinfo {title} {Differential Forms
  with Applications to the Physical Sciences}}}\ (\bibinfo  {publisher}
  {Dover},\ \bibinfo {address} {New York},\ \bibinfo {year} {1989})\BibitemShut
  {NoStop}%
\bibitem [{\citenamefont {Teixeira}\ and\ \citenamefont
  {Chew}(1999)}]{teixeira1999lattice}%
  \BibitemOpen
  \bibfield  {author} {\bibinfo {author} {\bibfnamefont {F.~L.}\ \bibnamefont
  {Teixeira}}\ and\ \bibinfo {author} {\bibfnamefont {W.}~\bibnamefont
  {Chew}},\ }\bibfield  {title} {\bibinfo {title} {Lattice electromagnetic
  theory from a topological viewpoint},\ }\href@noop {} {\bibfield  {journal}
  {\bibinfo  {journal} {J. Math. Phys.}\ }\textbf {\bibinfo {volume} {40}},\
  \bibinfo {pages} {169} (\bibinfo {year} {1999})}\BibitemShut {NoStop}%
\bibitem [{\citenamefont {Gross}\ and\ \citenamefont
  {Kotiuga}(2004)}]{kotiuga2004}%
  \BibitemOpen
  \bibfield  {author} {\bibinfo {author} {\bibfnamefont {P.~W.}\ \bibnamefont
  {Gross}}\ and\ \bibinfo {author} {\bibfnamefont {P.~R.}\ \bibnamefont
  {Kotiuga}},\ }\href@noop {} {\emph {\bibinfo {title} {Electromagnetic Theory
  and Computation: {A} Topological Approach}}}\ (\bibinfo  {publisher}
  {Cambridge University Press},\ \bibinfo {address} {Cambridge},\ \bibinfo
  {year} {2004})\BibitemShut {NoStop}%
\bibitem [{\citenamefont {He}\ and\ \citenamefont
  {Teixeira}(2007)}]{he2007differential}%
  \BibitemOpen
  \bibfield  {author} {\bibinfo {author} {\bibfnamefont {B.}~\bibnamefont
  {He}}\ and\ \bibinfo {author} {\bibfnamefont {F.~L.}\ \bibnamefont
  {Teixeira}},\ }\bibfield  {title} {\bibinfo {title} {Differential forms,
  {G}alerkin duality, and sparse inverse approximations in finite element
  solutions of {M}axwell equations},\ }\href
  {https://doi.org/10.1109/TAP.2007.895619} {\bibfield  {journal} {\bibinfo
  {journal} {IEEE Trans. Antennas Propag.}\ }\textbf {\bibinfo {volume} {55}},\
  \bibinfo {pages} {1359} (\bibinfo {year} {2007})}\BibitemShut {NoStop}%
\bibitem [{\citenamefont {Deschamps}(1981)}]{deschamps1981electromagnetics}%
  \BibitemOpen
  \bibfield  {author} {\bibinfo {author} {\bibfnamefont {G.~A.}\ \bibnamefont
  {Deschamps}},\ }\bibfield  {title} {\bibinfo {title} {Electromagnetics and
  differential forms},\ }\href@noop {} {\bibfield  {journal} {\bibinfo
  {journal} {Proc. IEEE}\ }\textbf {\bibinfo {volume} {69}},\ \bibinfo {pages}
  {676} (\bibinfo {year} {1981})}\BibitemShut {NoStop}%
\bibitem [{\citenamefont {Teixeira}(2013)}]{teixeira2013differential}%
  \BibitemOpen
  \bibfield  {author} {\bibinfo {author} {\bibfnamefont {F.~L.}\ \bibnamefont
  {Teixeira}},\ }\bibfield  {title} {\bibinfo {title} {Differential forms in
  lattice field theories: {An} overview},\ }\href@noop {} {\bibfield  {journal}
  {\bibinfo  {journal} {ISRN Math. Phys.}\ }\textbf {\bibinfo {volume}
  {2013}},\ \bibinfo {pages} {16} (\bibinfo {year} {2013})}\BibitemShut
  {NoStop}%
\bibitem [{\citenamefont {Teixeira}(2014)}]{teixeira2014lattice}%
  \BibitemOpen
  \bibfield  {author} {\bibinfo {author} {\bibfnamefont {F.~L.}\ \bibnamefont
  {Teixeira}},\ }\bibfield  {title} {\bibinfo {title} {Lattice {Maxwell's}
  equations},\ }\href@noop {} {\bibfield  {journal} {\bibinfo  {journal} {Prog.
  Electromagn. Res.}\ }\textbf {\bibinfo {volume} {148}},\ \bibinfo {pages}
  {113} (\bibinfo {year} {2014})}\BibitemShut {NoStop}%
\bibitem [{Note1()}]{Note1}%
  \BibitemOpen
  \bibinfo {note} {This is a good approximation for small $\Delta
  t$}\BibitemShut {NoStop}%
\bibitem [{\citenamefont {Tu}\ \emph {et~al.}(2013)\citenamefont {Tu},
  \citenamefont {Rowley}, \citenamefont {Luchtenburg}, \citenamefont
  {Brunton},\ and\ \citenamefont {Kutz}}]{tu2013dynamic}%
  \BibitemOpen
  \bibfield  {author} {\bibinfo {author} {\bibfnamefont {J.~H.}\ \bibnamefont
  {Tu}}, \bibinfo {author} {\bibfnamefont {C.~W.}\ \bibnamefont {Rowley}},
  \bibinfo {author} {\bibfnamefont {D.~M.}\ \bibnamefont {Luchtenburg}},
  \bibinfo {author} {\bibfnamefont {S.~L.}\ \bibnamefont {Brunton}},\ and\
  \bibinfo {author} {\bibfnamefont {J.~N.}\ \bibnamefont {Kutz}},\ }\bibfield
  {title} {\bibinfo {title} {On dynamic mode decomposition: theory and
  applications},\ }\href@noop {} {\bibfield  {journal} {\bibinfo  {journal}
  {arXiv:1312.0041}\ } (\bibinfo {year} {2013})}\BibitemShut {NoStop}%
\bibitem [{\citenamefont {Kutz}\ \emph {et~al.}(2016)\citenamefont {Kutz},
  \citenamefont {Brunton}, \citenamefont {Brunton},\ and\ \citenamefont
  {Proctor}}]{kutz2016dynamic}%
  \BibitemOpen
  \bibfield  {author} {\bibinfo {author} {\bibfnamefont {J.~N.}\ \bibnamefont
  {Kutz}}, \bibinfo {author} {\bibfnamefont {S.~L.}\ \bibnamefont {Brunton}},
  \bibinfo {author} {\bibfnamefont {B.~W.}\ \bibnamefont {Brunton}},\ and\
  \bibinfo {author} {\bibfnamefont {J.~L.}\ \bibnamefont {Proctor}},\
  }\href@noop {} {\emph {\bibinfo {title} {Dynamic mode decomposition:
  {Data-driven modeling of complex systems}}}}\ (\bibinfo  {publisher} {SIAM},\
  \bibinfo {year} {2016})\BibitemShut {NoStop}%
\bibitem [{\citenamefont {Kerschen}\ and\ \citenamefont
  {Golinval}(2002)}]{kerschen2002physical}%
  \BibitemOpen
  \bibfield  {author} {\bibinfo {author} {\bibfnamefont {G.}~\bibnamefont
  {Kerschen}}\ and\ \bibinfo {author} {\bibfnamefont {J.-C.}\ \bibnamefont
  {Golinval}},\ }\bibfield  {title} {\bibinfo {title} {Physical interpretation
  of the proper orthogonal modes using the singular value decomposition},\
  }\href@noop {} {\bibfield  {journal} {\bibinfo  {journal} {Journal of Sound
  and vibration}\ }\textbf {\bibinfo {volume} {249}},\ \bibinfo {pages} {849}
  (\bibinfo {year} {2002})}\BibitemShut {NoStop}%
\bibitem [{\citenamefont {Gavish}\ and\ \citenamefont
  {Donoho}(2014)}]{gavish2014optimal}%
  \BibitemOpen
  \bibfield  {author} {\bibinfo {author} {\bibfnamefont {M.}~\bibnamefont
  {Gavish}}\ and\ \bibinfo {author} {\bibfnamefont {D.~L.}\ \bibnamefont
  {Donoho}},\ }\bibfield  {title} {\bibinfo {title} {The optimal hard threshold
  for singular values is $4/\sqrt{3}$},\ }\href@noop {} {\bibfield  {journal}
  {\bibinfo  {journal} {IEEE Transactions on Information Theory}\ }\textbf
  {\bibinfo {volume} {60}},\ \bibinfo {pages} {5040} (\bibinfo {year}
  {2014})}\BibitemShut {NoStop}%
\bibitem [{\citenamefont {Gavish}\ and\ \citenamefont
  {Donoho}()}]{opt_thr_code}%
  \BibitemOpen
  \bibfield  {author} {\bibinfo {author} {\bibfnamefont {M.}~\bibnamefont
  {Gavish}}\ and\ \bibinfo {author} {\bibfnamefont {D.~L.}\ \bibnamefont
  {Donoho}},\ }\href {https://web.stanford.edu/~gavish/present_research.html}
  {\bibinfo {title} {Matlab code: The optimal hard threshold for singular
  values is $4/\sqrt{3}$}}\BibitemShut {NoStop}%
\bibitem [{\citenamefont {Jovanovi{\'c}}\ \emph {et~al.}(2014)\citenamefont
  {Jovanovi{\'c}}, \citenamefont {Schmid},\ and\ \citenamefont
  {Nichols}}]{jovanovic2014sparsity}%
  \BibitemOpen
  \bibfield  {author} {\bibinfo {author} {\bibfnamefont {M.~R.}\ \bibnamefont
  {Jovanovi{\'c}}}, \bibinfo {author} {\bibfnamefont {P.~J.}\ \bibnamefont
  {Schmid}},\ and\ \bibinfo {author} {\bibfnamefont {J.~W.}\ \bibnamefont
  {Nichols}},\ }\bibfield  {title} {\bibinfo {title} {Sparsity-promoting
  dynamic mode decomposition},\ }\href@noop {} {\bibfield  {journal} {\bibinfo
  {journal} {Physics of Fluids}\ }\textbf {\bibinfo {volume} {26}},\ \bibinfo
  {pages} {024103} (\bibinfo {year} {2014})}\BibitemShut {NoStop}%
\bibitem [{\citenamefont {Pan}\ and\ \citenamefont
  {Duraisamy}(2020)}]{pan2020structure}%
  \BibitemOpen
  \bibfield  {author} {\bibinfo {author} {\bibfnamefont {S.}~\bibnamefont
  {Pan}}\ and\ \bibinfo {author} {\bibfnamefont {K.}~\bibnamefont
  {Duraisamy}},\ }\bibfield  {title} {\bibinfo {title} {On the structure of
  time-delay embedding in linear models of non-linear dynamical systems},\
  }\href@noop {} {\bibfield  {journal} {\bibinfo  {journal} {Chaos: An
  Interdisciplinary Journal of Nonlinear Science}\ }\textbf {\bibinfo {volume}
  {30}},\ \bibinfo {pages} {073135} (\bibinfo {year} {2020})}\BibitemShut
  {NoStop}%
\bibitem [{\citenamefont {Koopman}(1931)}]{koopman1931hamiltonian}%
  \BibitemOpen
  \bibfield  {author} {\bibinfo {author} {\bibfnamefont {B.~O.}\ \bibnamefont
  {Koopman}},\ }\bibfield  {title} {\bibinfo {title} {Hamiltonian systems and
  transformation in hilbert space},\ }\href@noop {} {\bibfield  {journal}
  {\bibinfo  {journal} {Proceedings of the national academy of sciences of the
  united states of america}\ }\textbf {\bibinfo {volume} {17}},\ \bibinfo
  {pages} {315} (\bibinfo {year} {1931})}\BibitemShut {NoStop}%
\bibitem [{\citenamefont {Curtis}\ and\ \citenamefont
  {Alford-Lago}(2021)}]{PhysRevE.103.012201}%
  \BibitemOpen
  \bibfield  {author} {\bibinfo {author} {\bibfnamefont {C.~W.}\ \bibnamefont
  {Curtis}}\ and\ \bibinfo {author} {\bibfnamefont {D.~J.}\ \bibnamefont
  {Alford-Lago}},\ }\bibfield  {title} {\bibinfo {title} {Dynamic-mode
  decomposition and optimal prediction},\ }\href@noop {} {\bibfield  {journal}
  {\bibinfo  {journal} {Phys. Rev. E}\ }\textbf {\bibinfo {volume} {103}},\
  \bibinfo {pages} {012201} (\bibinfo {year} {2021})}\BibitemShut {NoStop}%
\bibitem [{\citenamefont {Dylewsky}\ \emph {et~al.}(2019)\citenamefont
  {Dylewsky}, \citenamefont {Tao},\ and\ \citenamefont
  {Kutz}}]{PhysRevE.99.063311}%
  \BibitemOpen
  \bibfield  {author} {\bibinfo {author} {\bibfnamefont {D.}~\bibnamefont
  {Dylewsky}}, \bibinfo {author} {\bibfnamefont {M.}~\bibnamefont {Tao}},\ and\
  \bibinfo {author} {\bibfnamefont {J.~N.}\ \bibnamefont {Kutz}},\ }\bibfield
  {title} {\bibinfo {title} {Dynamic mode decomposition for multiscale
  nonlinear physics},\ }\href {https://doi.org/10.1103/PhysRevE.99.063311}
  {\bibfield  {journal} {\bibinfo  {journal} {Phys. Rev. E}\ }\textbf {\bibinfo
  {volume} {99}},\ \bibinfo {pages} {063311} (\bibinfo {year}
  {2019})}\BibitemShut {NoStop}%
\bibitem [{\citenamefont {Kamb}\ \emph {et~al.}(2020)\citenamefont {Kamb},
  \citenamefont {Kaiser}, \citenamefont {Brunton},\ and\ \citenamefont
  {Kutz}}]{kamb2020time}%
  \BibitemOpen
  \bibfield  {author} {\bibinfo {author} {\bibfnamefont {M.}~\bibnamefont
  {Kamb}}, \bibinfo {author} {\bibfnamefont {E.}~\bibnamefont {Kaiser}},
  \bibinfo {author} {\bibfnamefont {S.~L.}\ \bibnamefont {Brunton}},\ and\
  \bibinfo {author} {\bibfnamefont {J.~N.}\ \bibnamefont {Kutz}},\ }\bibfield
  {title} {\bibinfo {title} {Time-delay observables for {Koopman}: Theory and
  applications},\ }\href@noop {} {\bibfield  {journal} {\bibinfo  {journal}
  {SIAM Journal on Applied Dynamical Systems}\ }\textbf {\bibinfo {volume}
  {19}},\ \bibinfo {pages} {886} (\bibinfo {year} {2020})}\BibitemShut
  {NoStop}%
\bibitem [{\citenamefont {Brunton}\ \emph {et~al.}(2017)\citenamefont
  {Brunton}, \citenamefont {Brunton}, \citenamefont {Proctor}, \citenamefont
  {Kaiser},\ and\ \citenamefont {Kutz}}]{brunton2017chaos}%
  \BibitemOpen
  \bibfield  {author} {\bibinfo {author} {\bibfnamefont {S.~L.}\ \bibnamefont
  {Brunton}}, \bibinfo {author} {\bibfnamefont {B.~W.}\ \bibnamefont
  {Brunton}}, \bibinfo {author} {\bibfnamefont {J.~L.}\ \bibnamefont
  {Proctor}}, \bibinfo {author} {\bibfnamefont {E.}~\bibnamefont {Kaiser}},\
  and\ \bibinfo {author} {\bibfnamefont {J.~N.}\ \bibnamefont {Kutz}},\
  }\bibfield  {title} {\bibinfo {title} {Chaos as an intermittently forced
  linear system},\ }\href@noop {} {\bibfield  {journal} {\bibinfo  {journal}
  {Nature communications}\ }\textbf {\bibinfo {volume} {8}},\ \bibinfo {pages}
  {19} (\bibinfo {year} {2017})}\BibitemShut {NoStop}%
\bibitem [{\citenamefont {Takens}(2006)}]{takens2006detecting}%
  \BibitemOpen
  \bibfield  {author} {\bibinfo {author} {\bibfnamefont {F.}~\bibnamefont
  {Takens}},\ }\bibfield  {title} {\bibinfo {title} {Detecting strange
  attractors in turbulence},\ }in\ \href@noop {} {\emph {\bibinfo {booktitle}
  {Dynamical Systems and Turbulence, Warwick 1980: proceedings of a symposium
  held at the University of Warwick 1979/80}}}\ (\bibinfo {organization}
  {Springer},\ \bibinfo {year} {2006})\ pp.\ \bibinfo {pages}
  {366--381}\BibitemShut {NoStop}%
\bibitem [{\citenamefont {Nayak}\ \emph
  {et~al.}(2021{\natexlab{c}})\citenamefont {Nayak}, \citenamefont {Kumar},\
  and\ \citenamefont {Teixeira}}]{nayak2021detecting}%
  \BibitemOpen
  \bibfield  {author} {\bibinfo {author} {\bibfnamefont {I.}~\bibnamefont
  {Nayak}}, \bibinfo {author} {\bibfnamefont {M.}~\bibnamefont {Kumar}},\ and\
  \bibinfo {author} {\bibfnamefont {F.}~\bibnamefont {Teixeira}},\ }\bibfield
  {title} {\bibinfo {title} {Detecting equilibrium state of dynamical systems
  using sliding-window reduced-order dynamic mode decomposition},\ }in\
  \href@noop {} {\emph {\bibinfo {booktitle} {AIAA Scitech 2021 Forum}}}\
  (\bibinfo {year} {2021})\ p.\ \bibinfo {pages} {1858}\BibitemShut {NoStop}%
\bibitem [{\citenamefont {Martínez-Cagigal}(2023)}]{colormap}%
  \BibitemOpen
  \bibfield  {author} {\bibinfo {author} {\bibfnamefont {V.}~\bibnamefont
  {Martínez-Cagigal}},\ }\href
  {https://www.mathworks.com/matlabcentral/fileexchange/69470-custom-colormap}
  {\bibinfo {title} {Custom colormap}} (\bibinfo {year} {2023}),\ \bibinfo
  {note} {{M}ATLAB Central File Exchange. Retrieved February 22,
  2023.}\BibitemShut {Stop}%

  
\bibitem [{Note2()}]{Note2}%
  \BibitemOpen
  \bibinfo {note} {The energy of a DMD mode is calculated by averaging the
  2-norm square of that particular DMD mode over all the time-samples inside
  the DMD training window.}\BibitemShut {Stop}%

\bibitem[{Note3()}]{Note3}
\BibitemOpen
\bibinfo {note} {{Note that the contemporary EMPIC algorithms such as geometric electromagnetic particle-n-cell algorithm or GEMPIC \cite{kraus2017gempic} and conformal EMPIC algorithms \cite{wang2016three} involves the ``grid to particle" and ``particle to grid" interpolation stages along with the particle pusher stage. Our DMD-EMPIC framework can be applied to these algorithms as well replacing the particle stages with DMD model of current density.}}
\BibitemShut {Stop}

\bibitem[{Note4()}]{Note4}
\BibitemOpen
\bibinfo {note} {{Note that there is a typo in section ``Computational complexity" of \cite{NAYAK2021110671}. The time complexity of truncated SVD of $N$ by $l$ matrix should be $\mathcal{O}(l^2N)$, not $\mathcal{O}(lN^2)$.}}
\BibitemShut {Stop}

\bibitem [{\citenamefont {Beaverstock}\ \emph {et~al.}(2015)\citenamefont
  {Beaverstock}, \citenamefont {Friswell}, \citenamefont {Adhikari},
  \citenamefont {Richardson},\ and\ \citenamefont
  {Du~Bois}}]{beaverstock2015automatic}%
  \BibitemOpen
  \bibfield  {author} {\bibinfo {author} {\bibfnamefont {C.}~\bibnamefont
  {Beaverstock}}, \bibinfo {author} {\bibfnamefont {M.}~\bibnamefont
  {Friswell}}, \bibinfo {author} {\bibfnamefont {S.}~\bibnamefont {Adhikari}},
  \bibinfo {author} {\bibfnamefont {T.}~\bibnamefont {Richardson}},\ and\
  \bibinfo {author} {\bibfnamefont {J.}~\bibnamefont {Du~Bois}},\ }\bibfield
  {title} {\bibinfo {title} {Automatic mode tracking for flight dynamic
  analysis using a spanning algorithm},\ }\href@noop {} {\bibfield  {journal}
  {\bibinfo  {journal} {Aerospace Science and Technology}\ }\textbf {\bibinfo
  {volume} {47}},\ \bibinfo {pages} {54} (\bibinfo {year} {2015})}\BibitemShut
  {NoStop}%
\bibitem [{\citenamefont {Hemati}\ \emph {et~al.}(2014)\citenamefont {Hemati},
  \citenamefont {Williams},\ and\ \citenamefont {Rowley}}]{hemati2014dynamic}%
  \BibitemOpen
  \bibfield  {author} {\bibinfo {author} {\bibfnamefont {M.~S.}\ \bibnamefont
  {Hemati}}, \bibinfo {author} {\bibfnamefont {M.~O.}\ \bibnamefont
  {Williams}},\ and\ \bibinfo {author} {\bibfnamefont {C.~W.}\ \bibnamefont
  {Rowley}},\ }\bibfield  {title} {\bibinfo {title} {Dynamic mode decomposition
  for large and streaming datasets},\ }\href@noop {} {\bibfield  {journal}
  {\bibinfo  {journal} {Physics of Fluids}\ }\textbf {\bibinfo {volume} {26}},\
  \bibinfo {pages} {111701} (\bibinfo {year} {2014})}\BibitemShut {NoStop}%

\bibitem [{\citenamefont {Decyk}\ and\ \citenamefont {Singh}(2014)}]{decyk2014particle}
  \BibitemOpen
  \bibfield  {author} {\bibinfo {author} {\bibfnamefont {Viktor K.}\ \bibnamefont
  {Decyk}}\ and\ \bibinfo {author} {\bibfnamefont {Tajendra V.}\ \bibnamefont {Singh}},\
  }\bibfield  {title} {\bibinfo {title} {{Particle-in-cell algorithms for emerging computer architectures}},\ }\href
  {https://doi.org/10.1016/j.cpc.2013.10.024} {\bibfield  {journal}
  {\bibinfo  {journal} {Computer Physics Communications}\ }\textbf {\bibinfo {volume} {185}},\
  \bibinfo {pages} {708--719} (\bibinfo {year} {2014})}\BibitemShut {NoStop}%

\bibitem [{\citenamefont {Sayadi}\ and\ \citenamefont {Schmid}(2016)}]{sayadi2016parallel}
  \BibitemOpen
  \bibfield  {author} {\bibinfo {author} {\bibfnamefont {Taraneh}\ \bibnamefont
  {Sayadi}}\ and\ \bibinfo {author} {\bibfnamefont {Peter J.}\ \bibnamefont {Schmid}},\
  }\bibfield  {title} {\bibinfo {title} {{Parallel data-driven decomposition algorithm for large-scale datasets: with application to transitional boundary layers}},\ }\href
  {https://doi.org/10.1007/s00162-016-0398-9} {\bibfield  {journal}
  {\bibinfo  {journal} {Theoretical and Computational Fluid Dynamics}\ }\textbf {\bibinfo {volume} {30}},\
  \bibinfo {pages} {415--428} (\bibinfo {year} {2016})}\BibitemShut {NoStop}%

\bibitem [{\citenamefont {Maryada}\ and\ \citenamefont {Norris}(2022)}]{maryada2022reduced}
  \BibitemOpen
  \bibfield  {author} {\bibinfo {author} {\bibfnamefont {KR}\ \bibnamefont
  {Maryada}}\ and\ \bibinfo {author} {\bibfnamefont {SE}\ \bibnamefont {Norris}},\
  }\bibfield  {title} {\bibinfo {title} {{Reduced-communication parallel dynamic mode decomposition}},\ }\href
  {https://doi.org/10.1016/j.jocs.2022.101599} {\bibfield  {journal}
  {\bibinfo  {journal} {Journal of Computational Science}\ }\textbf {\bibinfo {volume} {61}},\
  \bibinfo {pages} {101599} (\bibinfo {year} {2022})}\BibitemShut {NoStop}%

\bibitem [{\citenamefont {Erichson}\ \emph {et~al.}(2019)\citenamefont {Erichson}, \citenamefont {Mathelin}, \citenamefont {Kutz},\ and\ \citenamefont {Brunton}}]{erichson2019randomized}
  \BibitemOpen
  \bibfield  {author} {\bibinfo {author} {\bibfnamefont {N. Benjamin}\ \bibnamefont
  {Erichson}}, \bibinfo {author} {\bibfnamefont {Lionel}\ \bibnamefont {Mathelin}}, \bibinfo {author} {\bibfnamefont {J. Nathan}\ \bibnamefont {Kutz}}, \ and\ \bibinfo {author} {\bibfnamefont {Steven L.}\ \bibnamefont {Brunton}},\
  }\bibfield  {title} {\bibinfo {title} {{Randomized dynamic mode decomposition}},\ }\href
  {https://doi.org/10.1137/18M1224847} {\bibfield  {journal}
  {\bibinfo  {journal} {SIAM Journal on Applied Dynamical Systems}\ }\textbf {\bibinfo {volume} {18}},\
  \bibinfo {pages} {1867--1891} (\bibinfo {year} {2019})}\BibitemShut {NoStop}%


\bibitem [{\citenamefont {Na}\ \emph {et~al.}(2017)\citenamefont {Na}, \citenamefont {Omelchenko}, \citenamefont {Moon}, \citenamefont {Borges},\ and\ \citenamefont {Teixeira}}]{na2017axisymmetric}
  \BibitemOpen
  \bibfield  {author} {\bibinfo {author} {\bibfnamefont {Dong-Yeop}\ \bibnamefont
  {Na}}, \bibinfo {author} {\bibfnamefont {Yuri A.}\ \bibnamefont {Omelchenko}}, \bibinfo {author} {\bibfnamefont {Haksu}\ \bibnamefont {Moon}}, \bibinfo {author} {\bibfnamefont {Ben-Hur V.}\ \bibnamefont {Borges}}, \ and\ \bibinfo {author} {\bibfnamefont {Fernando L.}\ \bibnamefont {Teixeira}},\
  }\bibfield  {title} {\bibinfo {title} {Axisymmetric charge-conservative electromagnetic particle simulation algorithm on unstructured grids: {Application} to microwave vacuum electronic devices},\ }\href
  {https://doi.org/10.1016/j.jcp.2017.06.034} {\bibfield  {journal}
  {\bibinfo  {journal} {J. Comput. Phys.}\ }\textbf {\bibinfo {volume} {346}},\
  \bibinfo {pages} {295--317} (\bibinfo {year} {2017})}\BibitemShut {NoStop}%

\bibitem [{\citenamefont {Wang}\ \emph {et~al.}(2016)\citenamefont {Wang, Yue}, \citenamefont {Wang, Jianguo}, \citenamefont {Chen, Zaigao}, \citenamefont {Cheng, Guoxin},\ and\ \citenamefont {Wang, Pan}}]{wang2016three}
  \BibitemOpen
  \bibfield  {author} {\bibinfo {author} {\bibfnamefont {Yue}\ \bibnamefont
  {Wang}}, \bibinfo {author} {\bibfnamefont {Jianguo}\ \bibnamefont {Wang}}, \bibinfo {author} {\bibfnamefont {Zaigao}\ \bibnamefont {Chen}}, \bibinfo {author} {\bibfnamefont {Guoxin}\ \bibnamefont {Cheng}}, \ and\ \bibinfo {author} {\bibfnamefont {Pan}\ \bibnamefont {Wang}},\
  }\bibfield  {title} {\bibinfo {title} {{Three-dimensional simple conformal symplectic particle-in-cell methods for simulations of high power microwave devices}},\ }\href
  {Add-DOI-URL-Here} {\bibfield  {journal}
  {\bibinfo  {journal} {Computer Physics Communications}\ }\textbf {\bibinfo {volume} {205}},\
  \bibinfo {pages} {1--12} (\bibinfo {year} {2016})}\BibitemShut {NoStop}
{
\bibitem [{\citenamefont {Hesthaven}\ \emph {et~al.}(2023)\citenamefont {Hesthaven, Jan}, \citenamefont {Pagliantini, Cecilia},\ and\ \citenamefont {Ripamonti, Nicol{\`o}}}]{hesthaven2023adaptive}
  \BibitemOpen
  \bibfield  {author} {\bibinfo {author} {\bibfnamefont {Jan}\ \bibnamefont
  {Hesthaven}}, \bibinfo {author} {\bibfnamefont {Cecilia}\ \bibnamefont {Pagliantini}}, \ and\ \bibinfo {author} {\bibfnamefont {Nicol{\`o}}\ \bibnamefont {Ripamonti}},\
  }\bibfield  {title} {\bibinfo {title} {{Adaptive symplectic model order reduction of parametric particle-based Vlasov--Poisson equation}},\ }\href
  {} {\bibfield  {journal}
  {\bibinfo  {journal} {Mathematics of Computation}\ }\textbf {\bibinfo {volume} {93}},\
  \bibinfo {pages} {1153--1202} (\bibinfo {year} {2024})}\BibitemShut {NoStop}
}

\end{thebibliography}

%

\end{document}